\documentclass[seceq]{ptptex}

\usepackage{graphicx}
\usepackage{wrapft}
\usepackage{wrapfig}
\usepackage{algorithm}
\usepackage{algorithmic}
\usepackage{listings}

\newcommand{\beq}{\begin{eqnarray}}   \newcommand{\eeq}{\end{eqnarray}}
\newcommand{\nn}{\nonumber}
\newcommand{\bra}{\langle}  \newcommand{\ket}{\rangle}
 
  \newcommand{\tr}{\mbox{tr}}
  
\def\calD{{\cal D}}

\newcommand{\be}{\begin{equation}}
\newcommand{\ee}{\end{equation}\noindent}
\newcommand{\bea}{\begin{eqnarray}}
\newcommand{\eea}{\end{eqnarray}}

\newcommand{\maprightb}[1]{\smash{\mathop{
\hbox to 1cm{\rightarrowfill}}\limits_{#1}}}
\newcommand{\bc}{\begin{center}}
\newcommand{\ec}{\end{center}}

\newcommand{\matTwo}{\left(\begin{array}{rr}}
\newcommand{\matThree}{\left(\begin{array}{rrr}}
\newcommand{\emat}{\end{array}\right )}
\newcommand{\detTwo}{\left|\begin{array}{rr}}
\newcommand{\detThree}{\left|\begin{array}{rrr}}
\newcommand{\edet}{\end{array}\right |}

\newcommand{\muit}{\frac{\mu_I}{T}}

\newcommand{\Nred}{N_{\rm red}}
\newcommand{\ZNc}{Z_{3}}

\newcommand{\Tpc}{T_{\rm pc}}    
\newcommand{\Csw}{C_{\rm SW}}    

\newcommand{\mps}{m_{\rm ps}} \newcommand{\mV}{m_{\rm V}}
\newcommand{\LambdaQCD}{\Lambda_{\rm QCD}}

\title{
Towards extremely dense matter on the lattice%
}

\author{
Keitaro \textsc{Nagata}$^1$,
Shinji  \textsc{Motoki}$^2$,
Yoshiyuki \textsc{Nakagawa}$^3$,
Atsushi \textsc{Nakamura}$^{1, 4}$,
and Takuya \textsc{Saito}$^5$%
\\
(XQCD-J Collaboration)
}

\inst{
$^1$ 
Research Institute for Information Science and Education, 
Hiroshima University, \\
Higashi-Hiroshima 739-8527 Japan. \\
$^2$ KEK, Tsukuba, Ibaraki 305-0801, Japan. \\
$^3$ Graduate School of Science and Technology, Niigata University,\\
Niigata 950-2181, Japan. \\
$^4$ Graduate School of Sciences, Kyushu University, Fukuoka 812-8581, Japan.
\\
$^5$ Integrated Information Center, Kochi University, Kochi, 780-8520, Japan.
}

\abst{
QCD is expected to have a rich phase structure. 
It is empirically known to be difficult to access low temperature and nonzero 
chemical potential $\mu$ regions in lattice QCD simulations. 
We address this issue in a lattice QCD with the use of a dimensional reduction 
formula of the fermion determinant. 

We investigate spectral properties of a reduced matrix of the reduction formula. 
Lattice simulations with different lattice sizes show that the eigenvalues of the 
reduced matrix follow a scaling law for the temporal size $N_t$. 
The properties of the fermion determinant are examined using the reduction formula.
We find that as a consequence of the $N_t$ scaling law, the fermion determinant 
becomes insensitive to $\mu$ as $T$ decreases, and 
$\mu$-independent at $T=0$ for $\mu<m_\pi/2$.

The $N_t$ scaling law provides two types of the low temperature limit of the 
fermion determinant: (i) for low density and (ii) for high-density. 
The fermion determinant becomes real and the theory is free from the sign problem
in both cases.
In case of (ii), QCD approaches to a theory, where quarks interact only in 
spatial directions, and gluons interact via the ordinary Yang-Mills action. 
The partition function becomes exactly $Z_3$ invariant even in the presence of dynamical 
quarks because of the absence of the temporal interaction of quarks. 

The reduction formula is also applied to the canonical formalism and Lee-Yang 
zero theorem. 
We find characteristic temperature dependences of the canonical distribution and of 
Lee-Yang zero trajectory. 
Using an assumption on the canonical partition function, we discuss physical 
meaning of those temperature dependences and show that the change of the canonical 
distribution and Lee-Yang zero trajectory are related to the existence/absence 
of $\mu$-induced phase transitions.

}

\date{\today} 

\begin{document}
\maketitle
\tableofcontents
\section{Introduction}


QCD, first principle of strong interaction, has two important features; 
confinement and chiral symmetry breaking. 
They change its nature depending on circumstances, e.g., temperature 
($T$) and quark chemical potential ($\mu$), which leads to rich structure in the 
QCD phase diagram~\cite{Stephanov:2007fk,Fukushima:2010bq,Ohnishi:2011aa}. 
Hadrons and nuclei are formed at ordinary temperatures and chemical potentials. 
The quark-gluon plasma (QGP) is formed at high $T$, which is expected to be created 
in the early universe and also in experiments,  RHIC, LHC, etc. 
Several possibilities have been proposed for high density states of matter, which 
are realized in the core of neutron stars. 

Finite temperature properties of QCD have been uncovered by lattice QCD 
simulations~\cite{Aoki:2006we,Borsanyi:2010bp,Borsanyi:2010cj,Bazavov:2012ty,Borsanyi:2011sw,Umeda:2012er}, 
which is a computational approach implemented with Monte Carlo methods. 
For instance, 
recent simulations on finer lattices show that the deconfinement transition
is crossover~\cite{Aoki:2006we}, and occurs at $\Tpc=150-170$ MeV depending
on observables~\cite{Borsanyi:2010bp,Borsanyi:2010cj,Bazavov:2012ty}.

In contrast, the properties of QCD at nonzero $\mu$ have been difficult to 
study because of the notorious sign problem~\cite{deForcrand:2010ys}. 
The sign problem spoils the importance sampling at $\mu(\neq 0)$, and makes it 
unfeasible to generate gauge ensemble. 
Several approaches have been developed to avoid the sign problem and 
study nonzero-$\mu$ systems in lattice QCD simulations, see e.g., 
Refs.~\cite{Fodor:2001au,Muroya:2003qs,Schmidt:2006us,Philipsen:2005mj,Philipsen:2007rj}. 
The consistency between different approaches are shown for small $\mu/T$. 
For instance, it was shown~\cite{Kratochvila:2005mk,Fodor:2009ax} that 
several approaches of finite density lattice studies are consistent up to $\mu/T\sim 1$, 
for staggered fermions. For Wilson fermions, we showed the consistency of the 
Taylor expansion, multi-parameter reweighting (MPR) and imaginary chemical 
potential methods~\cite{Nagata:2012pc}. 
For status of lattice studies of the QCD phase diagram, see e.g, 
Refs.~\cite{Stephanov:2007fk,Philipsen:2008gf}. 
Extensive studies have been made for the finite density properties of QCD
in particular for the location of the QCD critical end point (CEP). 

The QCD phase diagram contains rich physics also 
at low temperatures~\cite{Fukushima:2010bq,Stephanov:2007fk,Ohnishi:2011aa}. 
One expectation is that towards large density region, QCD changes its state from 
hadron gas, nuclear liquid and color superconducting state. 
In addition, the discovery of a pulser with twice solar mass~\cite{Demorest:2010bx} 
calls the reliable equation of state based on QCD at low temperature and high density. 
The study of the low temperature and finite 
density states would be an important challenge for lattice QCD simulations. 

There are expectations of the existence of sign free regions at low temperatures. 
The finite density lattice QCD at low temperatures had been extensively 
investigated in the Glasgow method, see e.g. 
Ref.~\cite{Barbour:1999mc,Barbour:1997ej,Barbour:1997bh,Barbour:1991vs}, where it was shown the onset of the 
baryon density is given by $m_\pi/2$. 
Recently, the phase at low temperatures was investigated in a lattice 
simulation~\cite{Takeda:2011vd}.
It is empirically expected that the fermion determinant is 
independent of $\mu$ at $T=0$ up to $\mu<M_N/3$. 
However, the inclusion of chemical potentials can generally change the eigen 
spectrum of the Dirac operator and hadron spectrum. 
The $\mu$-independence at $T=0$ was, therefore, considered as a small puzzle 
and called ``Silver Blaze'' problem~\cite{Cohen:2003kd}. 
Cohen showed the $\mu$-independence at $T=0$ for isospin chemical 
potential $\mu_I$ case and Adams~\cite{Adams:2004yy} for quark chemical potential case. 
The orbifold equivalence for baryon chemical potential and isospin chemical potential 
also suggested~\cite{Cherman:2010jj,Hanada:2011ju}
that the phase of the fermion determinant is small up to $m_\pi/2$. 
The same result was also obtained by Splittorff and Verbaarschot 
by using the chiral perturbation theory (ChPT). 
On the other hand, the mechanism to extend the Silver Blaze region 
from $\mu=m_\pi/2$ to $\mu=M_N/3$ is not yet understood.
A sign-free region was also suggested for high density.
Hong derived a high density effective theory (HDET) of low energy modes 
in dense QCD $\mu \gg \LambdaQCD$~\cite{Hong:1999ru}. 
The positivity of the fermion action in HDET was later discussed by Hong and 
Hsu~\cite{Hong:2003zq,Hong:2002nn}. 
HDET was also studied in e.g., Refs.~\cite{Dougall:2007fp,Hands:2007by}. 

In this paper, we study the property of QCD at low temperature and 
finite density, using lattice QCD.
Particularly, the properties of the fermion determinant at low temperature 
and nonzero $\mu$ will be clarified.
The key idea to tackle this issue is a reduction formula for the fermion 
determinant. 
The formula is obtained by performing the temporal part of the determinant 
of a fermion matrix. 
The technique is analogous to the transfer matrix method in condensed matter 
physics, and provides several advantages in lattice simulations of QCD at 
finite density. 
The formula produces a matrix with a reduced dimension, which is called the 
reduced matrix. 
We will see that the reduced matrix plays an important role in the properties 
of the fermion determinant at low temperature and nonzero $\mu$. 

Some issues are discussed in the present paper.
In \S~\ref{Sec:secred}, we focus on the reduced matrix, and discuss 
its derivation, interpretation and spectral property, where 
the connection with important dynamics of QCD  is considered. 
In particular,  we will find a clear indication from 
lattice QCD simulations that the eigenvalues of the reduced matrix 
follow a scaling law of the temporal size $N_t$.
This section is partly a review of the reduction formula.

In \S~\ref{Sec:2012Mar04sec2}, the property of the fermion determinant is 
investigated in detail. 
We show that the quark determinant is insensitive to $\mu$ at low temperatures 
for small $\mu$. Using the relation between the eigenvalues of the reduced matrix
and pion mass, we will show that the $\mu$-independence continues up to $\mu=m_\pi/2$.
We will see that the $N_t$ scaling law leads to two expression of the low temperature 
limit of the fermion determinant both for low density and for high density. 
The low density expression corresponds to the $\mu$-independence at small $\mu$. 
Hence the $\mu$-independence at $T=0$ for $\mu<m_\pi/2$ is a consequence of the $
N_t$-scaling law of the eigenvalues of the reduced matrix.
The fermion determinant becomes real and the theory is free from the sign problem
both cases.

In the case of high density and low temperature limit, 
QCD approaches to a theory, where quarks interact only in spatial direction, 
and the gluon action is given by the ordinary Yang-Mills action. 
The corresponding partition function is exactly $Z_3$ invariant even in the 
presence of dynamical quarks because of the absence of the temporal interaction 
of quarks. Property, possible application and numerical feasibility are 
discussed. 

In \S~\ref{Sec:canonical}, the reduction formula is applied to 
the canonical formalism and Lee-Yang zero theorem. 
These methods are often used in analysis to find CEP and 
first order phase transition line at low temperatures.
The sign problem causes difficulties in these methods
~\cite{Ejiri:2005ts,Kratochvila:2005mk}.
For instance, Ejiri pointed out that the sign problem causes 
a fictitious signal for the finite size scaling analysis of 
the Lee-Yang zero closest to the positive real axis, which makes it 
difficult to distinguish crossover and first order transition.

We consider the fugacity expansion of the grand partition function, 
where $Z_{GC}(\mu)$ is expressed via the canonical partition functions 
$Z_n$ with quark number $n$. 
We show that both the $n$-dependence of $Z_n$ and the Lee-Yang zero trajectory
show characteristic change from high to low temperatures in a 
correlated manner. These behavior can be used to qualitatively
distinguish the crossover and first order phase transition.

\newcommand{\rforml}{reduction formula}

\section{Reduction Formula}
\label{Sec:secred}
In this section, we study the reduction formula of the fermion determinant. 
A basic idea of the reduction formula is to analytically carry out the 
temporal part of the fermion determinant, which reduces the rank of the determinant.
Since the chemical potential $\mu$ couples only to the temporal link 
variables, the reduction formula rearranges the fermion determinant 
regarding $\mu$, which  enables us to separate out $\mu$ from link variables. 
These characters offer some advantages in finite density simulations.

The reduction formula is not only a technical tool but also has
physical interpretation. 
The formula produces a matrix with a reduced dimension, which we refer to as the 
reduced matrix. Its eigenvalues characterize the $\mu$-dependence of the fermion determinant. 
The reduced matrix physically corresponds to a transfer matrix or a generalized 
Polyakov line, and its eigenvalues are related to a free energy of dynamical quarks.

In this section, the spectral properties of the reduced matrix are investigated 
in detail. 
The formulation is given in \S~\ref{sec:reduction_formulation}. 
The interpretation and overview of the reduced matrix are presented in 
\S~\ref{sec:red_physint}. 
Simulation setup is explained in \S~\ref{subsec3a}. 
Some important and fundamental properties of the reduced matrix 
are shown in \S~\ref{sec:red_pair}.
In \S~\ref{Sec:2012Mar04sec1}, we will find a clear indication of the 
$N_t$-scaling law of the eigenvalues of the reduced matrix. 
In \S~\ref{sec:red_gap}, we discuss the relation between the eigenvalues of the 
reduced matrix and pion mass.

\subsection{Formulation}
\label{sec:reduction_formulation}

The reduction formula was derived by Gibbs~\cite{Gibbs:1986hi}, for staggered fermions. 
Alternative derivation was found by Hasenfratz and Toussaint~\cite{Hasenfratz:1991ax}.
Later it has been applied to various studies, e.g. Glasgow 
method~\cite{Barbour:1999mc,Barbour:1997ej,Barbour:1997bh,Barbour:1991vs,Crompton:2001ws}, multi-parameter reweighting~\cite{Fodor:2001pe}, 
canonical formalism~\cite{deForcrand:2006ec,Kratochvila:2005mk,Kratochvila:2004wz,Kratochvila:2006jx}. 
The derivation for Wilson fermions needs a delicate treatment because the 
Wilson terms $(r\pm \gamma_4)$ are singular for $r=1$. 
Adams derived the formula  using zeta-regularization~\cite{Adams:2003rm}. 
Borici derived the formula by introducing a permutation matrix~\cite{Borici:2004bq}. 
Borici's method was further studied by Alexandru and Wenger~\cite{Alexandru:2010yb}
and two of present authors~\cite{Nagata:2010xi}. 
The formula for the Wilson fermion was later applied to studies of 
CEP~\cite{Li:2011ee} and thermodynamical quantities~\cite{Nagata:2012pc}.
The reduction formula in continuum theory was found by Adams 
using zeta-regularization, which was used to show the $\mu$-independence 
of the fermion determinant at $T=0$~\cite{Adams:2004yy}. 

Throughout the present paper, we consider the Wilson fermions. 
For the derivation for the staggered fermions, see e.g. Ref.~\cite{Muroya:2003qs}.
The grand partition function of $N_f$-flavor QCD at a temperature $T$ and quark chemical 
potential $\mu$ is given by 
\begin{align}
Z_{GC}(\mu, T) = \int \calD U \;[\det \Delta(\mu)]^{N_f} e^{-S_G}.
\label{Sep122011eq1}
\end{align}
In our simulations, a renormalization group(RG)-improved action is used for $S_G$.
The clover-improved Wilson fermion is employed for the fermion action
\begin{align}
\Delta(\mu) =  \delta_{x, x^\prime} &-\kappa \sum_{i=1}^{3} \left[
(r-\gamma_i) U_i(x) \delta_{x^\prime, x+\hat{i}} 
+ (r+\gamma_i) U_i^\dagger(x^\prime) \delta_{x^\prime, x-\hat{i}}\right] \nonumber \\
 &-\kappa \left[ e^{+\mu a} (r-\gamma_4) U_4(x) \delta_{x^\prime, x+\hat{4}}
+e^{-\mu a} (r+\gamma_4) U^\dagger_4(x^\prime) \delta_{x^\prime, x-\hat{4}}\right] \nonumber \\
&- \kappa  C_{SW} \delta_{x, x^\prime}  \sum_{\mu \le \nu} \sigma_{\mu\nu} 
F_{\mu\nu},
\label{Jul202011eq1}
\end{align}
where  $\kappa$ and $r$ are the hopping parameter and Wilson parameter, respectively.
The quark chemical potential $\mu$ couples to the fourth current 
$\bar{\psi}\gamma_4\psi$ in Euclidean path-integral, then a temporal hop 
accompanies a factor $e^{\pm \mu a}$~\cite{Hasenfratz:1983ba}.
$C_{SW}$ is a coefficient of the clover term. 
Note that the clover term is diagonal in the coordinate space, 
hence it does not cause any problem in the derivation of the reduction 
formula. 

The fermion determinant satisfies a $\gamma_5$-hermiticity relation
\begin{align}
(\det \Delta (\mu))^* = \det \Delta (-\mu^*).
\label{Eq:2011Dec24eq1}
\end{align}
It ensures that $\det \Delta(0)\in \mathbb{R}$. 
In the presence of nonzero real $\mu$, the fermion determinant is in general complex, 
which causes the sign problem. The $\det \Delta$ is real if $\mu$ is pure imaginary. 
This property has been used in the study of the QCD phase diagram in 
lattice QCD simulations
~\cite{Roberge:1986mm,deForcrand:2002ci,D'Elia:2009qz,D'Elia:2009tm,deForcrand:2010he,D'Elia:2002gd,D'Elia:2004at,D'Elia:2007ke,Cea:2012ev,Cea:2010md,Cea:2009ba,Cea:2007vt,Wu:2006su,Nagata:2011yf} and also in phenomenological studies, see e.g. Refs.~\cite{Kouno:2009bm,Morita:2011jva}. 

For the preparation of the reduction formula, we define 
the spatial part of the Wilson fermion matrix as $B$,
\begin{subequations}
\begin{align}
B =  \delta_{x, x^\prime} &-\kappa \sum_{i=1}^{3} \left[
(r-\gamma_i) U_i(x) \delta_{x^\prime, x+\hat{i}} 
+ (r+\gamma_i) U_i^\dagger(x^\prime) \delta_{x^\prime, x-\hat{i}}\right] \nonumber \\
&- \kappa  C_{SW} \delta_{x, x^\prime}  \sum_{\mu \le \nu} \sigma_{\mu\nu} 
F_{\mu\nu},
\end{align}
and two block-matrices
\begin{align}
\alpha_i &= \alpha^{ab, \mu\nu}(\vec{x}, \vec{y}, t_i) \nn \\
         &= c_- B^{ab, \mu\sigma}(\vec{x}, \vec{y}, t_i) \; r_{-}^{\sigma\nu} 
         -2  c_+  \kappa \; r_{+}^{\mu\nu} \delta^{ab} \delta(\vec{x}-\vec{y}), 
\\
\beta_i &= \beta^{ab,\mu\nu} (\vec{x}, \vec{y}, t_i) \nn \\ 
        &= c_+ B^{ac,\mu\sigma}(\vec{x}, \vec{y}, t_i)\; r_{+}^{\sigma\nu} 
U_4^{cb}(\vec{y}, t_i) -2 c_- \kappa \; r_{-}^{\mu\nu} \delta(\vec{x}-\vec{y}) 
U_4^{ab}(\vec{y}, t_i).
\label{Eq:2012Feb21eq1}
\end{align}%
\end{subequations}%
Here $r_\pm = (r \pm \gamma_4)/2$, which are projection operators in case of $r=1$. 
This property is used in the derivation of the formula.
$c_{\pm}$ are arbitrary scalar except for zero, which is introduced in 
so called permutation matrix $P = (c_- r_- + c_+ r_+ V e^{\mu a})$.
$\alpha_i$ describes a spatial hopping on a fixed time plane $t_i$, 
while $\beta_i$ describes a spatial hopping at $t_i$ and a temporal hopping to 
the next time slice. They are independent of $\mu$. 

A temporal matrix representation of the Wilson fermion matrix 
contains block-elements proportional to $r_\pm$, which are singular.
This fictitious singularity is avoided by calculating 
$\det (P\Delta)$~\cite{Borici:2004bq}. 
Then the determinant is obtained from $\det (P \Delta)/\det P$. 
Using the block matrices, the reduction formula is given by~\cite{Nagata:2010xi}
\begin{subequations}
\begin{align}
\det \Delta(\mu) & = (c_+ c_- )^{-N/2} \xi^{-\Nred/2}  C_0 \det\left( \xi +  Q \right), 
\label{May1010eq2}
\end{align}%
with 
\begin{align}%
Q   &= (\alpha_1^{-1} \beta_1) \cdots (\alpha_{N_t}^{-1} \beta_{N_t}), 
\label{Eq:2012Jan01eq3}\\
C_0 &= \left(\prod_{i = 1}^{N_t} \det(\alpha_i ) \right),
\label{Eq:2012Jan01eq4}%
\end{align}%
\label{Eq:2012Jan01eq5}%
\end{subequations}%
where $\xi=\exp(-\mu/T)$ is the fugacity. 
$N=4N_c N_s^3 N_t$ and $\Nred = N/N_t$ are the dimension of $\Delta$ and $Q$, 
respectively. The rank of $\alpha_i$ and $Q$ is reduced by $1/N_t$ compared to $N$. 
Instead, $Q$ is a dense matrix.
$Q$ and $C_0$ are independent of $\mu$.
Hence, the reduction formula separates the chemical potential from the gauge parts. 

To obtain $\det \Delta$, we need to evaluate $\det (Q+\xi)$. 
Here we calculate the eigenvalues $\lambda$ for $|Q-\lambda I|=0$. 
Once we obtain $\lambda$, the quark determinant is the analytic function of $\mu$. 
Then, values of $\det \Delta(\mu)$ are obtained for arbitrary values of $\mu$, 
which is an advantage, although the eigen problem requires large numerical cost. 
Alternative methods such as LU decomposition of $Q+\xi$ are available instead of solving 
the eigenvalue problem for $Q$. 
In this case, we need to perform the LU decomposition for each value of $\mu$. 

Once having the eigenvalues of $Q$, we obtain
\begin{subequations}
\begin{align}
\det \Delta (\mu) &=  C_0  \xi^{-\Nred/2}\prod_{n=1}^{\Nred} (\lambda_n + \xi)
\label{Nov292011eq1}\\
              &= C_0 \sum_{n=0}^{\Nred}c_n \xi^{n-\Nred/2}
              = C_0 \sum_{n=-\Nred/2}^{\Nred/2} c_n \xi^n,
\label{Jan1511eq1} 
\end{align}
\label{Jan1511eq3}%
\end{subequations}%
where we set $c_\pm =1$ for simplicity. 
In the second line, we redefine the index $n$ from the first expression
to the second one. 
The fermion determinant is described in two forms: 
a product form Eq.~(\ref{Nov292011eq1}), and a summation 
form Eq.~(\ref{Jan1511eq1}). The second one is nothing but a fugacity 
or winding number expansion of the quark determinant, 
where fugacity coefficients $c_n$ are polynomials of $\lambda_n$~\cite{Nagata:2010xi}. 
The consistency between the reduction formula for the Wilson fermions 
and staggered fermions can be found by considering the fugacity expansion form. 
The fugacity expansion of the fermion determinant is also obtained in 
other approaches, such as a domain decomposition technique~\cite{Danzer:2008xs}.

Here, we summarize some features of the reduction formula. 
The formula offers several advantages; 
\begin{itemize}
\item The dimension of the determinant is reduced by $1/N_t$, which 
reduces the numerical cost of the direct calculation of the fermion 
determinant. 
\item $\det \Delta(\mu)$ is the analytic function of $\mu$, which 
suppresses the numerical cost to evaluate values of $\det \Delta(\mu)$ 
for many values of $\mu$ and also provides an insight into 
the $\mu$-dependence of $\det \Delta(\mu)$. 
\item Increase of the numerical cost for low temperature(large $N_t$)
is also suppressed.
\end{itemize}
On the other hand, it has an applicable range; 
\begin{itemize}
\item A direct method is used for the eigenproblem of the reduced matrix. 
This requires large numerical cost. 
In particular, it is difficult to apply the reduction formula to 
large lattice size because of the limitation of a memory. 
\end{itemize}

\vspace{5mm}

\subsection{Physical interpretation}
\label{sec:red_physint}

\begin{figure}[htbp]
\begin{center}
\includegraphics[width=0.75\linewidth]{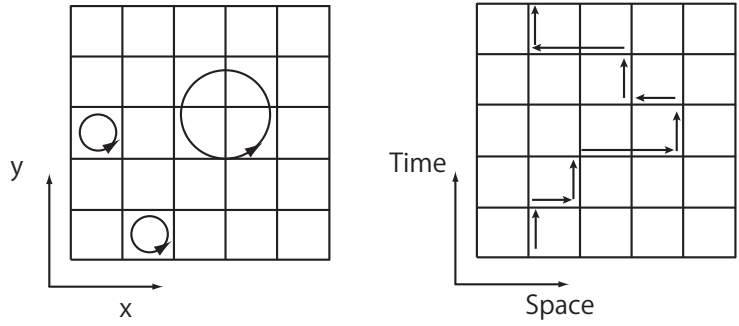}
\begin{minipage}{0.75\linewidth}
\caption{Schematic figures for $C_0$ and $Q$. 
The left panel describes spatial loops on a plane $t=t_i$. 
$C_0$ consists of those spatial loops. 
The right panel denotes $Q$, which describes propagations of quarks 
from $t=t_1$ to $t=t_{N_t}$. 
}\label{CanJun3010fig1}
\end{minipage}
\end{center}
\end{figure}
\begin{figure}[htbp]
\begin{center}
\includegraphics[width=0.60\linewidth]{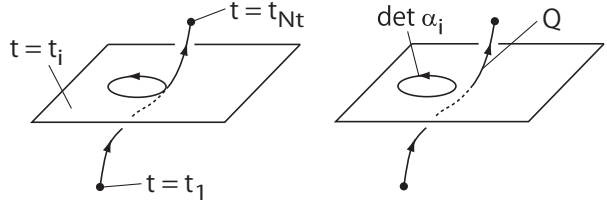}
\begin{minipage}{0.75\linewidth}
\caption{Schematic figures for the reduction procedure, where
the spatial loops $C_0$ are separated from $Q$.
}\label{CanJul0510fig1}
\end{minipage}
\end{center}
\end{figure}
In the reduced matrix $Q$, block elements $(\alpha_i^{-1} \beta_i)$ describe 
propagations of a quark from $t=t_i$ to $t=t_{i+1}$. 
$Q$ is the product of the block elements, and therefore it 
describes propagations from the initial $t=t_1$ to final time $t=t_{N_t}$ 
(depicted in the right panel of Fig.~\ref{CanJun3010fig1}). 
Accordingly, the reduction formula is analogous to the transfer matrix method, 
where block elements is interpreted as transfer matrices
~\cite{Gibbs:1986hi,Borici:2004bq,Alexandru:2010yb}. 
The matrix $Q$ is also understood as a generalized Polyakov line. 
If we pick up the temporal links in $Q$, it is written as
$Q = \cdots U_4(t_1) \cdots U_4(t_2) \cdots U_4(t_{N_t})$.
If the trace is taken, $\tr Q$ is similar to the Polyakov loop. 
It is known that the Polyakov loop describes the free energy of static quarks 
$\bra P \ket \sim e^{-F/T}$ with $P=\tr \prod_{i=1}^{N_t} U_4(t_i)$. 
However, $Q$ contains spatial hopping terms denoted as the dots, 
where quarks are not static. 
This suggests that the reduced matrix is related to the free energy of dynamical quarks. 

$C_0$ consists of the spatial loops, where each loop is in a fixed time slice, 
which accompanies no temporal hops, see the left panel of Fig.~\ref{CanJun3010fig1}
and Fig.~\ref{CanJul0510fig1}. Thus the spatial quark loops $C_0$ are separated 
from temporal hoppings and do not contribute to the $\mu$-dependence of the 
fermion determinant. 

Here, we summarize properties of the reduction formula. 
From $\gamma_5$-hermiticity, it follows ; 
\begin{subequations}
\begin{itemize}
\item $C_0$ is real. 
\item Two eigenvalues form a pair 
\begin{align}
\lambda_n, 1/\lambda_n^*.
\label{Eq:2012Jan01eq2}
\end{align}
This is because of a symmetry of $Q$, see e.g., Ref.~\cite{Alexandru:2010yb}. 
We give a simple proof of this relation in Appendix.~\ref{Sec:2012Mar03sec1}.
\item The coefficients of the positive and negative winding terms satisfy
\begin{align}
c_{-n}=c_n^*.
\end{align}
\item The product of all the eigenvalues is unity (followed from $c_{-n}=c_n^*$);
\begin{align}
\det Q = \prod_{n=1}^{\Nred} \lambda_n = \lambda_1 \lambda_2 \cdots \lambda_{\Nred} = 1.
\label{Jan0711eq5}
\end{align}
\item For the case $c_\pm = 1$, it is straightforward to show 
\begin{align}
\det \alpha_i = \det \beta_i.
\end{align}
\item
From Eq.(\ref{Eq:2012Jan01eq2}) and Eq.(\ref{Jan0711eq5}), 
we can separate all eigenvalues into two sets ($\{\lambda_k | |\lambda_k|>1\}$ 
and $\{\lambda_k | |\lambda_k|<1\}$ ). Then, the product of the 
normalized eigenvalues in each set is $\pm 1$,
\begin{align}
\prod_{k=1}^{\Nred/2} \frac{\lambda_k}{|\lambda_k|} = \pm 1. 
\end{align}
For later convenience, we introduce a notation $\lambda_L$ for 
$|\lambda|>1$ and $\lambda_S$ for $|\lambda|<1$. 

\end{itemize}
\end{subequations}

These properties are satisfied for configuration by configuration, 
independent of temperature. 

Although it is non-trivial to clarify the correspondence between the reduced matrix 
$Q$ and the original matrix $\Delta$, properties of QCD can be seen in the 
spectral property of $Q$; e.g.,
\begin{description}
\item[Confinement:] \hspace{0.5em} 
As we have mentioned, the matrix $Q$ is related to the Polyakov line. 
A correlation between the Polyakov loop and the eigenvalues of the reduced matrix 
was found in Ref.~\cite{Alexandru:2010yb}. 
We can find the confinement property is realized in the angular distribution 
of $\lambda$. 

\item[Chiral symmetry breaking :] \hspace{0.5em} 
It is known that the distribution of the eigenvalues form a gap near $|\lambda|=1$. 
Gibbs pointed out~\cite{Gibbs:1986hi} that the gap of the eigenvalue distribution 
is related to the pion mass. 
Hence, the behavior of the eigenvalues near the unit circle is related to 
the chiral symmetry breaking. 

\item[Hadron Spectroscopy:] \hspace{0.5em} 
Since the reduced matrix is related to the free energy of dynamical quarks, 
its eigenvalues are related to hadron spectrum.
Fodor, Szabo and T\'oth proposed to extract hadron masses from the 
eigenvalues of the reduced matrix based on thermodynamical 
approach~\cite{Fodor:2007ga}. 

\item[Low temperature behavior :] \hspace{1em} 
Because $Q$ is a product of $N_t$ block-matrices, then it is 
expected~\cite{Nagata:2010xi} that the eigenvalues of $Q$ follow a scaling law 
for the temporal lattice size $N_t$. 
The scaling law is useful to study the low $T$ behavior of the fermion determinant. 
For instance, we will see that the scaling law explains the $\mu$-independence 
of the fermion determinant at $T=0$ for $\mu<m_\pi/2$. 
\end{description}

\subsection{Simulation setup}
\label{subsec3a}
Before proceed to numerical simulations, simulation details are given here. 
Simulations were performed on four lattice setups; 
$(N_s^3\times N_t, m_{\rm PS}/m_{\rm V})$ = (i) $(8^3\times 4, 0.8)$ 
(ii) $(8^4, 0.8)$, (iii) $(10^3\times 4, 0.8)$ and (iv) $(8^3\times 4, 0.6)$.
The wide range of temperatures were covered by using (i) and (ii).
The setup (iii) and (iv) were used to investigate  
the finite size effect and quark mass dependence.
The detail of these simulations is as follows. 

\begin{itemize}
\item[(i)] We investigated $29$ values of $\beta$ in the interval 
$1.5\le \beta \le 2.4$, which covers the temperature range $T/\Tpc = 0.76 - 3$. 
The data were used in our previous studies on imaginary chemical 
potential approach~\cite{Nagata:2011yf} and thermodynamical 
quantities~\cite{Nagata:2012pc}.

\item[(ii)] This simulation was performed to examine lower 
temperature. We investigated $16$ values of $\beta$ in the interval 
$1.79 \le \beta \le 1.94$ corresponding to the temperature range 
$T/\Tpc=0.460-0.587$.

\item[(iii)] This setup was performed to study the finite size effects. 
We investigated $29$ values of $\beta$ in the interval $1.79\le \beta \le 1.94$, 
which covers the temperature range $T/\Tpc = 0.93 - 1.2$. 

\item[(iv)] This simulation was performed to investigate the quark mass dependence. 
We investigated four values of $\beta=1.6, 1.7, 1.95$ and $2.0$, 
which correspond to $T/\Tpc=0.86, 0.94, 1.48$ and $1.67$, respectively.

\end{itemize}

For all the four setup, the value of the hopping parameter $\kappa$ 
was determined for each $\beta$ by following lines of constant physics (LCP) in Ref.~\cite{Ejiri:2009hq}. 
The clover coefficient $\Csw$ was determined by using a result obtained 
in the one-loop perturbation theory : $\Csw = ( 1- 0.8412 \beta^{-1})^{-3/4}$. 
We also used the data in Ref.~\cite{Ejiri:2009hq} to obtain the values of the 
temperature from $\beta$. 

Gauge configurations were generated at $\mu=0$ with the hybrid Monte Carlo simulations.
The size of the molecular dynamics step was $\delta \tau =0.02$ for 
$m_{\rm PS}/m_{\rm V}=0.8$ and $\delta \tau = 0.01$ for $m_{\rm PS}/m_{\rm V}=0.6$ 
with unit length. 
The acceptance ratio was more than 90 \% for $N_s=8$ and 80 \% for $N_s=10$. 
HMC simulations were carried out for 11, 000 trajectories for each parameter set for 
$m_{\rm PS}/m_{\rm V}=0.8$, and 4, 000 trajectories $m_{\rm PS}/m_{\rm V}=0.6$.
It should be noted that the statistics are small for $m_{\rm PS}/m_{\rm V}=0.6$. 

The measurements were performed for each 20 HMC trajectories after thermalization. 
The eigenvalue calculations were performed for 400 configurations 
for heavy quark case, and 50 configurations for light quark case. 
Computational details of the eigenvalue calculation were 
explained in Ref.~\cite{Nagata:2012pc}.

\subsection{Pair nature, $\ZNc$ symmetry and temperature dependence}
\label{sec:red_pair}
Let us study the spectral properties of the eigenvalues of the reduced matrix. 
\begin{figure}[htpb] 
\includegraphics[width=5cm]{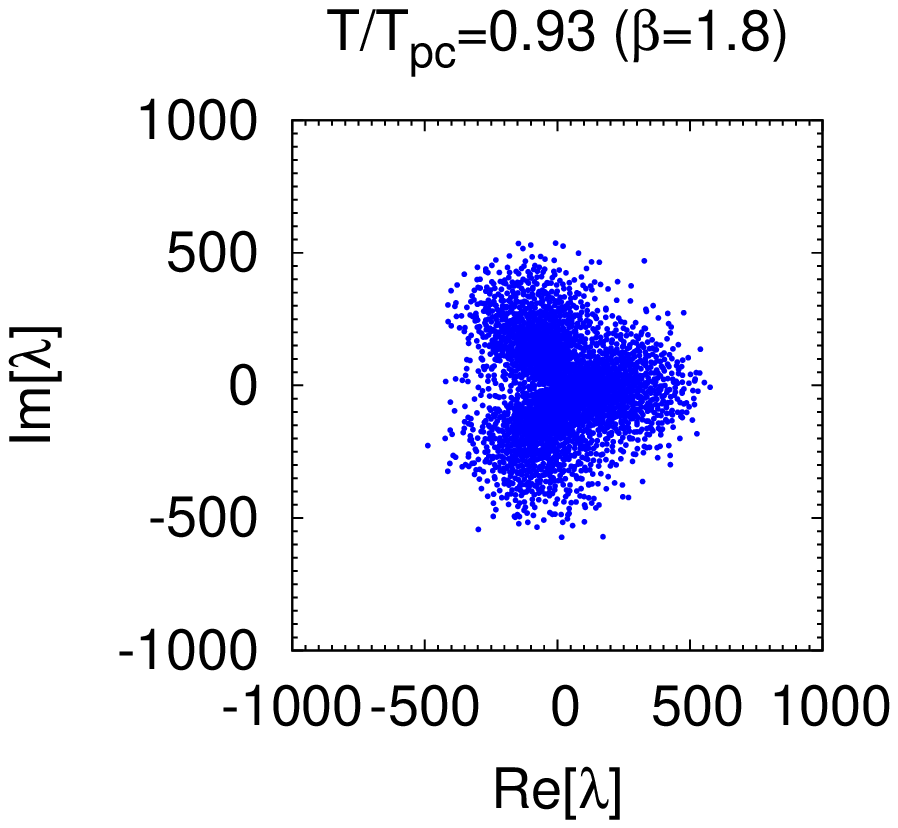}
\includegraphics[width=5cm]{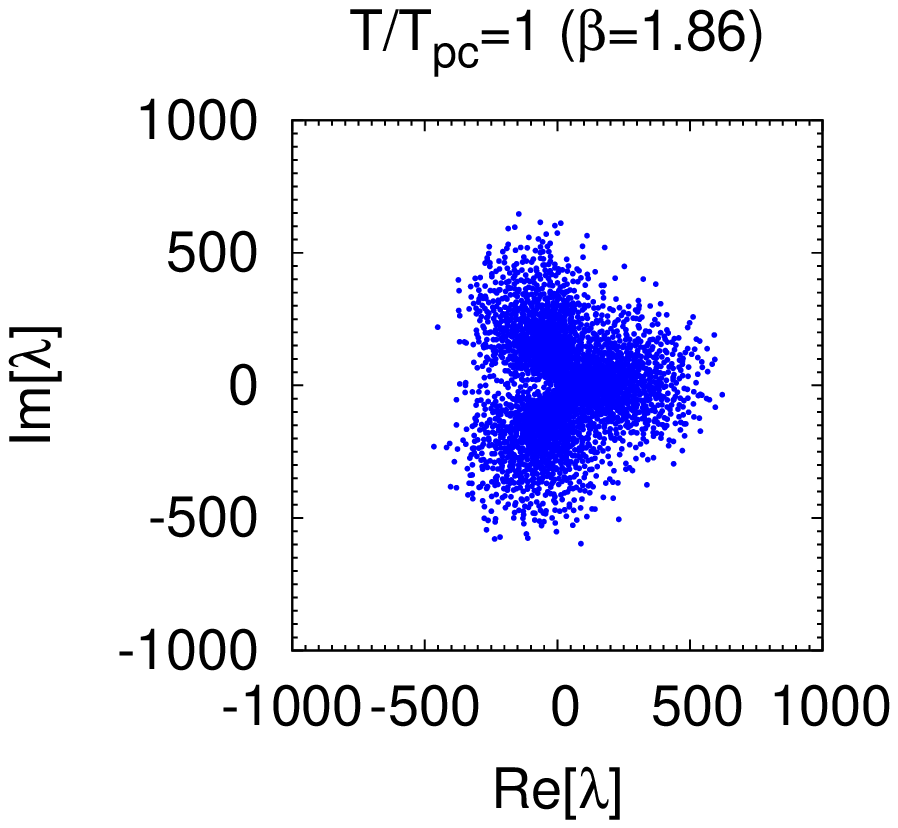}
\includegraphics[width=5cm]{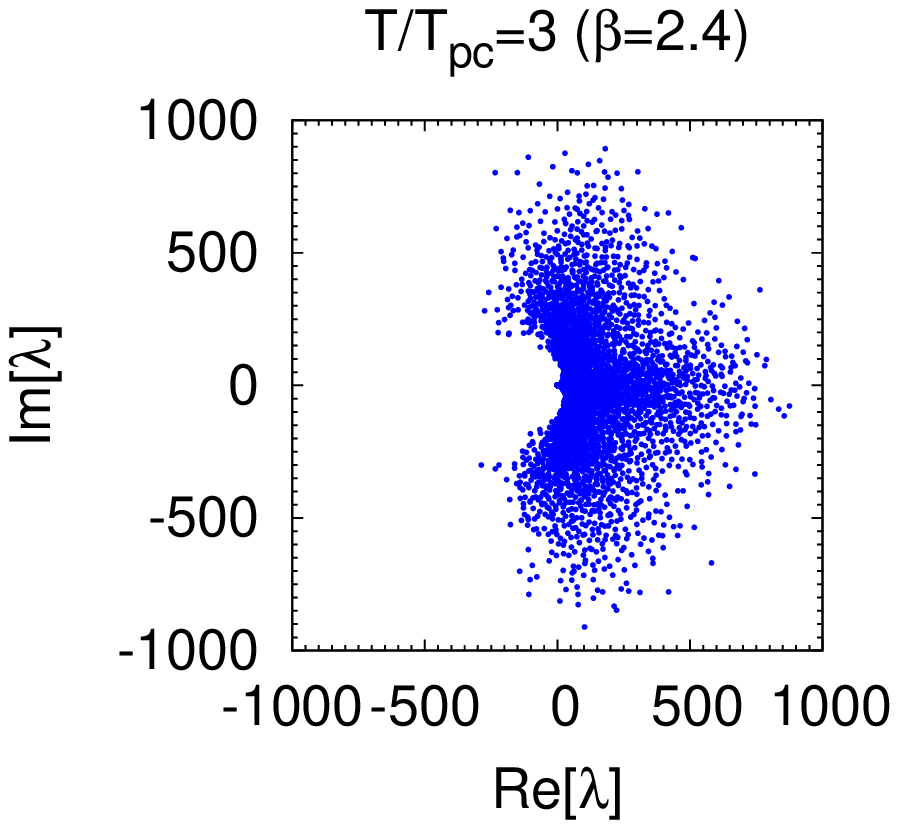}
\includegraphics[width=5cm]{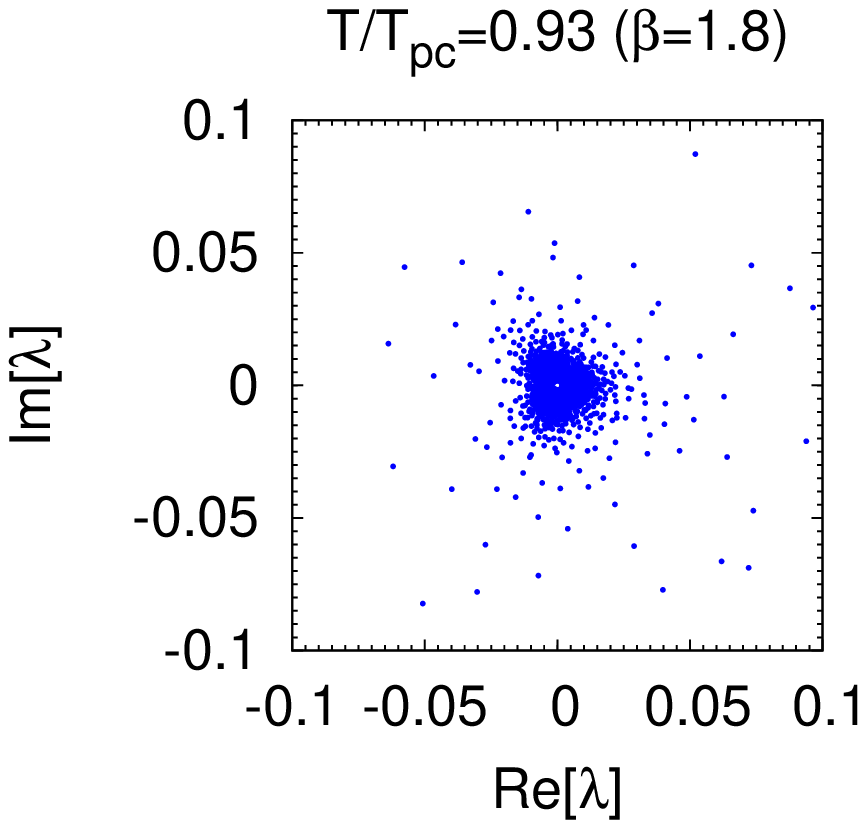}
\includegraphics[width=5cm]{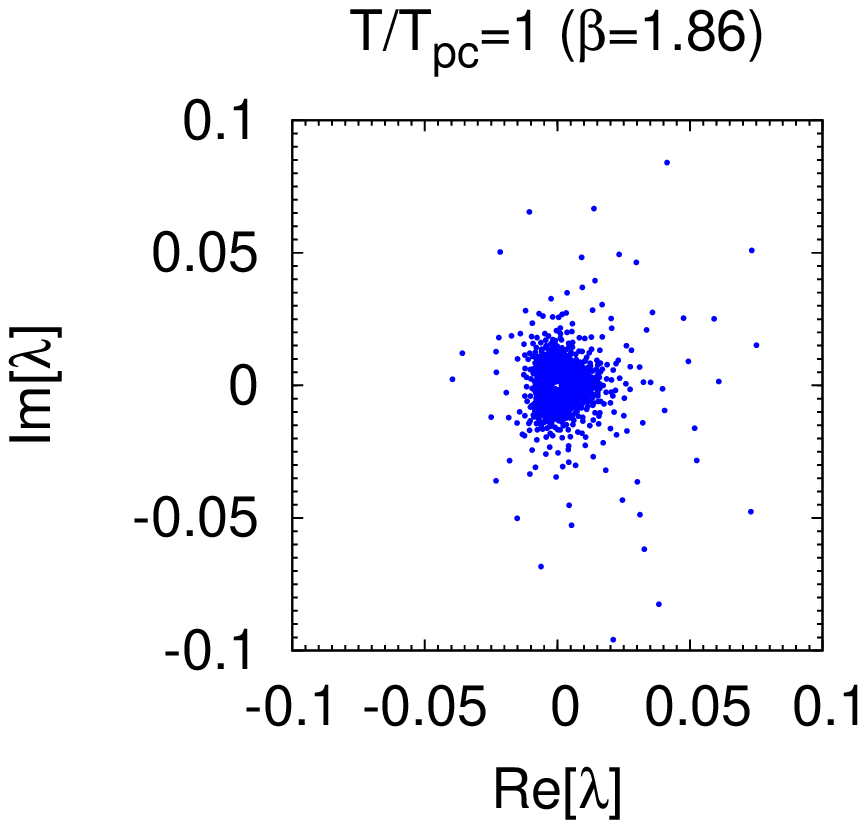}
\includegraphics[width=5cm]{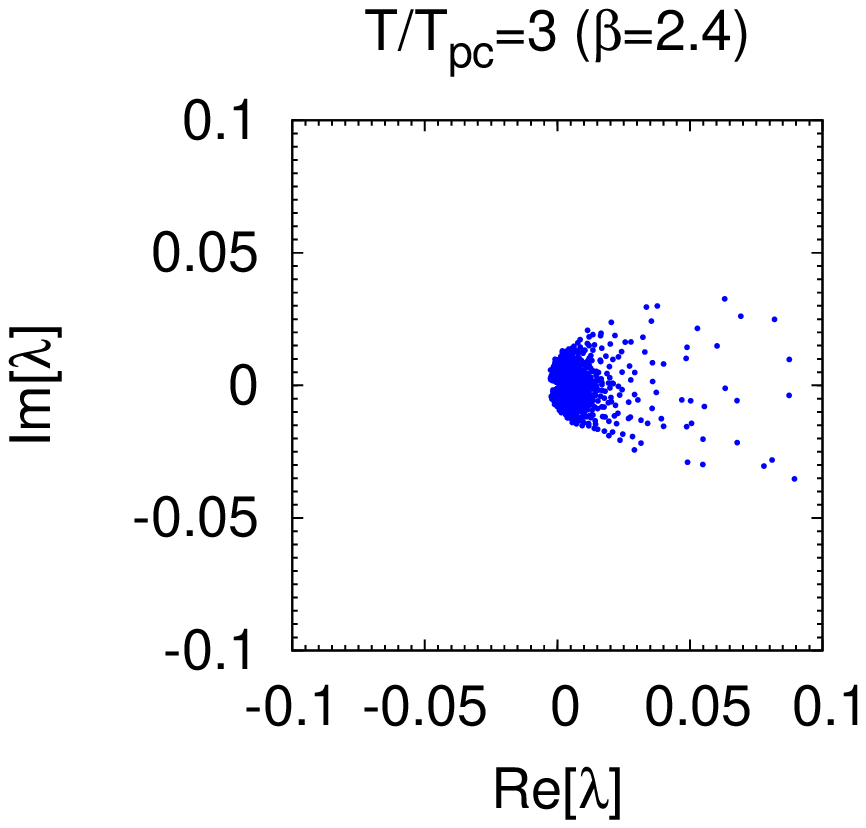}
\caption{Eigenvalue distribution on the complex $\lambda$ plane 
for three temperatures on the $8^3\times 4$ lattice with $m_{\rm PS}/m_{\rm V}=0.8$.
The top panels show the distribution of whole the eigenvalues, 
while the bottom panels magnify the region near the origin of the top panels.}
\label{Fig:2012Jan01fig1}
\end{figure} 

\begin{figure}[htpb] 
\begin{center}
\includegraphics[width=6.5cm]{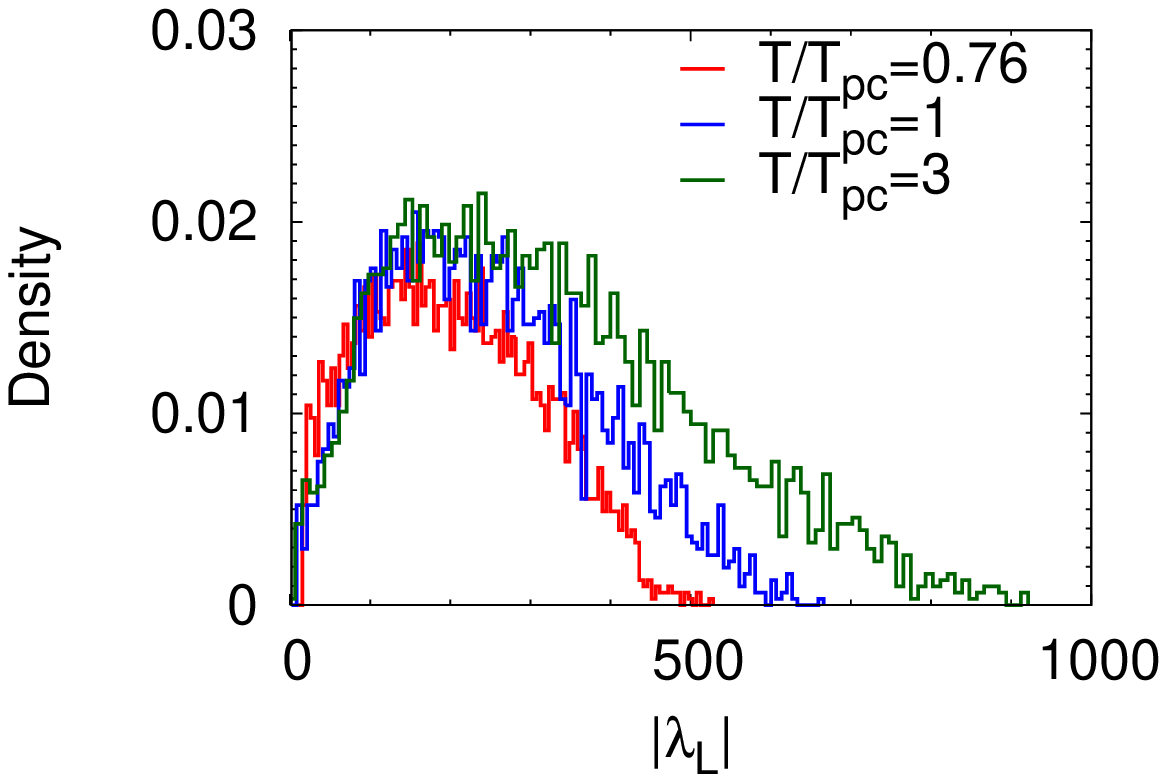}
\includegraphics[width=6.5cm]{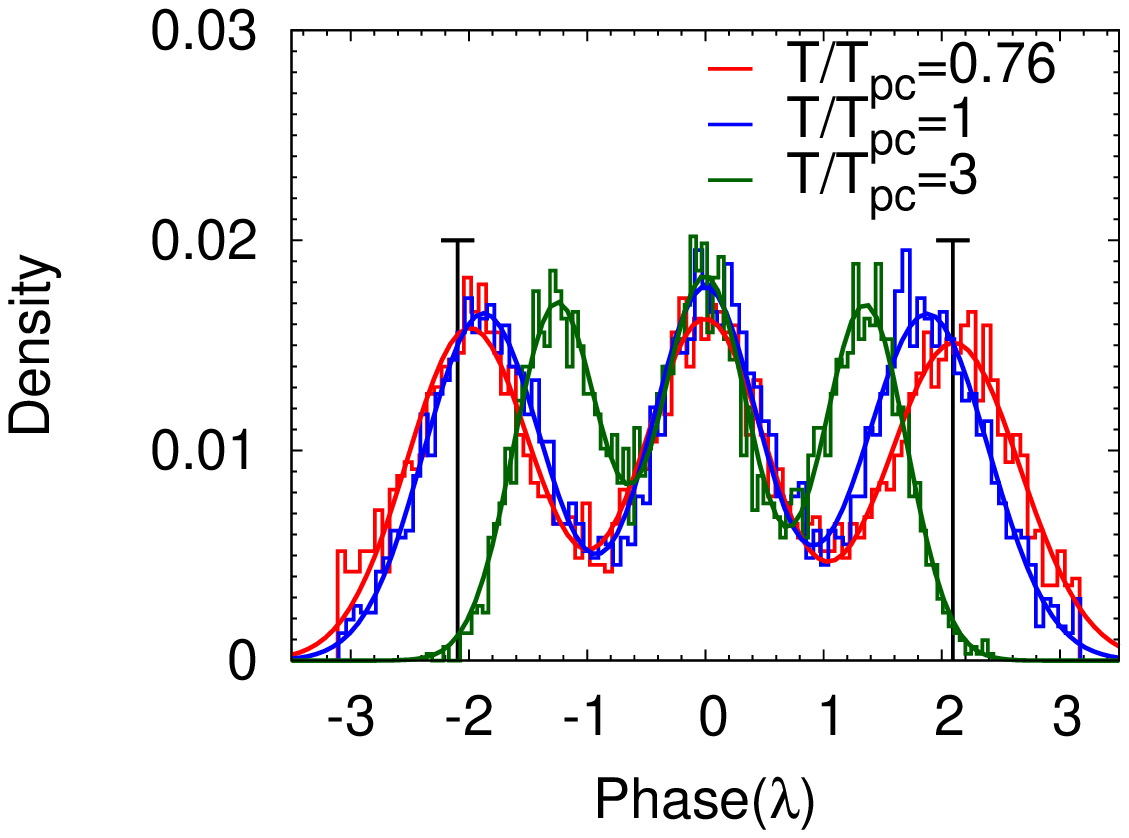}
\end{center}
\caption{The spectral density of the larger half of the eigenvalues $\lambda_n$
for three temperatures. The result for the smaller half of the eigenvalues is
obtained from the pair nature. Left : absolute value, right : phase of $\lambda_n$. 
In the right panel, thick lines are obtained from the three-Gaussian function :
$\rho(\theta)=\sum_{i=1, 2, 3} a_i \exp( - (\theta-c_i)^2/(2c_i^2))$, $(\theta=\arg(\lambda))$. The vertical lines denote $\arg(\lambda)=\pm 2\pi/3$.}
\label{Fig:2012Jan22fig2}
\end{figure} 
We look at typical behaviors of the eigenvalue distribution on 
the complex $\lambda$ plane in Fig.~\ref{Fig:2012Jan01fig1}, 
where results are shown for three temperatures. 
Each result is obtained from one configuration on the $8^3\times 4$ lattice. 
For each temperature, the eigenvalues are shown in two different scales; 
the top panels show the distribution of whole the eigenvalues in the 
complex plane, and bottom panels enlarge the region near the origin. 
These panels show two features of the reduced matrix, the pair nature and $\ZNc$-like property. 

First, we consider the angular (phase) distribution of the eigenvalues of 
the reduced matrix. 
The histogram of the angular distribution is shown in the right panel 
of Fig.~\ref{Fig:2012Jan22fig2}.
We find that the histogram of the angular distribution is well fitted with the three 
Gaussian function $\rho(\theta)=\sum_{i=1, 2, 3} a_i \exp( - (\theta-c_i)^2/(2c_i^2))$, 
where $\theta=\arg(\lambda)$. 
Three peaks are located at $\theta = 0, \pm 2\pi/3$ at $T\le \Tpc$, 
and two of them ($\theta = \pm 2\pi/3$) shift towards $\theta = 0$ 
as $T$ increases. 
Although $Z_3$ is explicitly broken in the presence of the quarks, 
the eigenvalue distribution is correlated to the Polyakov loop, 
as we have mentioned in the previous subsection. 
Indeed, Alexandru and Wenger found a correlation between the argument of the 
Polyakov loop and the location of the blank region of the eigenvalue
distribution at high $T$~\cite{Alexandru:2010yb}.
The confinement property of QCD may be seen in the angular distribution 
of the eigenvalues of the reduced matrix. 
The left panel of Fig.~\ref{Fig:2012Jan22fig2} shows the distribution 
of the magnitude of the large eigenvalues $\lambda_L$.
The distribution of $|\lambda_L|$ broadens at high temperatures.

Next, we consider a gap of the eigenvalues. 
According to the pair nature, the eigenvalues are split into two regions. 
The half of the eigenvalues are distributed in a region $|\lambda|<1$, 
and the other half in a region $|\lambda|>1$. 
There is the gap between the two regions where no eigenvalue exists, 
see the bottom panels of Fig.~\ref{Fig:2012Jan01fig1}. 
As we will see later, the gap is related to the pion mass. 


Let us consider a physical meaning of the two types of the eigenvalues: 
larger half ($|\lambda|>1$) and smaller half ($|\lambda|<1$). 
The pair nature of the eigenvalues is a consequence of $\gamma_5$-hermiticity, 
and therefore of the symmetry between the quark and anti-quark. 
As we have mentioned, $\tr\; Q$ is related to a generalized Polyakov loop 
and to free energies of an quark and anti-quark.
Qualitatively, the relation can be written as $\lambda \sim e^{-F/T}$ with 
a free energy $F$. 
Obviously $|\lambda|<1$ if $F>0$, while $|\lambda|>1$ if $F<0$. 
Hence, the smaller half of the eigenvalues correspond to quarks and 
the larger half correspond to anti-quarks. 

\begin{figure}[htpb] 
\includegraphics[width=7cm]{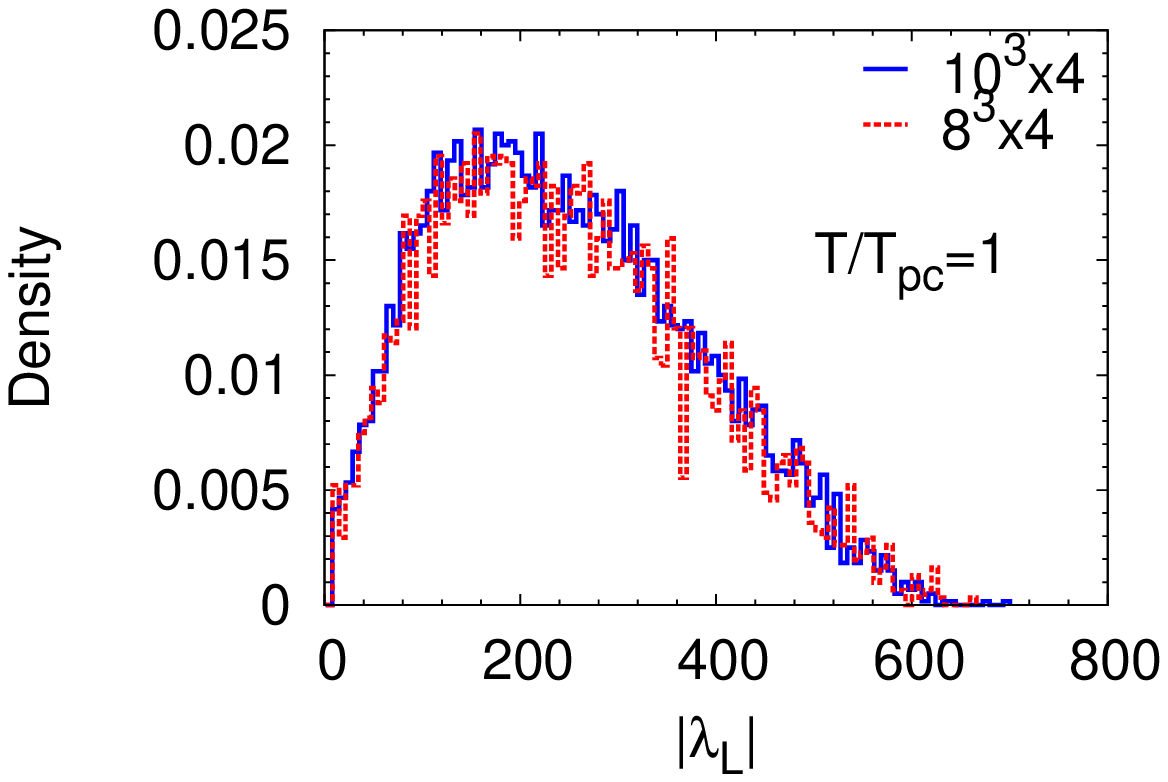}
\includegraphics[width=7cm]{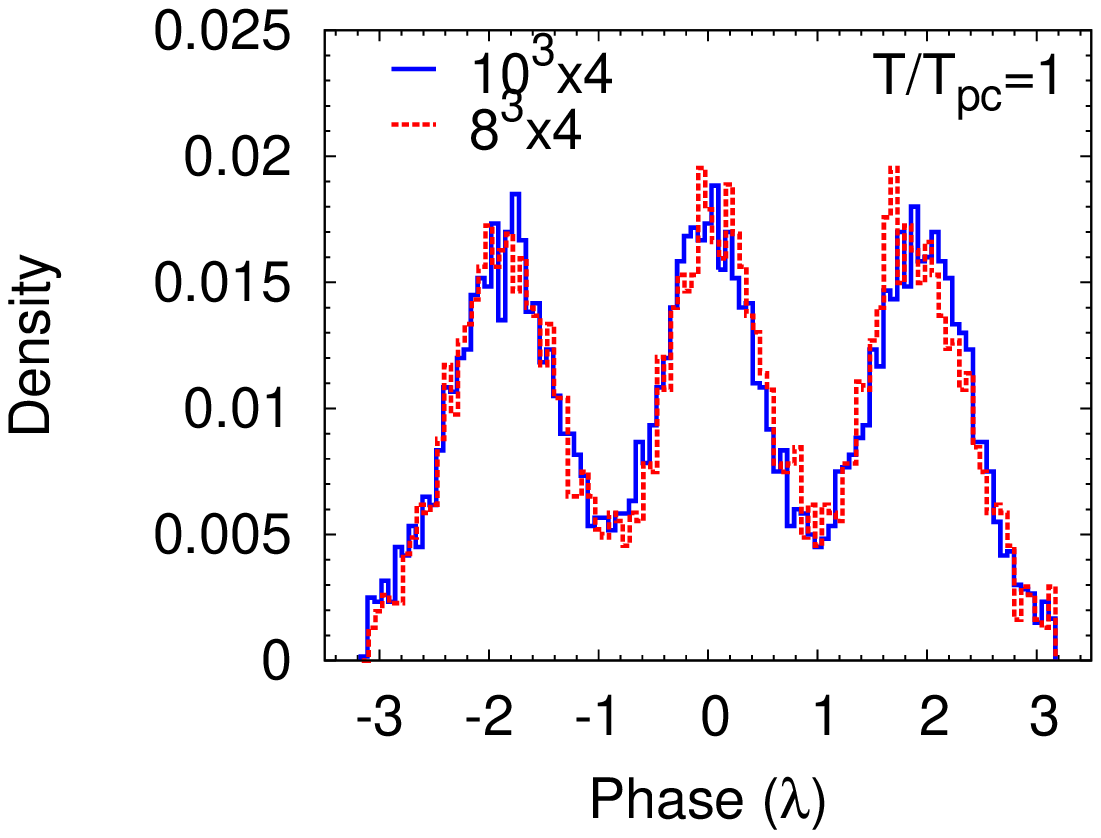}
\caption{Finite size effect on the spectral densities of the larger half of the 
eigenvalues. Left : absolute value, right: phase. 
In each panel, the blue-solid and red-dotted lines denote the results for 
$8^3\times 4$ and $10^3\times 4$. Note $V_s(N_s=10)/V_s(N_s=8)\sim 2$.
}
\label{Fig:2012Jan22fig1}
\end{figure} 
The volume dependence of the eigenvalues is shown in Fig.~\ref{Fig:2012Jan22fig1}. 
It turns out that the spectral density of the eigenvalues is almost insensitive to
the spatial lattice size $N_s$. 
The effect of increasing $N_s$ is small even for the gap and tail part of the 
distribution.
This insensitivity may be a consequence that long-range correlations are 
suppressed due to the heavy quark mass used in this work.
The volume dependence should be investigated in future simulations 
both with small quark masses and large lattice sizes.

Although the histogram of eigenvalues of $Q$ is insensitive to $N_s$, the 
fermion determinant depends on $N_s$. 
The fermion determinant can be written as 
$\det \Delta(\mu)\sim \exp ( \sum_{n=1}^{\Nred} \ln (\lambda_n + \xi))$, 
or spectral representation, 
\begin{align}
\det \Delta (\mu) = C_0 \exp \left( 2N_c V_s \mu/T + 4 N_c V_s \int d\lambda \rho(\lambda) \ln(\lambda+\xi) \right),
\label{Eq:2012Mar28eq2}
\end{align}
where $V_s=N_s^3$. $\rho(\lambda)$ is the spectral density on the complex $\lambda$ plane. 
The term $\ln (\lambda + \xi)$ is complex, and generates the phase of $\det \Delta(\mu)$.
As we have shown in Fig.~\ref{Fig:2012Jan22fig1}, $\rho(\lambda)$ is insensitive 
to $N_s$, and therefore the $\lambda$-integral is also insensitive to $N_s$. 
The imaginary part of the $\lambda$-integral, 
$\int d\lambda \rho(\lambda) \ln(\lambda+\xi)$, 
is also insensitive to $N_s$. 
This means the phase of $\det \Delta(\mu)$ is proportional to $V_s$, 
which shows the well-known fact~\cite{deForcrand:2010ys} that the sign problem becomes 
severe for large lattice volume.

Note that $N_t$ dependence is included in $\lambda$ itself, but does not 
appear in the overall factor of the exponent.  
Hence, the increase of $N_t$ does not increase the phase of the determinant. 
This suggests that the sign problem is not necessarily severe at low temperatures.
We will see later, the increase of $N_t$ makes the sign problem milder.

\begin{figure}[htpb] 
\includegraphics[width=7cm]{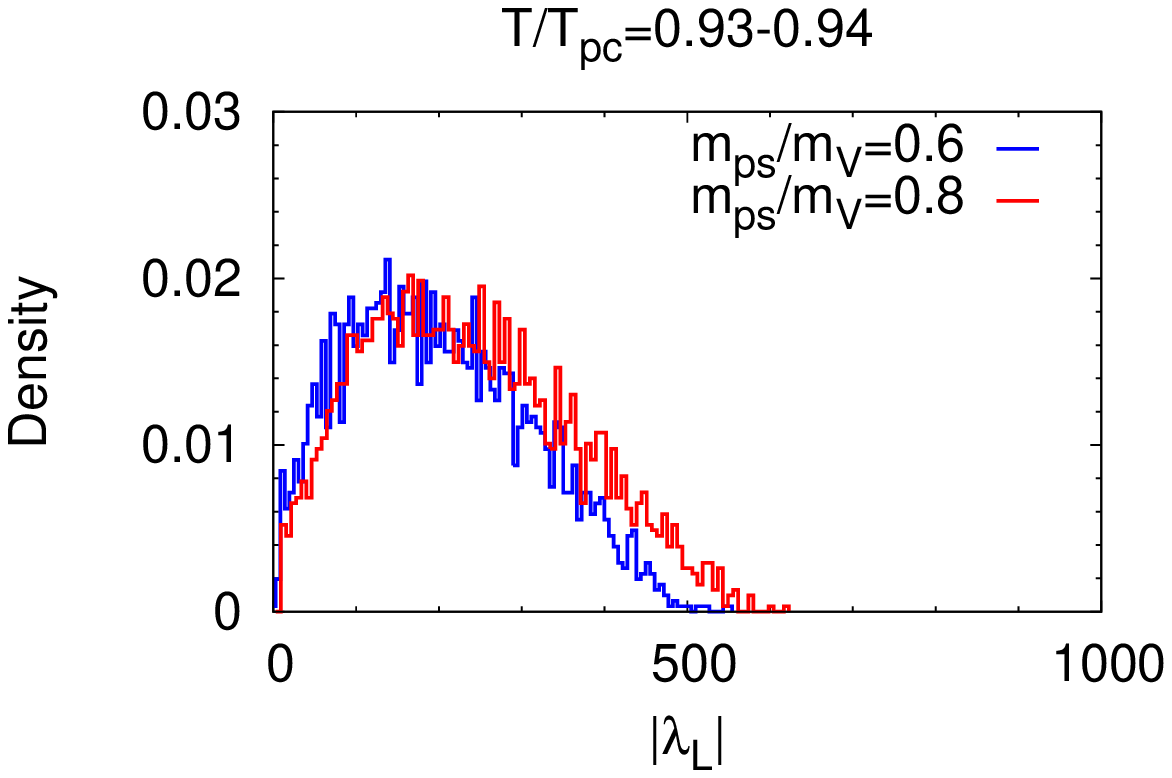}
\includegraphics[width=7cm]{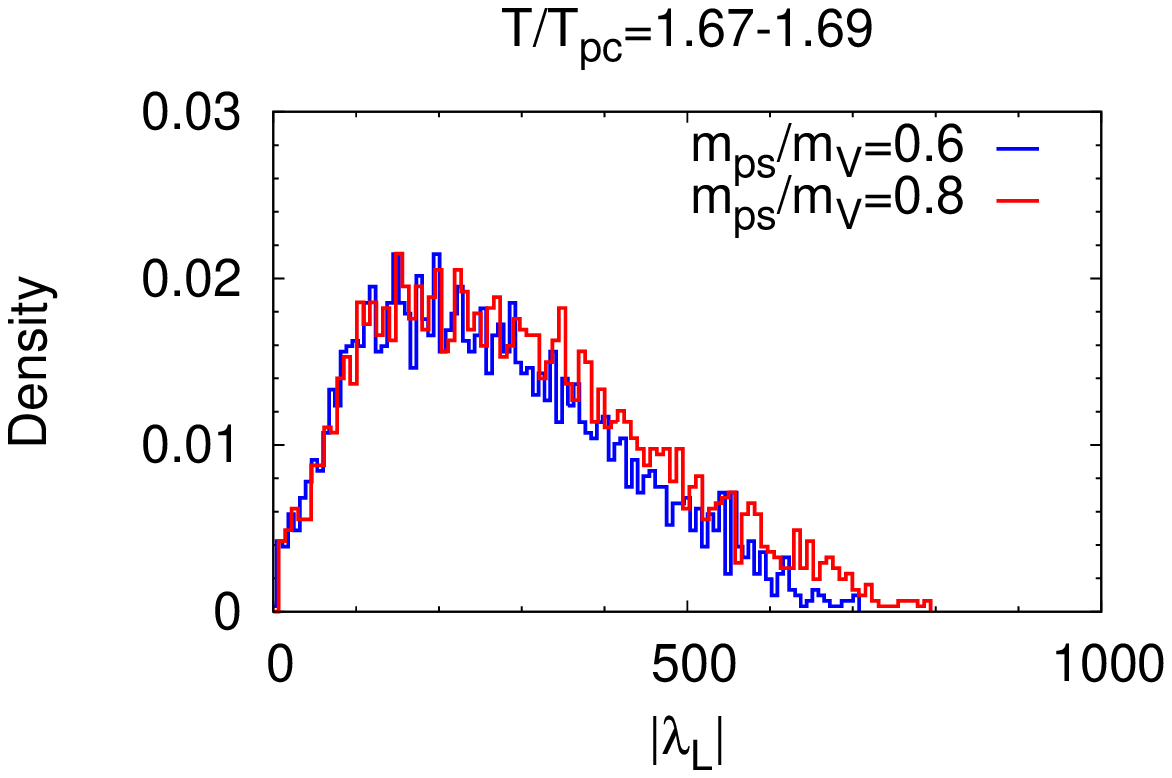}
\caption{The quark mass dependence of the spectral density. 
Here we consider the absolute values of the larger eigenvalues $|\lambda|>1$.
The parameter is $\beta=1.7$ for $\mps/\mV=0.6$ and $\beta=1.8$ for $\mps/\mV=0.8$.
The decrease of the quark mass narrows the width of the distribution. 
}
\label{Fig:2012Mar15fig1}
\end{figure} 
\begin{figure}[htpb] 
\includegraphics[width=7cm]{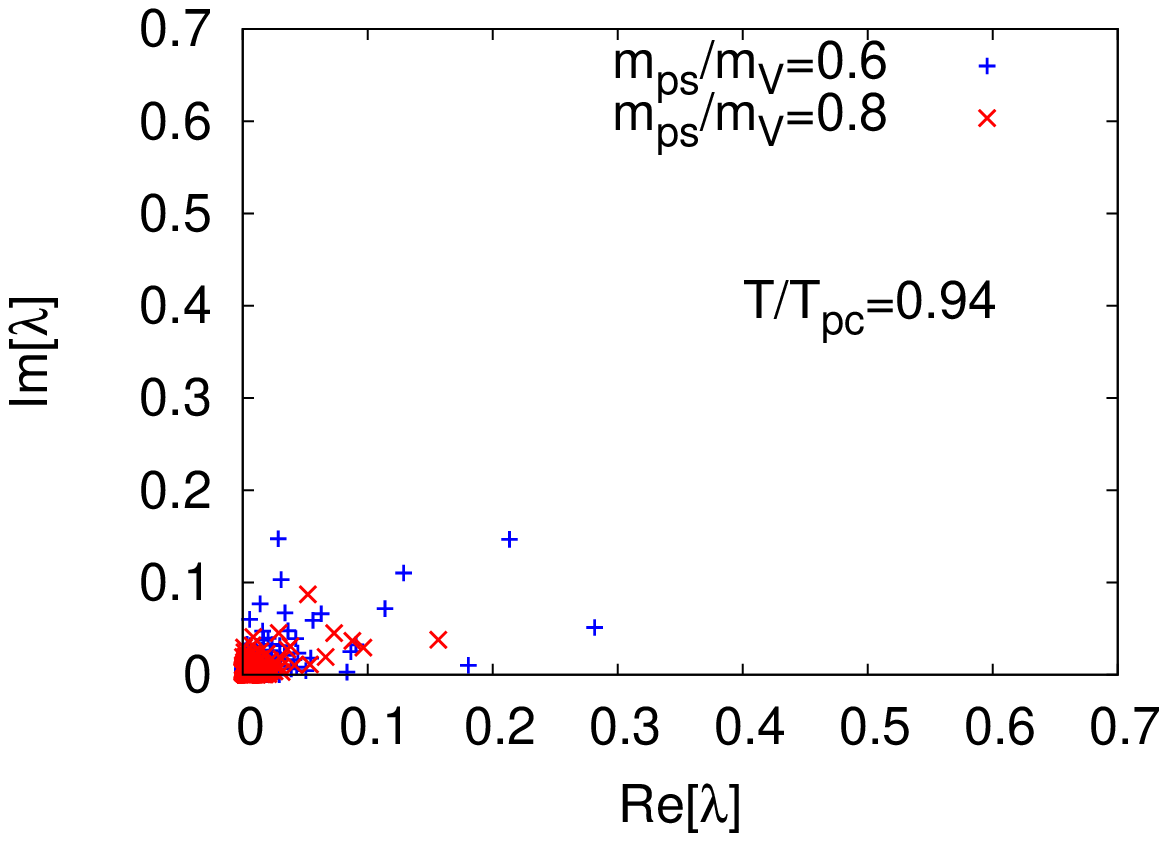}
\includegraphics[width=7cm]{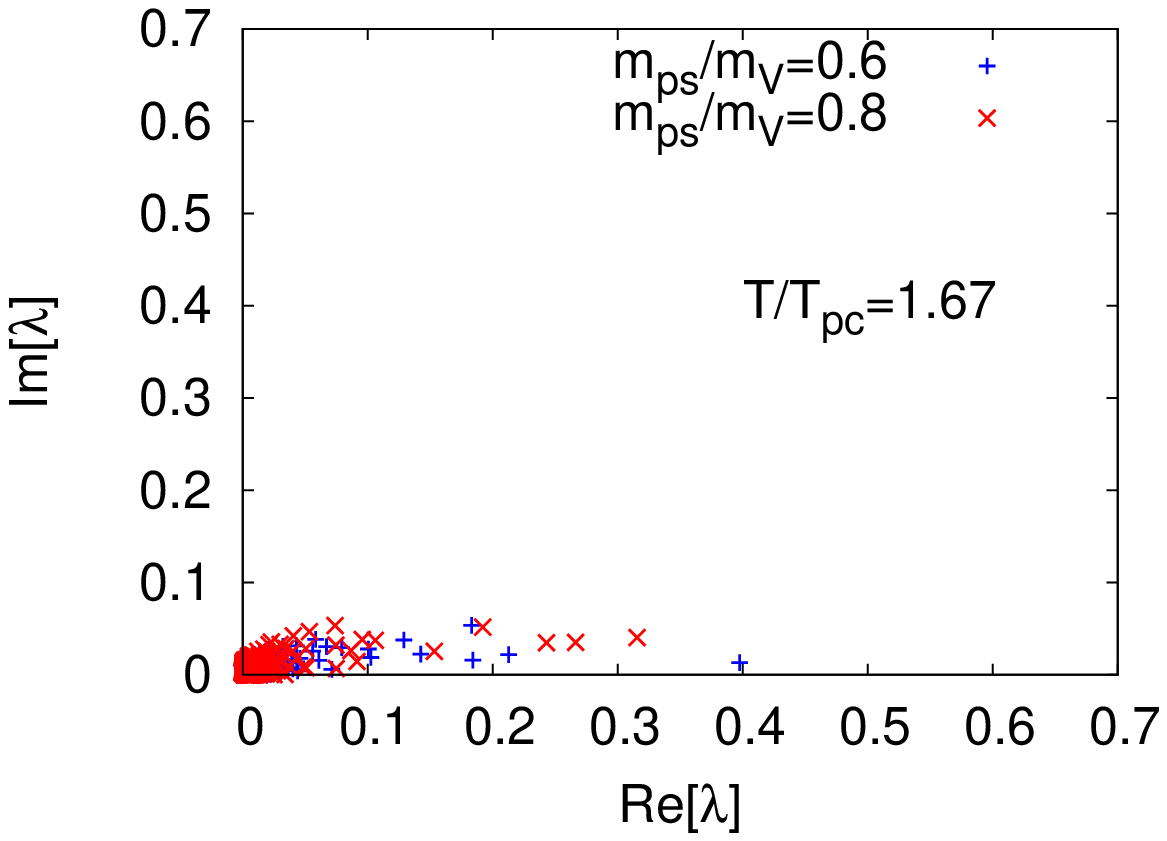}
\caption{The quark mass dependence of the scatter plot of the 
eigenvalues. The eigenvalues near the unit circle, which 
is related to the pion mass, is shown. Eigenvalues approach to 
the unit circle for small quark mass. 
}
\label{Fig:2012Mar15fig2}
\end{figure} 
The quark mass dependence of the eigenvalues of the reduced matrix is shown in 
Figs.~\ref{Fig:2012Mar15fig1} and \ref{Fig:2012Mar15fig2}.
The eigenvalues $\lambda$ depend on the quark mass for both the tail part and gap part. 
It is shown in Fig.~\ref{Fig:2012Mar15fig1} that the decrease of the quark mass narrows 
the eigenvalue distribution. 
Figure~\ref{Fig:2012Mar15fig2} shows the quark mass dependence of eigenvalues 
near the unit circle.
Qualitatively, eigenvalues approach to the unit circle as 
the quark mass becomes smaller. 
Gibbs pointed out~\cite{Gibbs:1986hi} for staggered fermions that eigenvalues 
outside the unit circle move away from the unit circle as the quark mass increases. 
The results in Figs.~\ref{Fig:2012Mar15fig1} and \ref{Fig:2012Mar15fig2} 
are consistent with the result in Ref.~\cite{Gibbs:1986hi}.
However, the quark mass dependence is considered without 
the ensemble average. 
We will consider the quark mass dependence of the average of the gap later. 

\subsection{Low-$T$ behavior and $N_t$-scaling law}
\label{Sec:2012Mar04sec1}

\begin{figure}[htbp] 
\includegraphics[width=7cm]{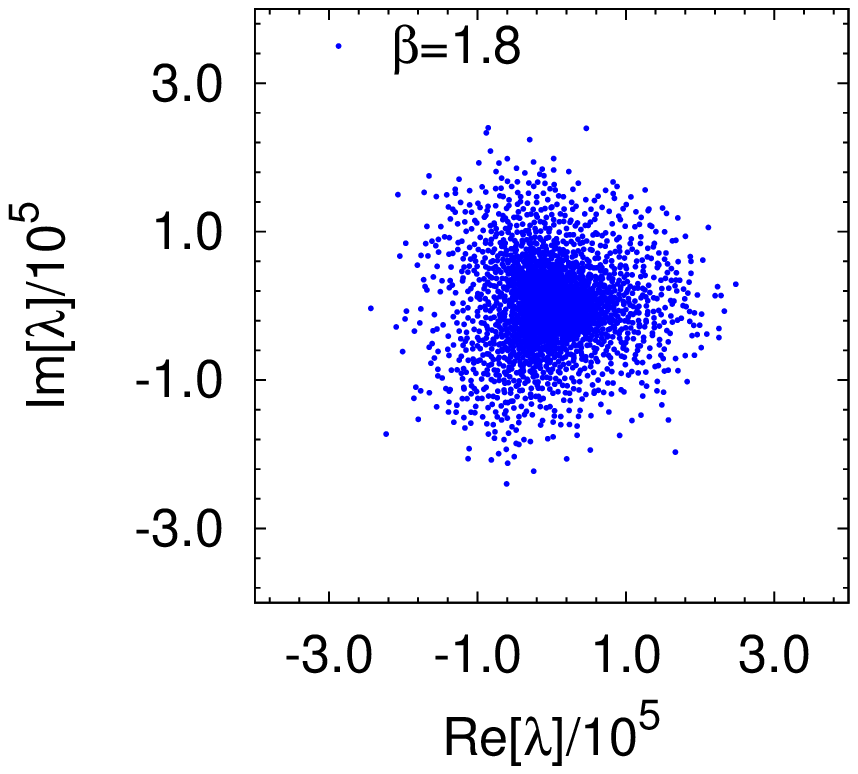}
\includegraphics[width=6.5cm]{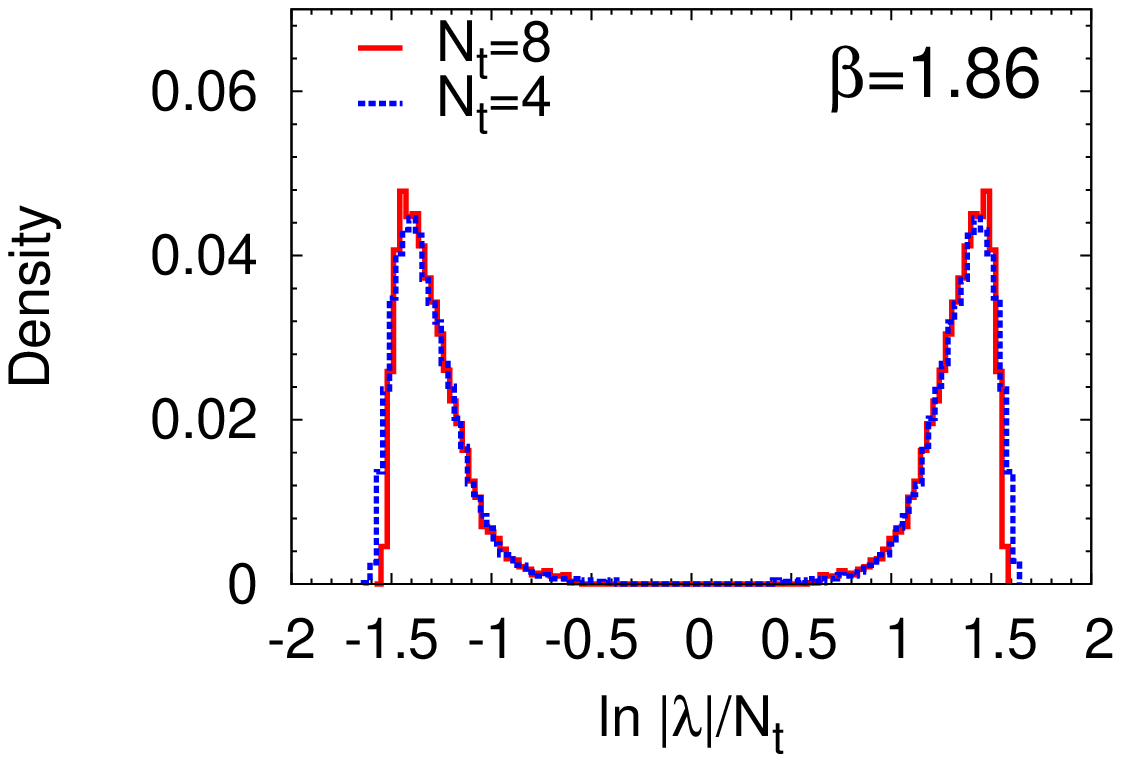}
\caption{
Left : The distribution of the large eigenvalues for $\beta=1.8$ on 
the $8^4$ lattice. See also the result on $8^3\times 4$ in Fig.~\ref{Fig:2012Jan01fig1}.
Right : Histogram of the eigenvalue distribution with scaled by $N_t$.  
The blue and red lines denote the results for $N_t=4$ and $8$, respectively. 
The agreement between them implies the scaling law $\lambda\sim l^{N_t}$. 
The two peaks are a consequence of the pair nature $\lambda$ and $1/\lambda^*$. 
}
\label{Fig:2012Jan01fig2}
\end{figure} 
Next, we investigate the $N_t$ dependence and the properties at lower temperature. 
The left panel of Fig.~\ref{Fig:2012Jan01fig2} shows the eigenvalue distribution 
for the $8^4$ lattice with $\beta=1.8$, which corresponds to $T/\Tpc\sim 0.5$. 
Since $T=1/(a N_t)$, the temperature in Fig.~\ref{Fig:2012Jan01fig2} is 
almost half compared to the case of Fig~\ref{Fig:2012Jan01fig1}. 
A major difference between $N_t=4$ and $N_t=8$ is the magnitude of the eigenvalues. 
As the temperature decreases, the smaller half of the eigenvalues become smaller 
and larger half of the eigenvalues become larger. 

The right panel of Fig.~\ref{Fig:2012Jan01fig2} shows the histogram of the absolute 
value of $\lambda$ scaled by $N_t$. 
The results for $N_t=4$ and $N_t=8$ agree well. 
This agreement implies that the eigenvalue distribution 
as a function of a variable $\ln |\lambda|/N_t$ is almost independent of $N_t$, 
which leads to the scaling law $|\lambda|\sim l^{N_t}$ with 
a quantity $\ln l\equiv \ln|\lambda|/N_t$. 
%
Note that $l$ depends on the lattice spacing $a$. 
We have pointed out this scaling law in the previous study~\cite{Nagata:2010xi}. 
The matrix $Q$ is given as the product of $A_i = \alpha_i^{-1} \beta_i, 
(i=1, \cdots N_t)$. 
Let us express $A_i = \bar{A} + \delta A_i$, where $\bar{A}$ is a matrix 
independent of time. 
$A_i$ is expected to depend on time moderately in equilibrium, i.e., 
$|A_i| > | \delta A_i|$.
An additional cancellation would occur in $\sum_i \delta A_i$, 
since $\delta A_i$ is a fluctuation from $\bar{A}$. 
If this argument holds, $Q$ is parameterize as $Q\sim \bar{A}^{N_t}$. 
The agreement observed in Fig.~\ref{Fig:2012Jan01fig2} clearly indicates 
this scaling behavior.

The right panel of Fig.~\ref{Fig:2012Jan01fig2} also shows the gap 
and pair nature of the eigenvalues. We focus on the gap in the next subsection.

\subsection{Gap and pion mass}
\label{sec:red_gap}
As we have mentioned, the reduced matrix describes the generalized Polyakov 
line and is related to the free energy of a dynamical quark. 
Combining several Polyakov lines, it is possible to construct states having 
quantum numbers of hadrons. 
Thus it is natural to expect that the eigenvalues of the reduced matrix 
have something to do with the hadron spectrum. 
Indeed, Gibbs pointed out~\cite{Gibbs:1986hi} a relation at $T=0$ between the 
pion mass and the eigenvalue gap. 
Later, Fodor, Szabo and T\'oth related the hadron spectrum 
to the eigenvalues of the reduced matrix, based on thermodynamical 
consideration~\cite{Fodor:2007ga}. 
In this subsection, we study the pion mass using an eigenvalue close to 
$|\lambda|=1$, which controls the gap. 

\begin{figure}[htbp]
\includegraphics[width=7cm]{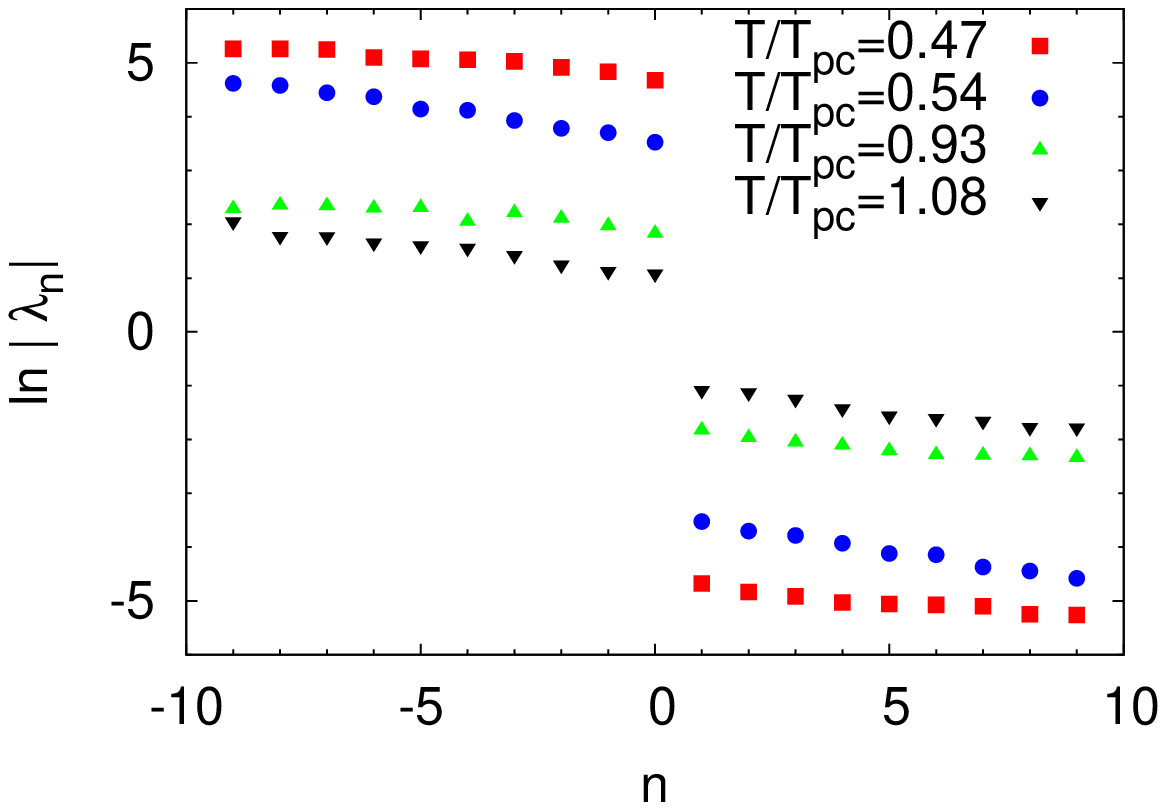}
\includegraphics[width=7cm]{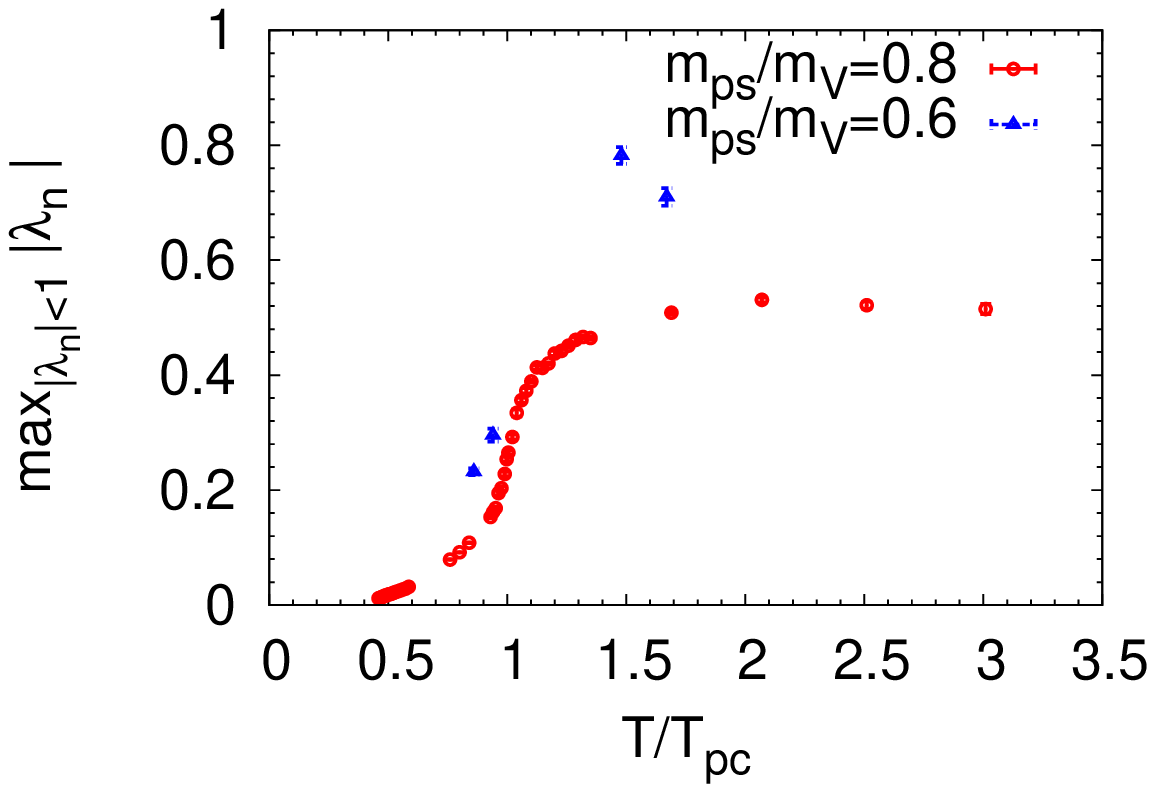}
\caption{Left panel: Eigenvalues near the unit circle $|\lambda|=1$. 
The parameters for each result are $(\beta, N_t)=(1.8, 8), (1.9, 8), (1.8, 4),$ 
and $(1.9, 4)$ for $T/\Tpc=0.47, 0.54, 0.93,$ and $1.08$, respectively. 
Each result is obtained from one configuration. 
Right panel : The largest eigenvalue in smaller half $\max_{|\lambda_n|<1} |\lambda_n|$,
which is the eigenvalue closest to the unity $|\lambda|=1$. The gap is defined as 
the deviation between the data and one.
}
\label{Fig:2012Jan04fig1}
\end{figure}
The temperature dependence of the gap is shown Fig.~\ref{Fig:2012Jan04fig1}. 
The left panel shows twenty eigenvalues close to unity, where
the result for each temperature is obtained from one configuration. 
It is shown that the gap shrinks as the temperature increases. 
To take into account the ensemble average, we focus on the maximum eigenvalue 
among the smaller half: $\max_{|\lambda|<1} \lambda$. 
The gap is given by the difference between unity and this quantity. 
The result is shown in the right panel, where the gap is large at low $T$
and decreases as $T$ increases.
In case of $\mps/\mV=0.8$, the gap is saturated at high $T$. 

The right panel of Fig.~\ref{Fig:2012Jan04fig1} contains the results for 
two values of $\mps/\mV$.
We find that the average value of $\max_{|\lambda|<1} |\lambda|$ is larger for smaller quark mass. 
This implies that the gap shrinks for smaller quark mass, which is 
consistent with Ref.~\cite{Gibbs:1986hi}.
However, the simulations for $\mps/\mV=0.6$ were done only for four 
temperatures and for the small number of HMC trajectories. 
Further simulations are needed to reveal the quark mass dependence 
of the eigenvalues of the reduced matrix. 

\begin{figure}[htbp]
\begin{center} 
\includegraphics[width=7cm]{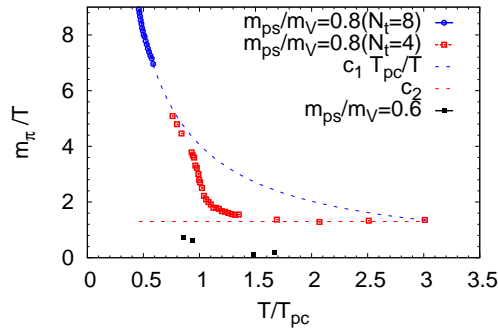} 
\caption{Pion mass obtained from Gibbs's definition. 
At low $T$, the curve is well fitted with $c_1 (T/\Tpc)^{-1}$ (the 
blue dashed line). The pion mass is extracted from the coefficient 
$c_1$ of the $1/T$ fit at low temperatures. 
}
\label{Fig:2012Feb23fig1}
\end{center}
\end{figure}
According to Gibbs~\cite{Gibbs:1986hi}, the pion mass is given by 
\begin{align}
a m_{\pi}= - \frac{1}{N_t} \max_{|\lambda_k|<1} \ln |\lambda_k|^2,
\label{Eq:2012Feb20eq1}
\end{align}
which is expected to be valid at $T=0$~\cite{Barbour:1997ej}. 
On the other hand, Fodor, Szabo and T\'oth derived a modified 
expression~\cite{Fodor:2007ga},
\begin{align}
a m_{\pi}= \lim_{N_t\to \infty} \left( -\frac{1}{N_t} \ln \left\langle \left| \sum \lambda_k \right|^2
\right\rangle \right).
\label{Eq:2012Feb20eq2}
\end{align}
The pion mass in Eq.~(\ref{Eq:2012Feb20eq1}) is shown in 
Fig.~\ref{Fig:2012Feb23fig1} as a function of $T/\Tpc$. 
The results are shown for $\mps/\mV=0.6$ and $0.8$. 
In case of $\mps/\mV=0.8$, $m_{\pi}/T$ is well fitted with 
$c_1(T/T_{pc})^{-1}$ at $T/\Tpc=0.5$ with $c_1=4.060(6)$. 
The pion mass is extracted from the low temperature behavior as 
$m_{\pi} = c_1 \Tpc$, which is approximately $4 \Tpc$. 
This large value is because of  the present simulation setup. 
At high $T$, $m_\pi$ is almost linearly proportional to $T$, which 
is probably due to thermal effects. 
The decreasing $\mps/\mV$ makes $m_\pi/T$ smaller. However, 
simulation data are not sufficient to obtain $c_1$. 

Hadron masses are extracted through the exponential damping in an 
usual method. In the present approach, the pion mass is extracted 
through $1/T$ behavior. 
It is important to note that the $1/T$ behavior is already obtained at $N_t=8$. 
Hence, this approach may be an alternative method for hadron spectroscopy, 
which was discussed in detail in Ref.~\cite{Fodor:2007ga}.

\section{Low temperature limit of QCD}
\label{Sec:2012Mar04sec2}
In this section, we consider the behavior of the fermion determinant at 
low temperatures. 
For small $\mu$, the fermion determinant becomes insensitive to $\mu$ 
as $T$ decreases, and becomes almost independent of $\mu$ at enough low $T$.
We explain this behavior by using the properties of the eigenvalues of the 
reduced matrix. 

We first consider $T$- and $\mu$-dependence of a reweighting factor. 
Then, we consider the low temperature limit with the use of the $N_t$-scaling law 
of the eigenvalues of the reduced matrix. 
We will derive two expressions for low temperature limit of the quark determinant; 
one is for small $\mu$ and the other is for large $\mu$. 
The low density expression shows the $\mu$-independence of the fermion 
determinant at $T=0$ for $\mu<m_\pi/2$.
We will conclude that the $\mu$-independence at $T=0$ for small $\mu$ is 
the consequence of the $N_t$ scaling law of the eigenvalues of the reduced matrix. 
The other expression is its high-density counter part. 
We discuss the physical meaning and criterion for these limits.

We discuss the high density limit of the fermion determinant. 
In low temperature and high density limit, 
QCD approaches to a theory, where quarks interact
in spatial directions with the ordinary Yang-Mills type of the gauge action. 
The corresponding partition function is $Z_3$ invariant even in the presence 
of dynamical quarks. 
The fermion determinant becomes real and the theory is free from the sign problem.

\subsection{Fluctuation of the fermion determinant}
\label{sec3b}
In this subsection, we consider the quark determinant at low temperature 
and finite $\mu$. 
\begin{figure}
\includegraphics[width=7cm]{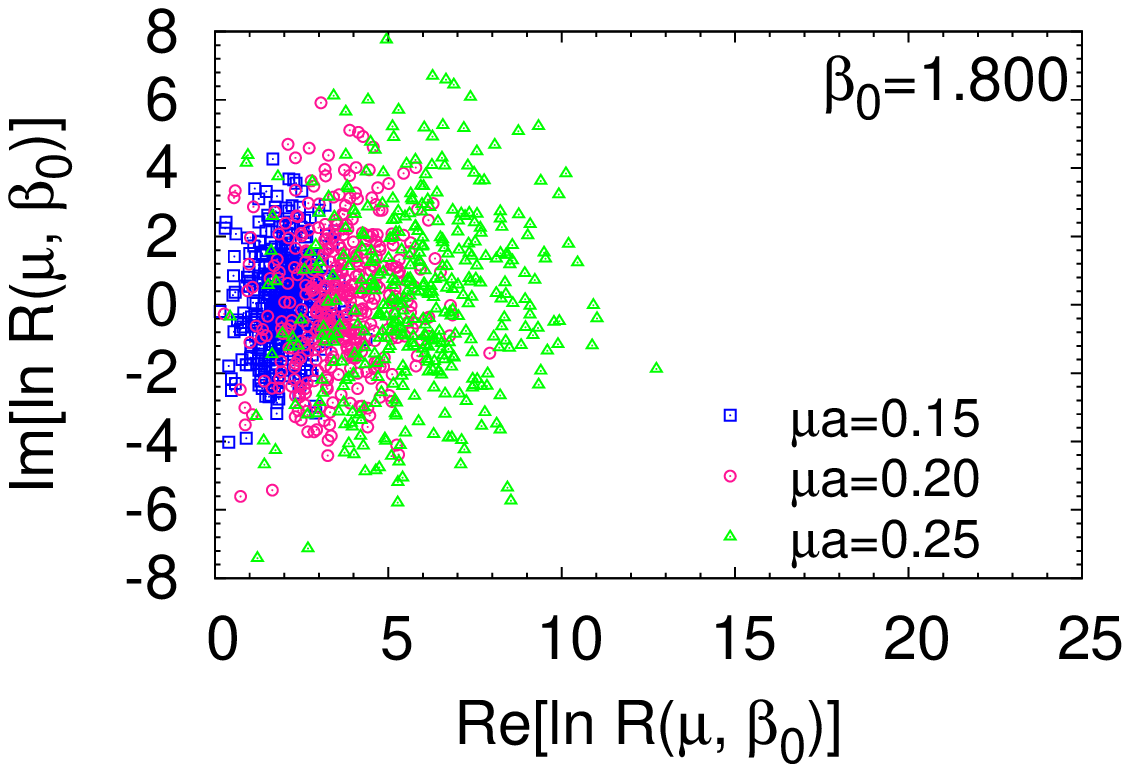}
\includegraphics[width=7cm]{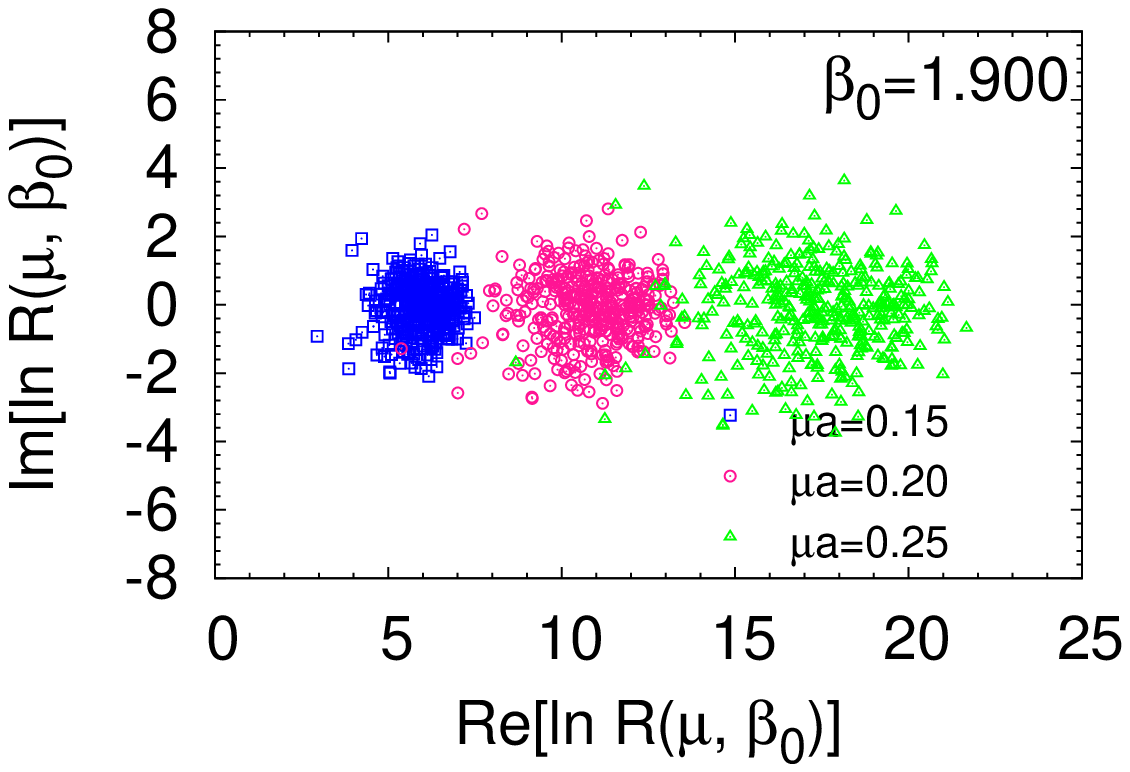}
\caption{The scatter plot of the reweighting factor on the complex plane. 
Left panel : hadron phase ($T/\Tpc=0.93$). Right panel : QGP phase($T/\Tpc=1.08$). 
The figures are taken from Ref.~\cite{Nagata:2012pc}. 
The results are obtained for $8^3\times 4$ lattice with $\mps/\mV=0.8$.
}
\label{Oct142011Fig1}
\end{figure}
Figure~\ref{Oct142011Fig1} is the scatter plot of 
$\ln R(\mu, \beta_0)_{(0,\beta_0)}=N_f\ln \det \Delta(\mu)/\det \Delta(0)$, 
which is so called reweighting factor. 
This shows how the quark determinant develops as $\mu$ is varied from a 
simulation point $(\mu=0,\beta_0)$, where gauge configurations 
are generated. Here we fix a temperature ($\beta=\beta_0$). 
The left and right panels show the results for $T/\Tpc\sim 0.93 (\beta_0=1.8)$ and 
$1.08(1.9)$, respectively. 
The horizontal and vertical axes are the real and imaginary 
parts of $\ln R$, i.e., the exponent of $|R|$ and 
the phase of $R$. 

\begin{figure}[htbp]
\includegraphics[width=7cm]{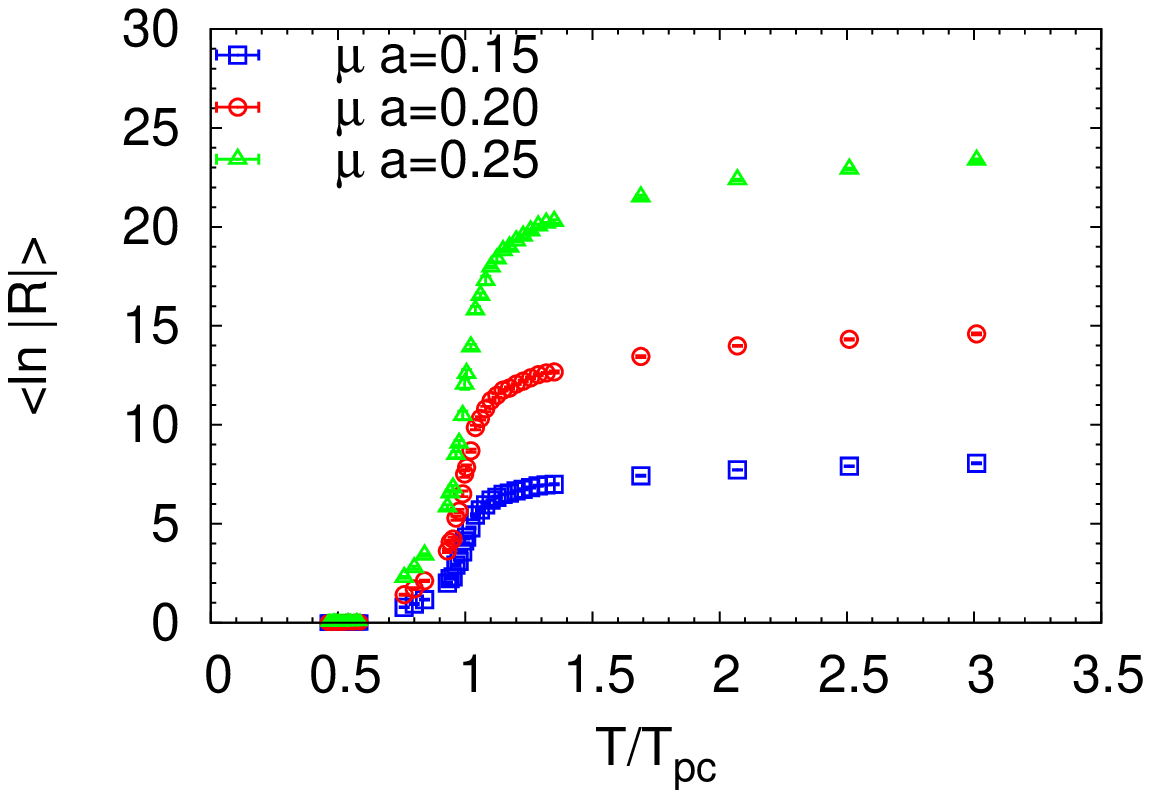}
\includegraphics[width=7cm]{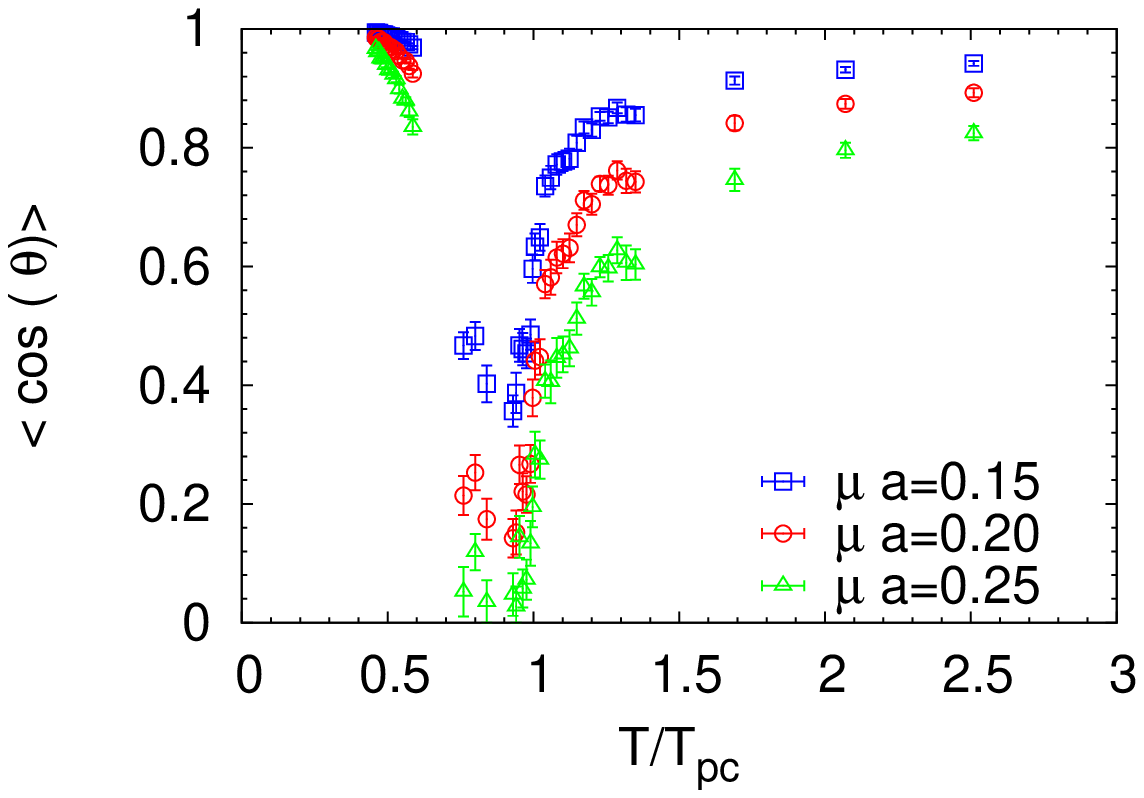}
\caption{Left : the average of $\ln |R|=N_f\ln 
|\det \Delta(\mu)/\det \Delta(0)|$, right : the average 
phase factor defined by $\bra \cos \theta\ket$, where $\theta = \arg (R)$.
The results are obtained for $8^3\times 4$ lattice with $\mps/\mV=0.8$.
}
\label{Fig:2012Mar28fig1}
\end{figure}
The results are shown in Fig.~\ref{Fig:2012Mar28fig1} now as 
functions of $T/\Tpc$ for three values of $\mu$. 
The magnitude of the reweighting factor is very small at 
low temperature, and rapidly increases near $\Tpc$. 
As $T$ decreases, the average of $\ln |R|$ approaches to zero at least 
up to $\mu a = 0.25$. This behavior means $|\det \Delta(\mu) / \det \Delta(0)|\sim 1$, 
and therefore the fermion determinant is insensitive to $\mu$ at low 
temperatures $T/\Tpc \sim 0.5$.
In the right panel, we plot the average phase factor 
$\bra \cos \theta \ket$ with $\theta = \arg(R)$. 
The average phase factor is close to one at high and low temperatures, 
and has minimum at slightly below $\Tpc$. 
The sign problem is very severe in the vicinity of $\Tpc$. 
At high temperatures, the average phase factor increases with increasing $\mu$. 
On the other hand, the $\mu$-dependence of the average phase factor 
is small at low temperatures. The average phase factor approaches to 
one for the three values of $\mu$. 

\begin{figure}[htbp]
\includegraphics[width=7cm]{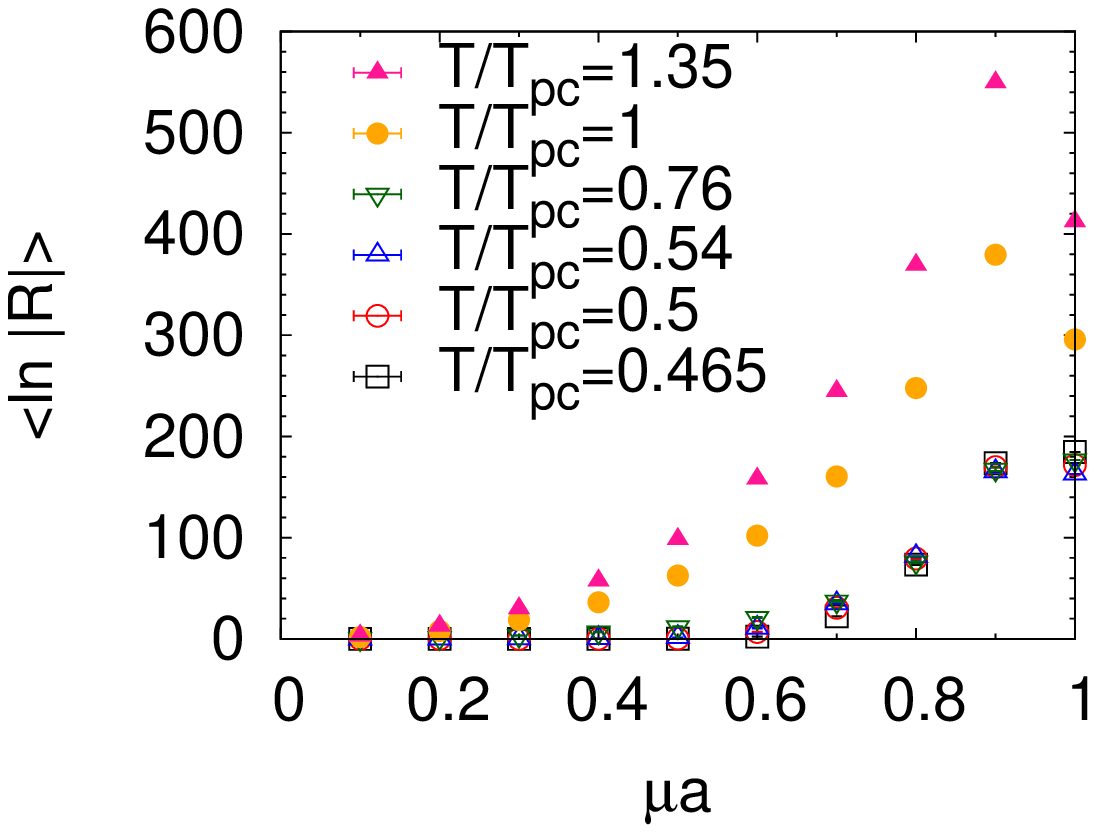}
\includegraphics[width=7cm]{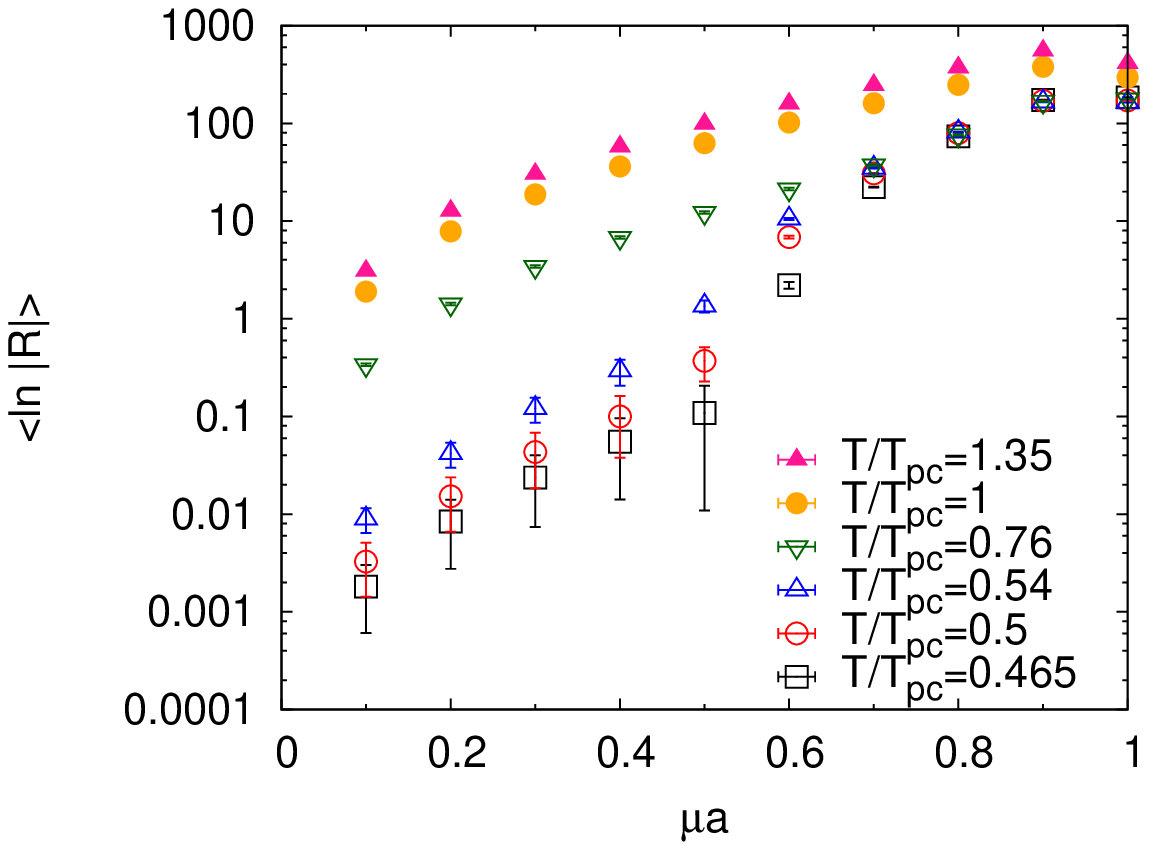}
\caption{The quark chemical potential($\mu$) dependence of the average of 
$\ln |R|=N_f\ln |\det \Delta(\mu)/\det \Delta(0)|$.}
\label{Fig:2012Mar28fig2}
\end{figure}
We found that the fermion determinant is insensitive to $\mu$ at low temperatures. 
However, the low-$T$ behavior of the fermion determinant depends on $\mu$. 
Next, we consider the reweighting factor as a function of $\mu$ 
in Fig.~\ref{Fig:2012Mar28fig2}.  
At high $T$($T\ge 0.76 \Tpc$), the average value of $\ln |R|$ smoothly 
increases as $\mu$ goes to larger. 
A discontinuous change is found at $\mu a=0.5\sim 0.6$ 
at low $T$($T\le 0.54 \Tpc$). 
$|R|$ is approximately unity for small $\mu$, and starts to increase 
at $\mu a = 0.5\sim 0.6$, see the right panel.
The onset of the $\mu$-dependence of the fermion determinant is 
about $\mu a = 0.5 \sim 0.6(\mu/T = 4.0 \sim 4.8)$. 
Using the value of the pion mass obtained in the 
previous section, it corresponds to $\mu = 0.5 m_\pi \sim 0.6 m_\pi$. 
This is consistent with a well known result that the fermion determinant is 
independent of $\mu$ for $\mu<m_\pi/2$ at $T=0$, see e.g. 
Refs~\cite{Barbour:1997bh,Cohen:2003kd,Adams:2004yy,Splittorff:2007zh}.

Phenomenologically, it is expected that the $\mu$-independence at $T=0$ 
continues up to $\mu=M_N/3$, where $M_N$ is the mass of the nucleon.
This problem was raised long time ago in the Glasgow method, 
see e.g. Ref.~\cite{Barbour:1997bh}. 
There would be several reasons for the discrepancy between the critical value of 
$\mu$ and $M_N/3$.
For instance, the importance sampling at $\mu=0$ may cause this discrepancy, 
because the quark chemical potential is equivalent to the isospin chemical potential 
at $\mu=0$. 

\begin{figure}[htbp]
\includegraphics[width=7cm]{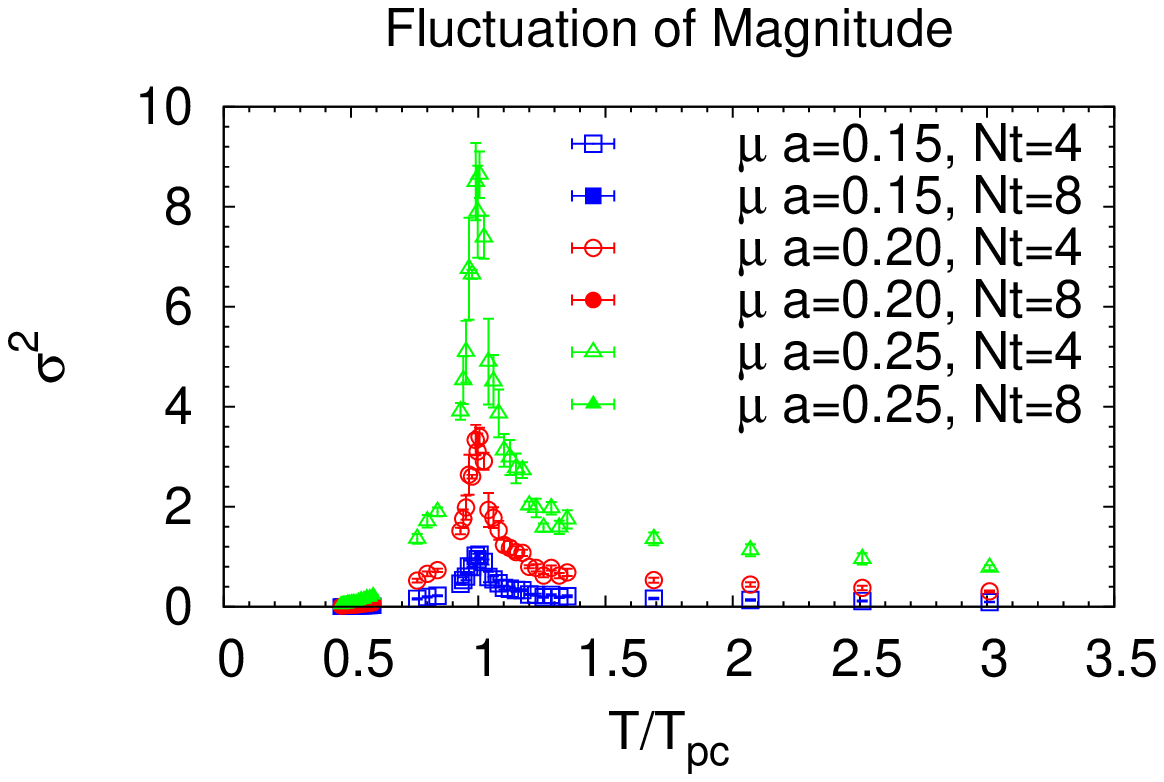}
\includegraphics[width=7cm]{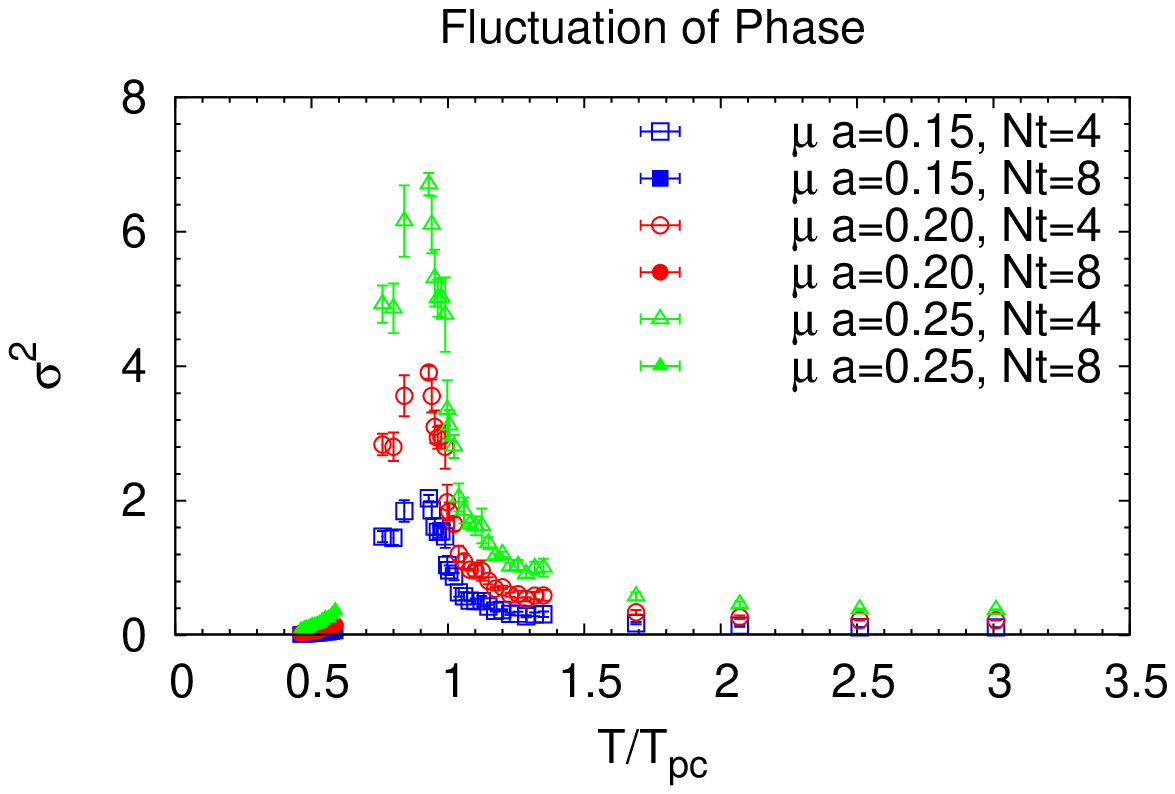}
\caption{The fluctuation of the reweighting factor as a function of $T/\Tpc$. 
$\sigma^2= \frac{1}{n}\sum_i (x_i-\bar{x})^2, \bar{x}=\frac{1}{n}\sum_i x_i$, 
where $x={\rm Re}[\ln R(\mu, \beta_0)_{(0,\beta_0)}]$ for left panel and $x={\rm Im}[\ln R(\mu, \beta_0)_{(0,\beta_0)}]$ 
for right panel. The results for $N_t=4$ are taken from 
Ref.~\cite{Nagata:2012pc}.}
\label{QDet2011Oct21Fig1}
\end{figure}
The deviation of the reweighting factor, $\sigma^2$, is shown in Fig.~\ref{QDet2011Oct21Fig1}.  
The deviation of the reweighting factor reaches the maximum near 
the crossover transition point $\Tpc$ both in the magnitude (left panel) 
and the phase (right panel), 
and decreases as the temperature is away from $\Tpc$. 
The peak becomes prominent as $\mu$ increases. 
In Fig.~\ref{Oct142011Fig1}, we showed the reweighting factor for
low $T$($T/\Tpc=0.93$) and high $T$($1.08$).
In the vicinity of $\Tpc$, 
gauge configurations visit low-$T$ states and high-$T$ states, which results in 
the peak of the fluctuation. 
The real part of $\ln |R|$ reaches the maximum at $\Tpc$, while the 
imaginary part reaches the maximum at slightly below $\Tpc$. 
Hence the sign problem is most severe at slightly below $\Tpc$, 
see also the right panel of Fig.~\ref{Fig:2012Mar28fig1}. 

\begin{figure}[htbp]
\includegraphics[width=7cm]{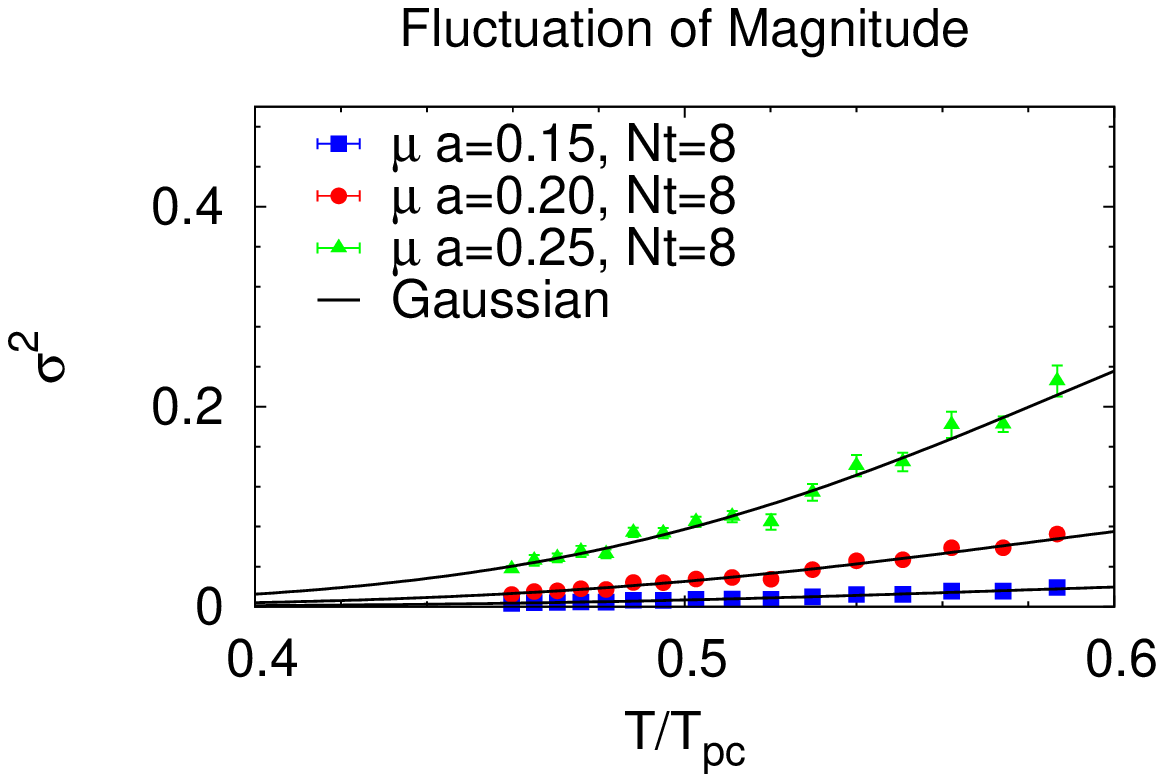}
\includegraphics[width=7cm]{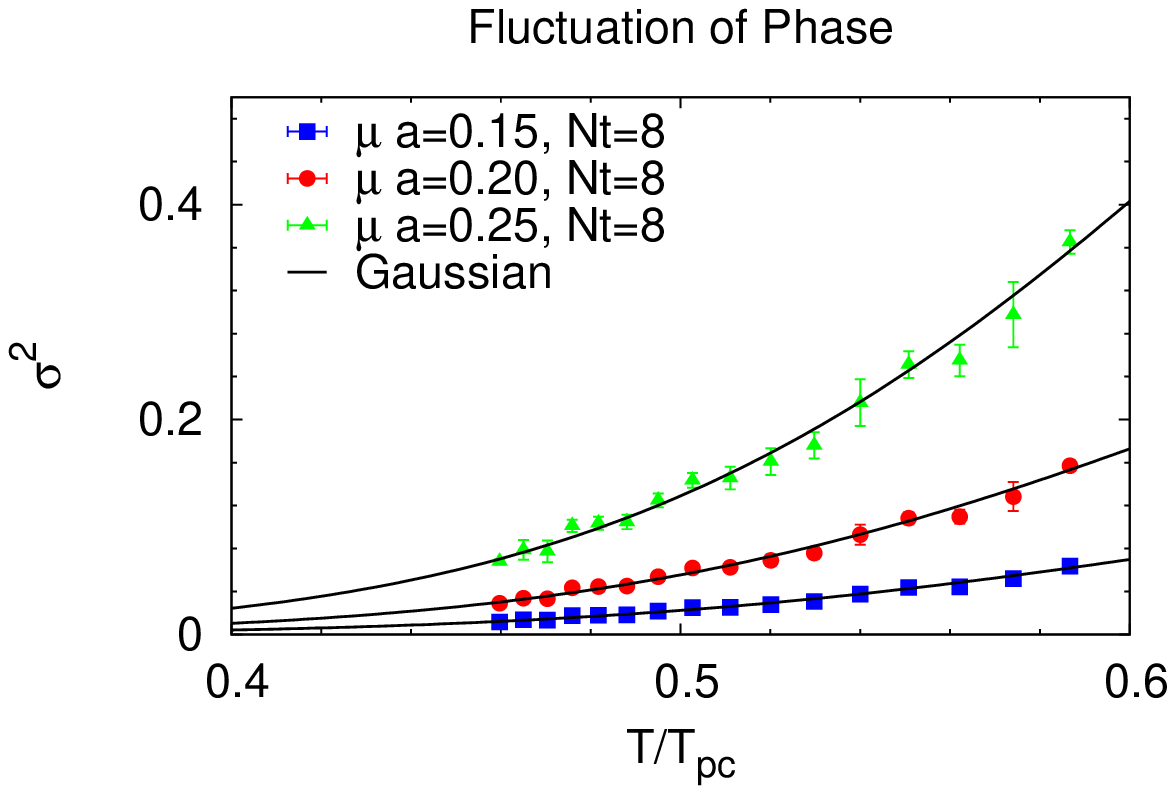}
\caption{The fluctuation of the quark determinant as a function of $T/\Tpc$ and 
the Gaussian fit. See also Fig.~\ref{QDet2011Oct21Fig1}.}
\label{Fig:2012Mar14Fig1}
\end{figure}

In Fig.~\ref{Fig:2012Mar14Fig1}, we focus on the low temperature region.
As $T$ decreases, the deviation rapidly decreases for both real and 
imaginary parts. 
We find that the low temperature behavior is well fitted with the Gaussian function, 
which is shown in the solid lines in Fig.~\ref{Fig:2012Mar14Fig1}. 
This implies that fluctuation of the quark determinant is 
exponentially suppressed as the temperature decreases 
for small $\mu$. 

We have seen that the fermion determinant is insensitive to $\mu$ at low 
temperatures for $\mu<m_\pi/2$. 
As we have discussed in \S~\ref{Sec:2012Mar04sec1}, the eigenvalues follow 
the $N_t$-scaling law. Let us consider an large eigenvalue $\lambda$ ($|\lambda|>1$) 
and describe its counterpart as $1/\lambda^*$. 
As increasing $N_t$ or decreasing $T$, the large eigenvalue $\lambda$ becomes 
lager and the smaller eigenvalue $1/\lambda^*$ smaller. 
If $\mu$ is fixed at a small value, $\xi$ is $O(1)$. 
This leads to the scale separation $|1/\lambda^*| \ll \xi \ll |\lambda|$.  
The contribution of each eigenvalue to the fermion determinant is approximated 
as $\xi + \lambda \sim \lambda$ and $\xi + 1/\lambda^* \sim \xi$. 
This causes the $\mu$-independence of the fermion determinant at low 
temperatures. We will turn back to this point in the next subsection.

The average phase factor $\bra \cos \theta\ket$ is close to one at low 
temperatures as well as high temperatures. 
Does it mean the feasibility of MC simulations at low $T$ ? 
At high $T$, although the fluctuation is small, the $\mu$ dependence of the 
fermion determinant is non vanishing, while the $\mu$-dependence almost 
disappears at low $T$. 
It is already known that the neglecting phase leads to the phase transition 
at $\mu=m_\pi/2$. 
The phase of the determinant plays an important role to go beyond $m_\pi/2$.  
The careful analysis would be required to evaluate the phase of the determinant, 
which may cause another difficulty at low temperature simulations. 

\begin{figure}
\includegraphics[width=7cm]{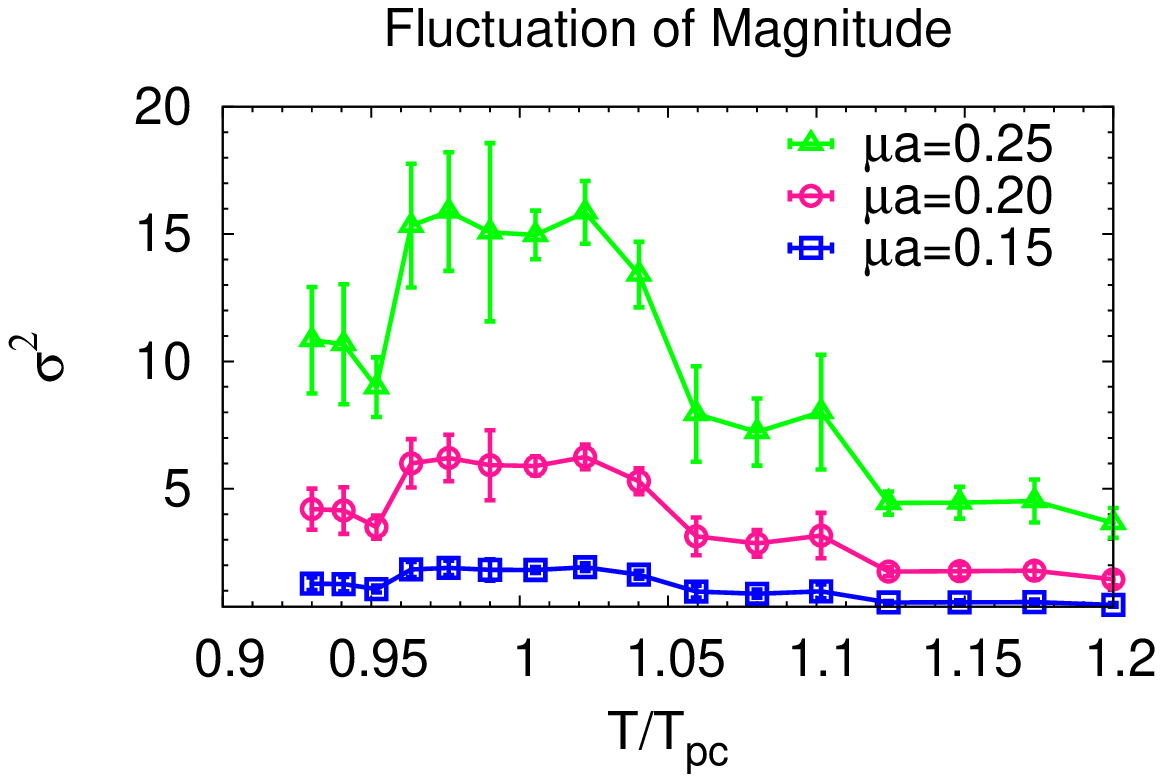}
\includegraphics[width=7cm]{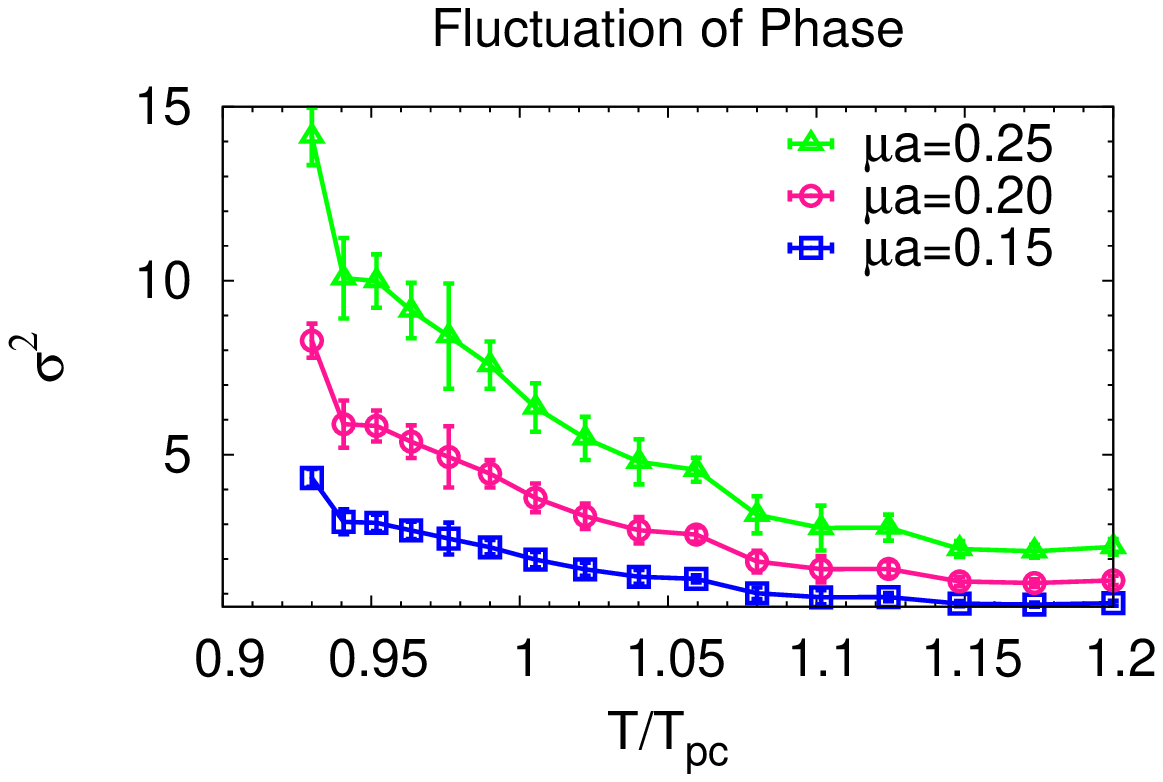}
\caption{The fluctuation of the quark determinant on $10^3\times 4$. 
See also the results for $N_s=8$ in Fig.~\ref{QDet2011Oct21Fig1}.
$\sigma$ is given in the caption of Fig.~\ref{QDet2011Oct21Fig1}. 
}
\label{Nov092011Fig1}
\end{figure}
Note that the above results are obtained for fixed spatial lattice 
size $N_s$. As we have mentioned, the fermion determinant depends 
on $N_s$, although the eigenvalues of the reduced matrix are 
insensitive to $N_s$. 
Finally in this subsection, we consider the volume dependence of the 
reweighting factor. 
The result for $N_s=10$ is shown in Fig.~\ref{Nov092011Fig1}.
See also the result for $N_s=8$ in Fig.~\ref{QDet2011Oct21Fig1}.
Increasing $N_s$ from $8$ to $10$, the maximum value of the deviation 
is almost twice both in the left and right panels, 
which is approximately equal to the volume ratio 
$V_s(N_s=10)/V_s(N_s=8)=10^3/8^3 \sim 2$. 

A question arises on the phase of the fermion determinant at the low temperature 
limit $N_t\to \infty$ and thermodynamical limit $N_s\to \infty$. 
As we have shown, decreasing temperature suppresses the phase, 
while increasing the spatial volume enhances the phase. 
What happens for the phase at the low temperature limit and thermodynamical limit ? 
The result depends in general on the order of taking these two limits.
In this paper, we gave only a result in Fig.~\ref{Fig:2012Mar28fig1}
which was obtained for a fixed lattice volume. 
We will extend the present study for 
various sets of lattice sizes $N_s$ and $N_t$ in the next step
\footnote{
Splittorff and Verbaarschot investigated the average phase factor at low temperature
in chiral perturbation theory~\cite{Splittorff:2007zh}, and
discussed this problem.}.

\subsection{Low temperature limit of the fermion determinant}

According to the $N_t$-scaling law, we parameterize the larger half of 
the eigenvalues as $\lambda_n = l_n^{N_t}e^{i\theta_n}$. 
Using the pair nature, the smaller half of the eigenvalues 
are described as $1/\lambda_n^*=l_n^{-N_t}e^{i\theta_n}$. 
Now, the quark determinant is parameterize as
\begin{align}
\det \Delta & =   C_0\xi^{-\Nred/2} \prod_{n=1}^{\Nred/2}( \xi + l_n^{N_t} e^{i\theta_n} ) 
\prod_{n=1}^{\Nred/2}( \xi + l_n^{-N_t} e^{i\theta_n}), 
\label{Eq:2012Jan05eq1}
\end{align}
where  we set $c_\pm =1$ for simplicity. 
In Eq.~(\ref{Eq:2012Jan05eq1}), we divide the product into two parts
and replace the smaller eigenvalues with $1/\lambda_n^*$. 

In the low temperature limit $T= 1/(a N_t)\to 0, (N_t\to \infty)$, 
large and small eigenvalues follow
\begin{align}
l_n^{N_t}\to \infty, (l_n)^{-N_t} \to 0. 
\end{align} 
Since the fugacity $\xi = \exp(-\mu a N_t)$ also decreases with the same exponent 
as the small eigenvalues, $\det \Delta(\mu)$ in the low-$T$ limit depends on $\mu$. 
\begin{description}
\item[I) Fixed $\mu/T$.] \hspace{1em}
In this case, the fugacity $\xi$ is constant, and the smaller eigenvalues decrease
faster than $\xi$: $\xi \gg l_n^{-N_t}$. 
Then, the quark determinant is reduced to  
\begin{align}
\det \Delta & =  C_0 \prod_{n=1}^{\Nred/2}\lambda_n, 
\label{Eq:2012Jan27eq1}
\end{align}
where the product is taken over the larger half of the eigenvalues $|\lambda_n|>1$. 
$\det \Delta(\mu)$ is independent of $\mu$, and therefore there is no sign problem 
in this limit. 

\item[II) Fixed $\mu a$.] \hspace{1em} 
In this case, the fugacity decreases with the same exponent as the smaller 
eigenvalues. 
This case is further classified into two cases corresponding to the magnitude 
relation of $ \exp (\mu a)$ and $l_n$. 
We introduce a typical magnitude $\bar{l}$, which is used to compare 
the eigenvalues and $\exp(\mu a)$. 
This may be the average of $l_n$ or smallest one of $l_n$ depending on the 
distribution of eigenvalues. We will turn back to this point later. 

\begin{description}
\item[a) $\exp(\mu a) < \bar{l}$.] \hspace{1em}
In this case, the smaller eigenvalues decrease faster than the 
fugacity; $\xi \gg (l_n)^{-N_t}$. We obtain
\begin{align}
\det \Delta & =  C_0 \prod_{n=1}^{\Nred/2}\lambda_n. 
\end{align}
This is equal to Eq.~(\ref{Eq:2012Jan27eq1}), which implies that 
the quark determinant remains unchanged up to $\mu a < \ln \bar{l}$ in 
the low temperature limit. 
Namely, $\det \Delta(\mu)$ is independent of $\mu$ at 
low temperature and small $\mu$ regions. 
This is nothing but the Silver Blaze phenomena discussed above~\cite{Adams:2004yy}.

This is analogous to a situation in Fermi statistics. 
If $T$ is smaller than a lowest excitation energy of a system, then the inclusion 
of small $\mu$ can excite no quark in excited energy levels, the system remains 
to stay at the lowest energy state. 
This can lead the system to be independent of $\mu$. 
We discuss this point in Ref.~\cite{Nagata:2012ad}.

\item[b) $\exp(\mu a) > \bar{l}$.] \hspace{1em}
In this case, the fugacity decreases faster than 
the smaller eigenvalues; $\xi \ll (l_n^*)^{-N_t}$.  
Then, we obtain
\begin{align}
\det \Delta & =  \xi^{-\Nred/2} C_0 \det Q \nn \\
    &= \xi^{-\Nred/2} \prod_{i=1}^{N_t} \det (\beta_i) \nn \\
    &= \xi^{-\Nred/2} \prod_{i=1}^{N_t} \det(B^{ac,\mu\sigma}(\vec{x}, \vec{y}, t_i)\; r_{+}^{\sigma\nu} -2  \kappa \; r_{-}^{\mu\nu} \delta(\vec{x}-\vec{y}) ).
\label{Eq:2012Jan27eq2}
\end{align}
where we first use Eq.~(\ref{Eq:2012Jan01eq5}), then substitute
Eq.~(\ref{Eq:2012Feb21eq1}). In the last line, we use $\det U =1$. 
In Eq.~(\ref{Eq:2012Jan27eq2}), $B(t_i)$ contains 
spatial hops in the $i$-th time slice, but does not contain 
any temporal hopping terms. 
The $\mu$ dependence comes only from the overall factor 
$\xi^{-\Nred/2}=\exp(2 N_c N_s^3  \mu/T)$. 
This is the highest order term in the fugacity expansion and means 
that all the states are occupied by quarks. 

As we have discussed, $\det Q=1$ and $C_0$ is real, therefore 
Eq.~(\ref{Eq:2012Jan27eq2}) is real and free from the sign problem. 
The fermion determinant is real in the low temperature limit both 
for large and small $\mu$. 
However, the fermion determinant is given by the different expressions 
at large and small $\mu$, which may suggest that two different states 
exist at $T=0$. 

\end{description}
\end{description}

One may think to determine the critical value of $\mu_0 a=\ln \bar{l}$. 
However the determination of $\bar{l}$ is nontrivial task because of 
the finite width of the distribution of the eigenvalues. 
If we employ the eigenvalue closest to $1$ for $\bar{l}$, then 
$\bar{l}^{-N_t} =  \max_{|\lambda_n|<1} |\lambda_n|$. 
Using Eq.~(\ref{Eq:2012Feb20eq1}), we obtain 
\begin{align}
\ln \bar{l} = \mu a =  a m_{\pi}/2.
\label{Eq:2012Mar28eq1}
\end{align}
The fermion determinant is independent of $\mu$ for $\mu<m_\pi/2$. 
This value of $\bar{l}$ corresponds to the phase transition point for 
a pion condensation observed in the phase quench simulations. 
If we employ the average value of the larger half of 
$\lambda_n$ for $\bar{l}$, 
we approximately obtain $\bar{l}\sim (200)^{1/4}$. 
We obtain 
$\mu a \sim \ln \bar{l} = \frac{1}{4} \ln 200 \sim 1.32$
which is much larger than $m_\pi/2$ and beyond the lattice cutoff $\mu a = 1$. 

\begin{figure}[htpb] 
\begin{center}
\includegraphics[width=6cm]{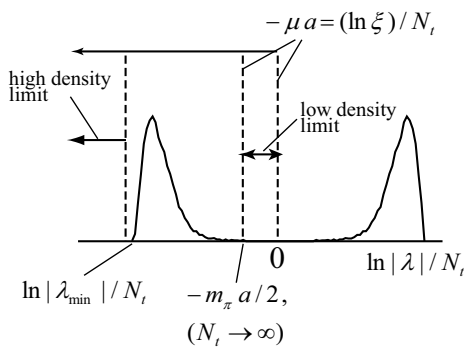}
\caption{Schematic figure for the low temperature limit. 
The solid line denotes the histogram of $\ln |\lambda|/N_t$, 
see also Fig.~\ref{Fig:2012Jan01fig2}.
Dotted vertical lines denote the behavior of $-\mu/T = \ln \xi$ with increasing $\mu$. 
}
\label{Fig:2012Jan27fig1}
\end{center}
\end{figure} 
The criterion $\bar{l}$ may be different for the low density limit 
and high density limit, depending on the eigenvalue distribution at $T=0$. 
The situation of the low temperature limit is shown in a schematic figure 
Fig.~\ref{Fig:2012Jan27fig1}.
If $\mu$ is smaller than $m_\pi/2$, the fugacity is located within the gap. 
Taking $N_t\to \infty$ leads to $|1/\lambda^*| \ll \xi \ll |\lambda|$,
(for $|\lambda|>1$), which corresponds to (II-a).
The fermion determinant is independent of $\mu$ in this case. 
Increasing $\mu$, the fugacity becomes comparable to the smaller half of the 
eigenvalues, which causes the $\mu$ dependence. 
If $\mu$ goes beyond a certain value, which is probably given by 
the minimum eigenvalue, then the low temperature limit (II-b) is 
obtained. The fermion determinant has a trivial $\mu$-dependence in this 
case. 
According to the above discussion, the criterion would be given by 
\begin{align}
&\xi > \max_{|\lambda|<1} |\lambda|, &\mbox{for (II-a)}, \\
&\xi < \min |\lambda|, &\mbox{for (II-b)}.
\end{align}
The criterion for (II-a) is related to the pion mass. 
The criterion for (II-b) may also have a similar interpretation. 
As we have discussed in \S~\ref{sec:red_physint}, the eigenvalues of 
the reduced matrix is related to the free energy of quarks. 
According to the discussion there, the minimum eigenvalue is 
related to the highest energy state of a quark. 
Hence, the low temperature limit (II-b) is obtained if $\mu$ is 
sufficiently larger than the highest energy state of a quark. 


A question arises if the high density limit reflects a real physics of QCD 
or just a consequence of lattice artifacts, because the highest energy state 
probably depends on the lattice spacing $a$ and is a consequence of UV-cutoff. 
The $a$-dependence of the minimum eigenvalue is understood from 
the left panel of Fig.~\ref{Fig:2012Jan22fig2}.  
The increase of $T$ means the decrease of $a$, which 
follows from $T=(a N_t)^{-1}$.  
It turns out that the maximum eigenvalue $\lambda_{\rm max}$ 
becomes larger as $a$ decreases. 
This means that the minimum eigenvalue $\lambda_{\rm min}$ 
becomes smaller with decreasing $a$ because of the relation 
$\lambda_{\rm min}= 1/\lambda_{\rm max}^*$. 
The minimum eigenvalue is described as $\lambda_{\rm min} \sim e^{-F/T}$ 
assuming the eigenvalues correspond to free energies of a quark. 
With decreasing $a$, $\lambda_{\rm min}$ becomes smaller, which means 
$F$ becomes larger. 
Thus, the highest energy state depends on the lattice spacing $a$. 
Investigations of the eigenvalue distribution for finer lattices 
would clarify if the limit (II-b) corresponds to the low temperature 
and high density limit of QCD. 

It is also important to consider the dependence of $\lambda_{\rm max}$ and 
$\lambda_{\rm min}$ on the quark mass and lattice volume as well as the 
lattice spacing. 
In the present work, the value of $\lambda_{\rm max}$ depends on the quark mass (Fig.~\ref{Fig:2012Mar15fig1}), 
while it is insensitive to the lattice volume $N_s^3$ (Fig.~\ref{Fig:2012Jan22fig1}).

\subsection{QCD at low temperature and high density}
In this subsection, we study further the low temperature and high density limit (II-b). 
Even though the limit (II-b) may be the consequence of the lattice cutoff, 
it is meaningful to consider this case. 
For instance, it can be used to generate gauge configurations at high 
density regions. 
So far, there is no case where direct MC simulations are feasible and 
the quark number density is high. If gauge configurations are generated 
at high density regions by direct MC simulations, they may provide 
valuable information, e.g., for multi-ensemble reweighting.

Using Eq.~(\ref{Eq:2012Jan27eq2}), we obtain the low-temperature and 
high-density limit of the QCD partition function
\begin{subequations}
\begin{align}
\lim_{T \to 0} Z_{GC}(\mu, T)  &= e^{2 N_f N_c N_s^3  \mu/T} \int \calD U  
\left(\det \Delta(\mu)|_{T\to 0}\right)^{N_f} e^{-S_G}, 
\label{Eq:2012Feb23eq1}
\\
\det \Delta(\mu, T)|_{T\to 0} &= \prod_{i=1}^{N_t} 
\det\left(B^{ac,\mu\sigma}(\vec{x}, \vec{y}, t_i)\; 
r_{+}^{\sigma\nu} -2 \kappa \; 
r_{-}^{\mu\nu} \delta(\vec{x}-\vec{y})\right), 
\label{Eq:2012Feb23eq2}%
\end{align}%
\label{Eq:2012Feb23eq3}%
\end{subequations}%
where the definitions of $B$, $r_\pm$ etc were given 
in \S~\ref{sec:reduction_formulation}.

In Eq.~(\ref{Eq:2012Feb23eq3}), the gluon part remains unchanged and is given 
by the ordinary Yang-Mills action, while the fermionic part is different from 
the ordinary QCD action. 

Quarks interact only in spatial directions, where no interaction exists 
in the temporal direction.

Equation (\ref{Eq:2012Feb23eq2}) is also different from the naive spatial fermion matrix $B$, 
but contains a constant term $2\kappa$ with the projection operator $r_\pm = (r \pm \gamma_4)/2$. 
The quark chemical potential appears only in the bulk factor 
$\exp(2 N_f N_c N_s^3 \mu/T)$, which gives the quark number density, 
\begin{subequations}
\begin{align}
\bra n \ket &= 2 N_f N_c            , \;\;\mbox{ (lattice unit)} \\
            &\propto 2 N_f N_c \mu^3, \mbox{(physical unit)}. 
\end{align}
\end{subequations}

Since Eq.~(\ref{Eq:2012Feb23eq2}) contains no temporal hopping term, 
the partition function is $\ZNc$ symmetric even in the presence of 
dynamical quarks. 
Naively, it is expected that a deconfinement transition occurs if baryon 
number density is large enough to cause the overlap of baryon's wave 
functions, where effective degrees of freedom would be quarks rather than 
hadrons. 
If this is the case, $\ZNc$ symmetry would be an exact symmetry of QCD 
in extremely high-dense matter. 

Another high density limit was proposed in Ref.~\cite{Blum:1995cb} with an 
approximation that the quark mass and chemical potential are simultaneously made 
large. In the approximation, the quark mass depends on the chemical potential. 
In the present case, the fine tune of the quark mass is not needed in taking the 
low-$T$ and high-$\mu$ limit. 

Equation~(\ref{Eq:2012Feb23eq3}) is realized at extremely large $\mu$.  
It would be very difficult to access such a high density region in 
experiments. 
Nevertheless there are several interests in the high density limit (II-b). 
Theoretically, it is interesting to consider the nature of QCD at 
high density regions : confinement, chiral symmetry and color superconductivity. 
In lattice QCD simulations, the knowledge on important configurations at high 
density regions would be valuable information. 
For instance, such configurations can be used in multi-ensemble reweighting 
method, or 
they may be used for a reweighting from high density regions in order to find a 
QCD phase transition line at low temperatures. 

It is important to consider the numerical feasibility. 
As we have mentioned, whole the determinant is real, and therefore sign free. 
Although it is free from the sign problem, it needs large $N_T$ to 
take the low temperature limit, which requires a large numerical cost. 
This increase of the computational time may be suppressed by using the 
property of the fermion determinant in the low temperature limit. 
In the low-$T$ and high-$\mu$ limit, 
the fermion determinant is expressed as the product of the $N_t$ block determinants. 
If each block determinant is real, it may be possible to evaluate the block 
determinant by the Gaussian integral with the pseudo-fermion field with smaller 
dimension. 
Instead, the Gaussian integral appears $N_t$ times. 
The real positivity of the block determinant is necessary to implement this idea. 
We leave the proof of this expectation in future works.

\renewcommand{\mod}{{\rm mod}}
\renewcommand{\Im}{{\rm Im}}
\renewcommand{\Re}{{\rm Re}}
\section{The partition function on the complex plane}
\label{Sec:canonical}
The determination of the confinement/deconfinement phase boundary is an important issue 
in the study of the QCD phase diagram.
A canonical formalism and Lee-Yang zero analysis are useful approaches to 
identify a phase transition point, and have been investigated in 
Refs.~\cite{Hasenfratz:1991ax,Kratochvila:2005aaa,Kratochvila:2004wz,Kratochvila:2005mk,Kratochvila:2006jx,Ejiri:2008xt,Li:2008fm,Alexandru:2007bb,Alexandru:2005ix,Meng:2008hj,Li:2007bj,Alexandru:2010yb,Li:2011ee,Danzer:2012vw}. 
However, some difficulties were pointed out~\cite{Kratochvila:2005mk,Ejiri:2005ts}. 
For instance, the sign problem causes a fictitious signal in Lee-Yang zero analysis, 
which makes it difficult to distinguish a physical phase transition and fictitious 
signal~\cite{Ejiri:2005ts}.

In this section, we consider the canonical formalism and Lee-Yang zero theorem, 
where the reduction formula is useful. 
The purpose of this section is to propose a method to identify a $\mu$-induced 
phase transition by using a temperature dependence of canonical partition 
functions and of Lee-Yang zero trajectories. 
The method provides a qualitative way to distinguish a crossover and 
first order phase transition, and would be useful in case that the sign 
problem causes a difficulty in ordinary methods such as finite size 
scaling of the Lee-Yang zero near a positive real axis. 

Canonical partition functions $Z_n$ with the quark number $n$ are obtained 
from the fugacity expansion form of the fermion determinant. 
Then, we consider the temperature dependence of two quantities: the canonical 
distribution $Z_n$ and trajectory of Lee-Yang zeros. 
They show characteristic changes as the temperature decreases from 
high $T$ to low $T$. 
We discuss the relation between their temperature dependences and the existence 
of a $\mu$-induced phase transition. 


In \S~\ref{sec:canonicalsec1}, we give a brief overview of the 
canonical formalism and Lee-Yang zero analysis. 
Fugacity coefficients of the fermion determinant are calculated in 
\S~\ref{sec:canonicalsec2}.
Canonical partition functions are calculated in \S~\ref{sec:canonicalsec3}, 
where we employ the Glasgow method~\cite{Barbour:1997bh}. 
Lee-Yang zeros are calculated in \S~\ref{sec:canonicalsec4}.

\subsection{Brief overview}
\label{sec:canonicalsec1}

According to statistical mechanics, a grand canonical partition function 
is described as a superposition of canonical partition functions, 
\begin{align}
Z(T,\mu) = \sum_{n=1}^N Z_n(T) \xi^n, 
\label{Eq:2012Feb29eq2}
\end{align}
where $Z_n$ describes a canonical partition function with a 
fixed particle number $n$, and $N$ is the maximum number of the particle. 
In thermodynamical limit, $N\to \infty$.  

Several methods are available for the determination of phase transition points.  
For instance, a phase transition point is identified by the convergence radius of 
Eq.~(\ref{Eq:2012Feb29eq2}), or the finite size scaling analysis of a Lee-Yang zero $Z(\mu)=0$,  
the Maxwell construction for an S-shape in a $\mu$-$n$ diagram, etc.

The grand partition function Eq.~(\ref{Eq:2012Feb29eq2})
can be considered as a polynomial of $\xi$ at a given temperature. 
Phase transitions result from discontinuities of derivatives of the free energy.
In general, $Z(\xi)$ contains the $N$-roots on the complex $\xi$ plane.  
Since the canonical partition function $Z_n$ is real positive 
for all $n$, no root exists for real positive values of 
$\xi$.  Hence, all the roots are distributed somewhere on the complex $\xi$
plane except for the positive real axis.
Lee and Yang showed ~\cite{Yang:1952be,Lee:1952ig} that in the case a phase transition occurs, zeros of the grand partition function approach to the 
positive real axis in the thermodynamical limit $V\to \infty$. 
Thus, the phase transition is described by zeros of the grand partition 
function, which are called the Lee-Yang zeros. 
The order of the phase transition is distinguished by considering the trajectory
formed by zeros near the positive real axis~\cite{Blythe:2003aaa}. 
In Ref.~\cite{Blythe:2003aaa} the application to non-equilibrium systems was 
also discussed. 
Stephanov investigated the properties of Lee-Yang zero of QCD by 
using the scaling and universality~\cite{Stephanov:2006dn}. 
In lattice QCD simulations, the thermodynamical singularities of Lee-Yang 
zeros were investigated in e.g. Refs.~\cite{Barbour:1991vs,Fodor:2001pe,Fodor:2004nz,Ejiri:2005ts,Ejiri:2009bs}. 

\subsection{Fugacity expansion of the fermion determinant}
\label{sec:canonicalsec2}
We start from the fugacity expansion of $\det \Delta(\mu)$,  
\begin{align}
\det \Delta(\mu)= C_0 \sum_{n=-\Nred/2}^{\Nred/2} c_n \xi^n,
\end{align}
which is defined in Eq.~(\ref{Jan1511eq1}). 

Before seeing numerical results, we discuss the numerical procedure for the calculation of 
$c_n$. 
There are two difficulties in the determination of $c_n$; numerical precision and 
overflow/underflow due to the applicable range of the double precision.

First, the expansion of Eq.~(\ref{Nov292011eq1}) requires an enormous amount of 
calculations because of the larger number $\Nred$, which easily causes a loss of precision. 
A recursive algorithm with a recurrence relation was used to expand Eq.~(\ref{Nov292011eq1}). 
A Vandermonde matrix approach was not effective because of the large 
dimension and of the existence of close eigenvalues. 
Divide and conquer algorithms are best in respect to the number of the 
calculation steps. 
Second, $c_n$ span a wide range of magnitude, from order $O(1)$ to $O(e^{4 N_c N_s^3})$. 
A huge number such as $O(e^{4 N_c N_s^3})$ exceeds the maximum value of 
double precision, where an ordinary double precision variable is not applicable. 
Numerical libraries such as FMLib~\cite{Web:FMLIB} are available 
for this calculation. Other libraries have been also used in
literature, see e.g., Ref.~\cite{Alexandru:2010yb}. 
However, the use of libraries may become an obstructive factor of fast computation.
Therefore, we developed a new variable~\cite{Nagata:2010xi}. 
These procedures to calculate $c_n$ are explained in Appendix.~\ref{App:WideRangeNum}. 

\begin{figure}[htpb] 
\begin{center}
\includegraphics[width=6cm]{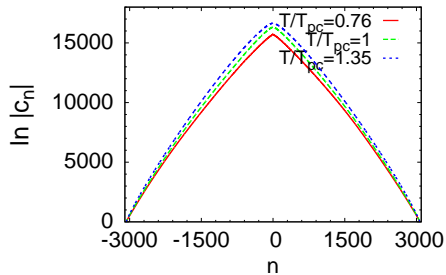}
\caption{The magnitude of all the fugacity coefficients for various 
temperature. The lattice size is $8^3\times 4$. Data is taken from one 
configuration for each temperature.}
\label{Fig:2012Jan31fig1}
\end{center}
\end{figure} 
\begin{figure}[htpb] 
\begin{center}
\includegraphics[width=4.5cm]{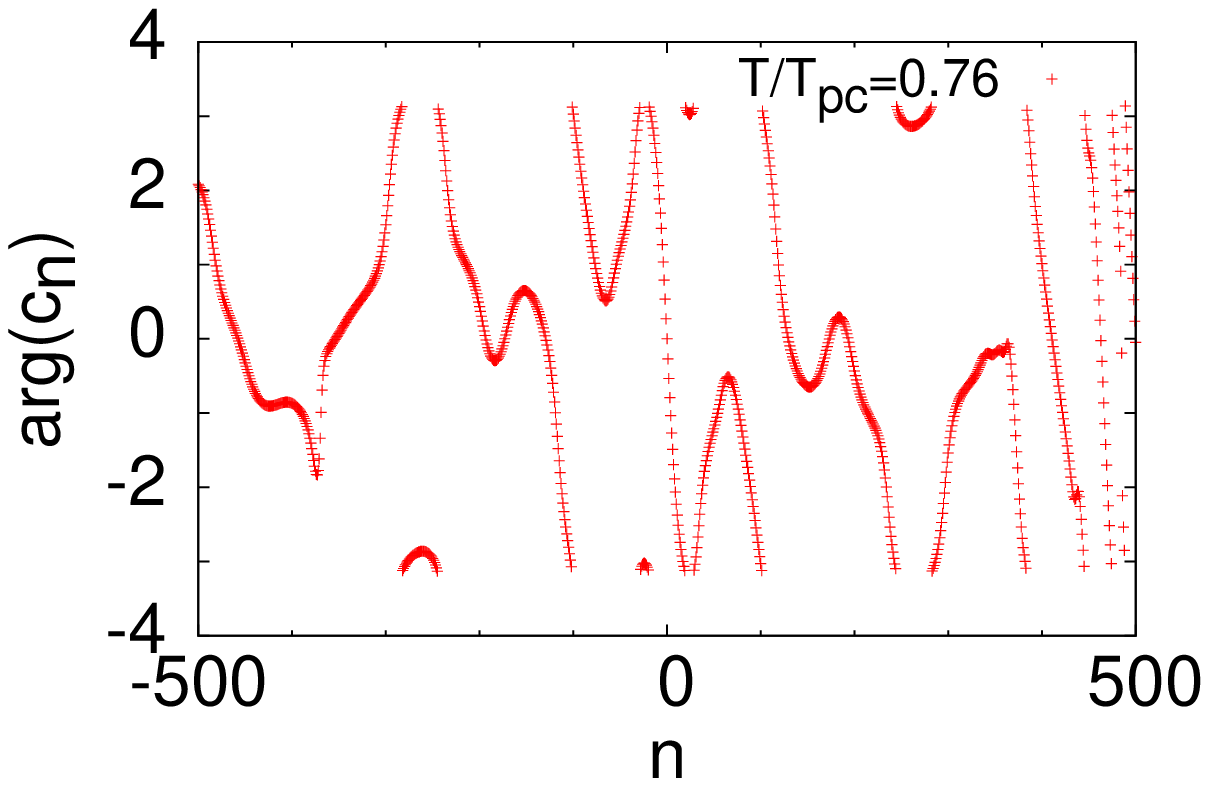}
\includegraphics[width=4.5cm]{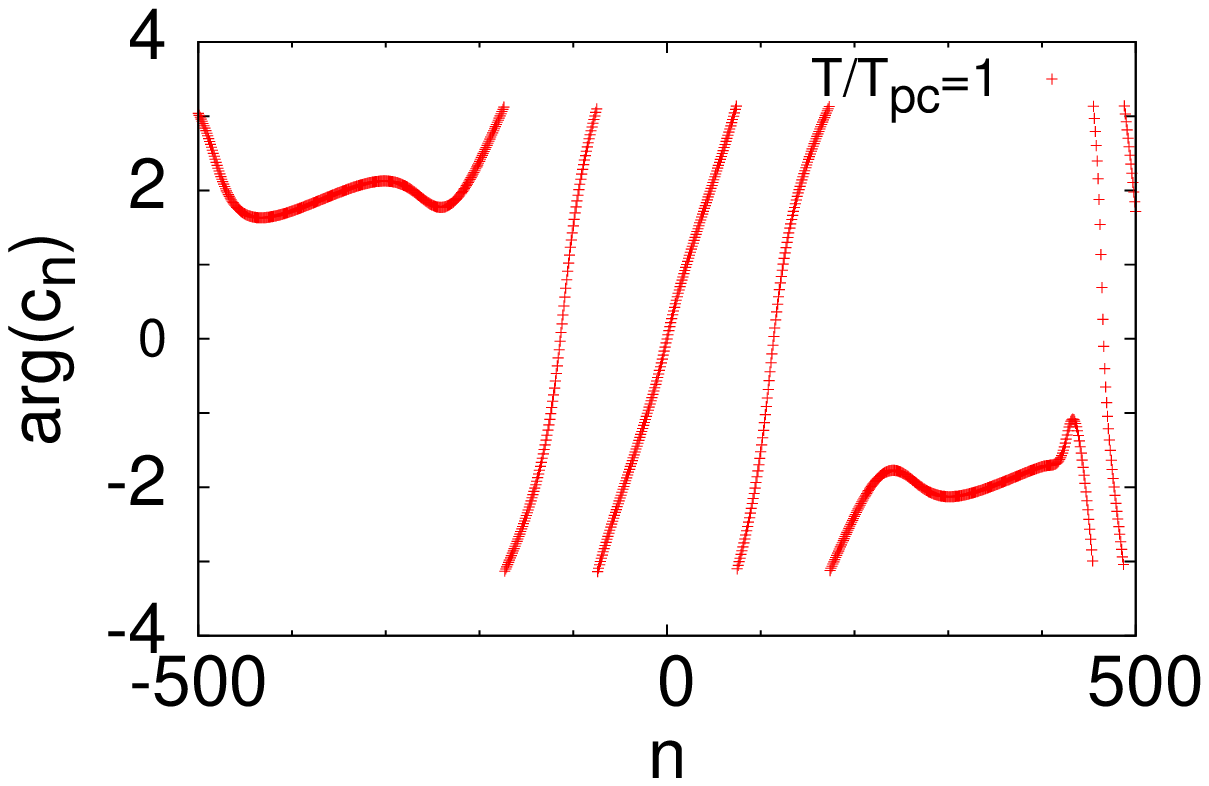}
\includegraphics[width=4.5cm]{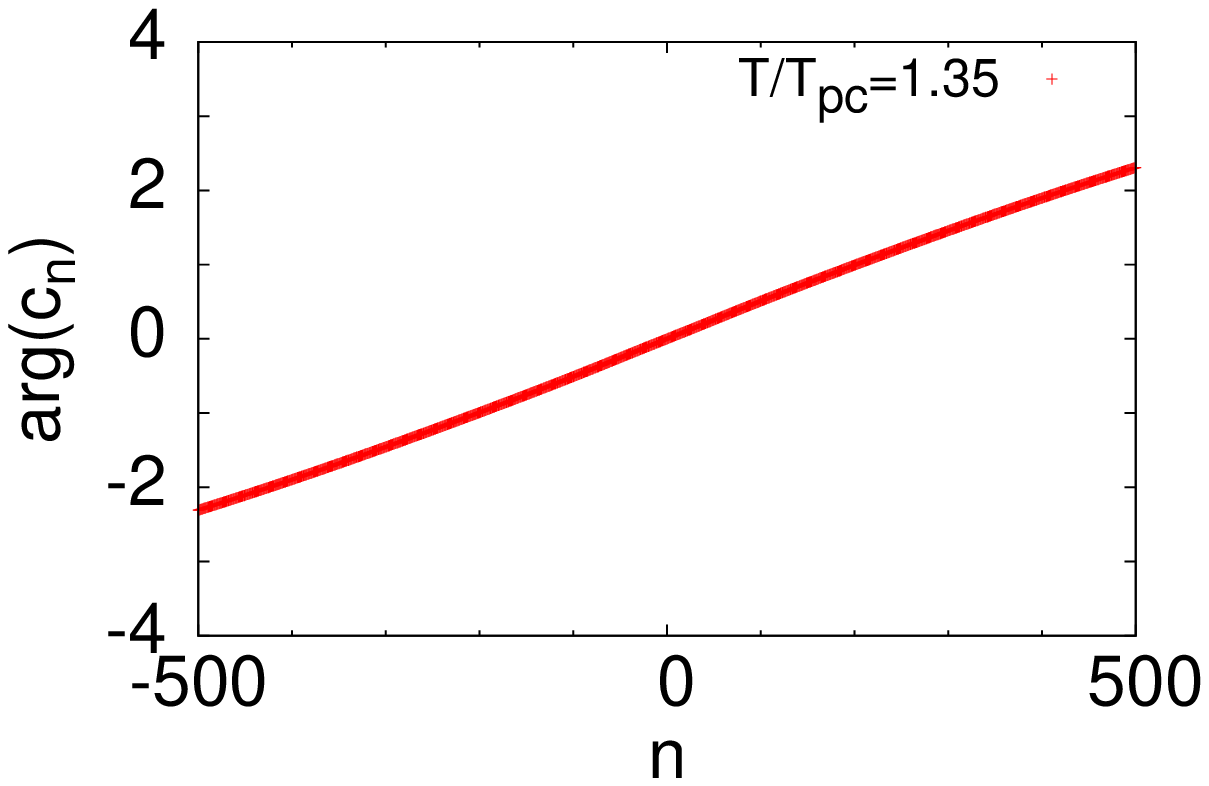}
\caption{The argument of the fugacity coefficients for various 
temperature. Data are shown for small quark number sector. 
The lattice size is $8^3\times 4$. Data is taken from one 
configuration.}
\label{Fig:2012Jan27fig2}
\end{center}
\end{figure} 
Now we proceed to the numerical results of $c_n$. 
Figures~\ref{Fig:2012Jan31fig1} and ~\ref{Fig:2012Jan27fig2} show
the modulus and argument of the fugacity coefficients $c_n$ for
a gauge configuration. 
The simulation setup was given in the section \ref{subsec3a}. 
The magnitude $|c_n|$ spans over the wide range of order from $O(1)$ 
to $O(e^{10^4})$. At $\mu=0$, $\det \Delta(\mu)$ is dominated by 
several $c_n$ near $n=0$. 
The argument of $c_n$ shows complicated $n$ dependence, as shown in 
Fig.~\ref{Fig:2012Jan27fig2}, where $\arg(c_n)$ is defined for 
$-\pi \le \arg(c_n) \le \pi$. 
Qualitatively, $\arg(c_n)$ tends to oscillate with higher frequency 
as the temperature becomes lower. 
Calculating $c_n$ for 400 configurations, we found that $|c_n|$ was stable 
for the change of configurations. 
On the other hand, $\arg(c_n)$ rapidly changes for configuration by configuration,  
which leads to the cancellation of $c_n$ in the ensemble average. 
This cancellation becomes more severe at low temperatures. 
We will see this point in the next subsection. 

The fugacity coefficients $c_n$ satisfy the relation $c_{-n}^* = c_n$, 
as a consequence of the $\gamma_5$ hermiticity. Hence, positive and 
negative $n$ describe the winding number of a quark and an anti-quark
around the temporal cylinder, respectively.
This is realized as the reflection symmetry $|c_{-n}|=|c_{n}|$ 
with respect to $n=0$ for the absolute value, which is well satisfied 
see Fig.~\ref{Fig:2012Jan31fig1}. 
The relation for the phase, which is given by $\arg(c_{-n}) = - \arg(c_n)$, 
was numerically satisfied for several hundreds $c_n$ near 
$n \sim 0$ and near $n\sim \pm \Nred/2$. 
For instance, in the left panel of Fig.~\ref{Fig:2012Jan27fig2}, 
$\arg(c_{-n}) = - \arg(c_n)$ for $n=0\sim 300$, while $\arg(c_{-n}) \neq - \arg(c_n)$ for $n\sim 500$. 
Fortunately, the quark determinant is dominated by coefficients near $n=0$ 
for $\mu=0$. 

\subsection{Canonical partition function}
\label{sec:canonicalsec3}
The grand canonical partition function of QCD with $N_f$ flavors can 
be also expanded in powers of the fugacity, 
\begin{align}
Z_{GC}(\mu, T) = \sum_{n=-N_q}^{N_q} Z_{C}(n, T) \xi^{n},
\label{Eq:2012May15eq1}
\end{align}
where $N_q=N_f \Nred/2 = 2N_f N_c N_s^3$ is the maximum quark number
which can be put on the $N_s^3$ lattice. 
$Z_C(n, T)$ is a canonical partition function with a fixed quark number $n$. 
Using the Fourier transformation~\cite{Kratochvila:2004wz,Kratochvila:2005mk,Kratochvila:2006jx}, 
the canonical partition function is obtained 
\begin{align}
Z_C(n) = \frac{1}{2\pi/3} \int_{-\pi/3}^{\pi/3} d\phi e^{-in \phi} Z_{GC}(\mu=i\mu_I), 
\label{Eq:2012Feb29eq1}
\end{align}
where $\phi=\mu_I/T$. We have used the Roberge-Weiss periodicity. 

In this work, we employ the Glasgow method, which is based on the reweighting 
in $\mu$ and reduction formula, for the calculation of the canonical partition 
functions. 
It was pointed out that it suffers from the overlap problem~\cite{Fodor:2001au}.
In this work, we focus on the properties of the canonical partition functions 
and Lee-Yang zeros as a first step, and leave the improvement of the overlap 
for future works. 

Substituting the reduction formula into Eq.~(\ref{Eq:2012Feb29eq1}), 
the canonical partition function is given by 
\begin{align}
Z_n \equiv Z_{C}(n) = \left\langle \frac{C_0^2 d_n}{(\det \Delta(0))^2}\right \rangle_0.
\end{align}
Here $d_n$ are the fugacity coefficients of the two-flavor 
determinant, and $\bra \cdot \ket_0$ denotes an ensemble average for 
gauge configurations generated at $\mu=0$.   
$d_n$ is determined by using the recursive algorithm for 
$\left(\xi^{-\Nred/2}\prod_{n=1}^{\Nred}(\xi+\lambda_n)\right)^{N_f} = \sum_{n=0}^{N_f \Nred} d_n \xi^n$. 
Although it is possible to obtain $d_n$ in terms of $c_n$, this method was slower than
 the above procedure.

\begin{figure}[htpb] 
\begin{center}
\includegraphics[width=6cm]{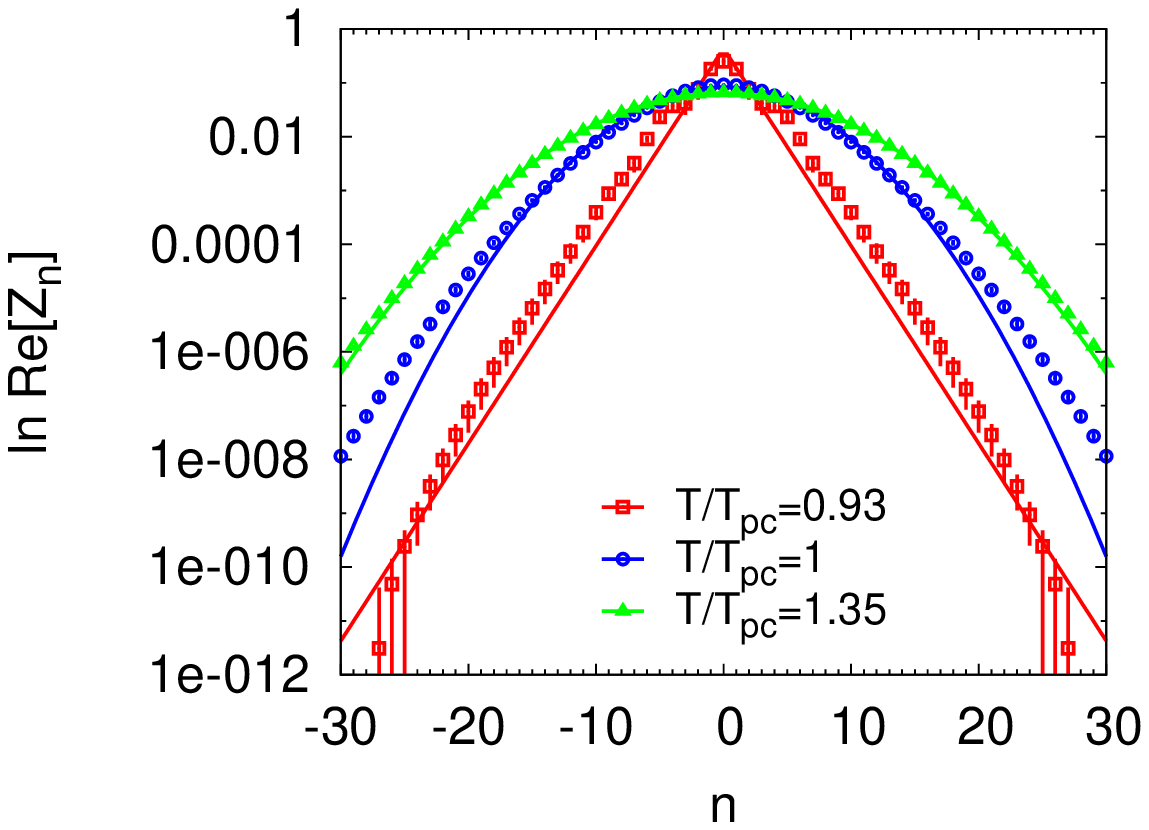}
\includegraphics[width=6cm]{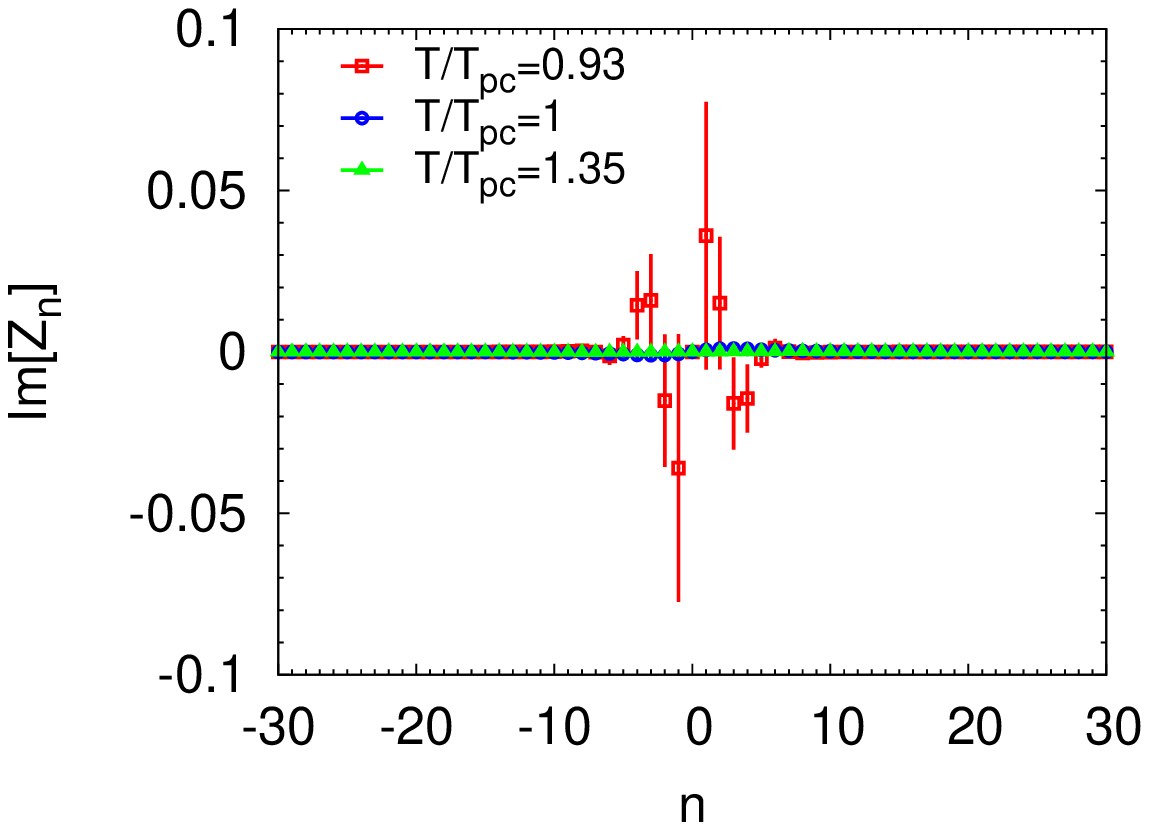}
\caption{The canonical partition functions $Z_n$ for three temperatures. 
Left : real part, right: imaginary part. 
In the left panel, the solid lines are $\exp(- a_L |n|)$ for 
$T/\Tpc=0.93$ and $\exp ( - a_H n^2)$ for $T/\Tpc=1$ and $1.35$. 
The results are obtained by using the Glasgow method with the use of 400 
sets of the eigenvalues of the reduced matrix. 
}\label{Fig:2012Feb03fig1}
\end{center}
\end{figure} 

The canonical partition functions $Z_n$ are shown up to $|n|=30$ 
for three temperatures in Fig.~\ref{Fig:2012Feb03fig1}. 
$Z_n$ must be real positive, i.e., $\Im [Z_n]=0$. 
The results for high temperatures satisfy this condition. 
Even at low temperature, $\Im [Z_n]$ is zero within errorbars for most $n$. 
However,  the errorbars are large, which is caused by the 
fluctuation of the phase of the fugacity coefficients $c_n$ for 
configuration by configuration, which is the sign problem in 
canonical approaches~\cite{deForcrand:2006ec}. 
The signal-to-noise ratio for $\Re [Z_n]$ becomes small for large $n$. 
This is probably because of the importance sampling at $\mu=0$.  
The average value of the quark number density is zero at $\mu=0$, and 
therefore configurations generated at $\mu=0$ have less overlap 
with large quark number sectors. 
The canonical partition functions are obtained up to about $|n|=30$ 
in the present simulation setup. 
The calculation of $Z_n$ for larger $n$ needs an improvement of overlap 
or increase of statistics. 
However, it should be noted that the real part of the canonical partition 
functions show a characteristic temperature dependence even for $|n|<30$. 


The canonical partition function $\Re[Z_n]$ exponentially decreases as $|n|$ 
increases for all the three temperatures. 
The width of the distribution of $\Re[Z_n]$ becomes broad at high temperature, 
which is consistent with the increase of the effective degrees of freedom at high $T$. 
The $n$-dependence of $\Re [Z_n]$ changes at high temperatures ($T/\Tpc=1, 1.35$) 
and low temperature $(T/\Tpc=0.93)$. 
Qualitatively, $\Re [Z_n]$ is approximated by a Gaussian function $\exp(-a_H n^2)$ for $T\ge \Tpc$, 
and to a function $\exp(-a_L |n|)$ for $T< \Tpc$, where $a_{H,L}$ are parameters. 
To be precise, the result agrees with the Gaussian function for $T/\Tpc=1.35$ up to $|n|=30$,  
while there is a deviation from the Gaussian function for $|n|=20\sim 30$ for $T/\Tpc=1$. 
There is also deviation from the function $\exp(-a_L |n|)$ for $T/\Tpc=0.93$. 
In case of $\exp(-a_L |n|)$, the partition function becomes a geometric series. 

Triality nonzero terms $\mod(n,3)\neq 0$ do not vanish, which
is a consequence of the importance sampling at $\mu=0$. 
Those terms can be eliminated by using the Roberge-Weiss periodicity~\cite{Roberge:1986mm}.  
We discuss this point in Appendix.~\ref{Sec:App_triality}.
The triality nonzero terms change neither the trajectory of 
Lee-Yang zeros nor the convergence radius of the fugacity polynomial 
as long as $Z_n$ with zero- and nonzero-triality is described by the same function of $n$. 
Triality nonzero terms change the density of Lee-Yang zeros on its trajectory. 
However, this effect vanishes if thermodynamical limit is taken. 
Hence, we keep the triality nonzero terms, which does not cause 
any problem in the following discussion.
Here it is interesting to note that the low-temperature high density limit of 
QCD may suggest the importance of the other $Z_3$ sectors. 
If all the $Z_3$ sectors are visited in MC simulations, the triality
nonzero terms would disappear. 

The $n$-dependence of the canonical partition functions change from 
high temperatures to low temperatures. 
Now, we discuss the relation between this change of the $n$-dependence of 
$\Re[Z_n]$ and a finite density phase transition. 
We found that $\Re[Z_n]$ is approximately given by 
\begin{align}
\Re[Z_{C}(n)] \sim 
\left\{\begin{matrix}
 e^{- a_{\rm L} |n|}, & (T<\Tpc), \\
 e^{- a_{\rm H} n^2}, & (T\ge\Tpc).
\end{matrix}\right.
\label{Eq:2012Mar24eq2}
\end{align}
Assuming that these distributions hold for larger $n$ and that 
the imaginary part is sufficiently small, the convergence radius $r$ of 
Eq.~(\ref{Eq:2012May15eq1}) is given by 
\begin{align}
r^{-1} =\lim_{n\to \infty} \left| \frac{Z(n+1)}{Z(n)} \right| = 
\left\{\begin{matrix}
 e^{ - a_L}, & (T< \Tpc), \\
 0 , & (T \ge \Tpc).
\end{matrix}\right.
\label{Eq:2012Mar24eq1}
\end{align}
The Gaussian function for $T\ge \Tpc$ suggests no phase transition, 
while the function $e^{-a|n|}$ for $T<\Tpc$ suggests the existence of a 
phase transition at finite $\mu$. 
Note that the result for $T=\Tpc$ shows the Gaussian behavior, 
which is consistent with the absence of a $\mu$-induced phase transition at $T=\Tpc$.
Thus, the shape change of the canonical distribution is related 
to exisitence or absence of a $\mu$-induced phase transition.

The deviation from Eq.~(\ref{Eq:2012Mar24eq2}) observed for $T=\Tpc$ and $T=0.93\Tpc$ 
may be important in the study of CEP. 
The shape of the canonical distribution including the large-$n$ part and 
the deviation from Eq.~(\ref{Eq:2012Mar24eq2}) can contribute to higher-order moments, 
such as skewness and kurtosis. 
In the present analysis, we employed the one-parameter reweighting, where 
the overlap problem becomes severe for large $n$ parts of $Z_n$.  
The improvement of the overlap is needed in order to apply the above discussion 
to cases where the tail part of the canonical distribution is important. 

\subsection{Lee-Yang zeros}
\label{sec:canonicalsec4}
In this subsection, we consider the Lee-Yang zeros. 
Several methods are available for the calculation of Lee-Yang zeros. 
For instance, it is possible to search zeros of the left hand side of Eq.~(\ref{Eq:2012May15eq1}). 
It is also possible to solve roots of the fugacity polynomial, 
which is the right hand side of Eq.~(\ref{Eq:2012May15eq1}). 
Here, we adopt the latter approach by using the canonical partition 
functions obtained in the previous subsection. 

In order to calculate roots of the fugacity polynomial,  
We consider the truncation of the fugacity polynomial 
in order to calculate roots of the right hand side of Eq.~(\ref{Eq:2012May15eq1}), 
because the order of the fugacity polynomial $N_q$ is large. 
In general, it is not allowed to truncate the polynomial in the vicinity of 
phase transition points. 
However, the truncated polynomial can reproduce the {\it trajectory} of 
Lee-Yang zeros in the case of the geometric series. 
First, we discuss this point. 

We divide the fugacity polynomial into small $n$ and large $n$ parts
\begin{align}
Z_{GC}(\mu) = \sum_{|n|\le M} Z_{C}(n) \xi^{n}  + \sum_{|n|\ge M} Z_{C}(n) 
\xi^{n} 
\label{Eq:2012Mar02eq1}
\end{align}
where $M$ is an integer. 
Although $\Re[Z_n]$ rapidly decreases as $|n|$ increases, the sum of the 
higher-order terms significantly contribute to $Z_{GC}(\mu)$ in the vicinity of the 
convergence radius, i.e., phase transition points. 

At high $T$, the higher-order terms do not affect the convergence 
property because of the Gaussian shape. 
At low $T$, the $e^{-a |n|}$-shape provides the nonzero convergence radius, 
where the sum of higher-order terms is significant near transition points.
However, if $\Re[Z_n]=\exp(-a |n|)$ even for large $n$, 
the trajectory formed by Lee-Yang zeros does not depend on 
the maximum order of the fugacity polynomial. 
This is a consequence of the geometric series. 
For instance, considering  $1+x+x^2+\cdots + x^N=0$, 
its roots lie on the unit circle. 
The order $N$ changes the density of the roots on the circle, 
but does not change the trajectory formed by roots.

In fact, the fugacity polynomial with $\Re[Z_n]=\exp(-a |n|)$ is different 
from a naive geometric series, because it contains both the quark and 
anti-quark components. However, as we will show below, the 
trajectory of the roots for the quark and anti-quark sectors are separated from 
inside and outside of $|\xi|=1$ because of the symmetry between 
quarks and anti-quarks. 

\begin{figure}[htpb] 
\begin{center}
\includegraphics[width=6.5cm]{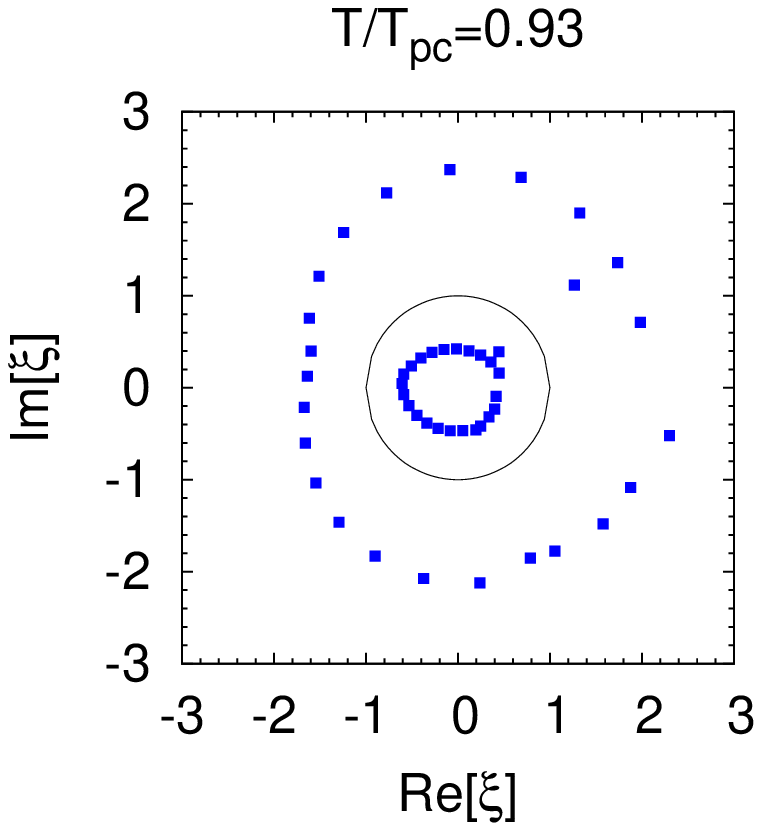}
\includegraphics[width=6.5cm]{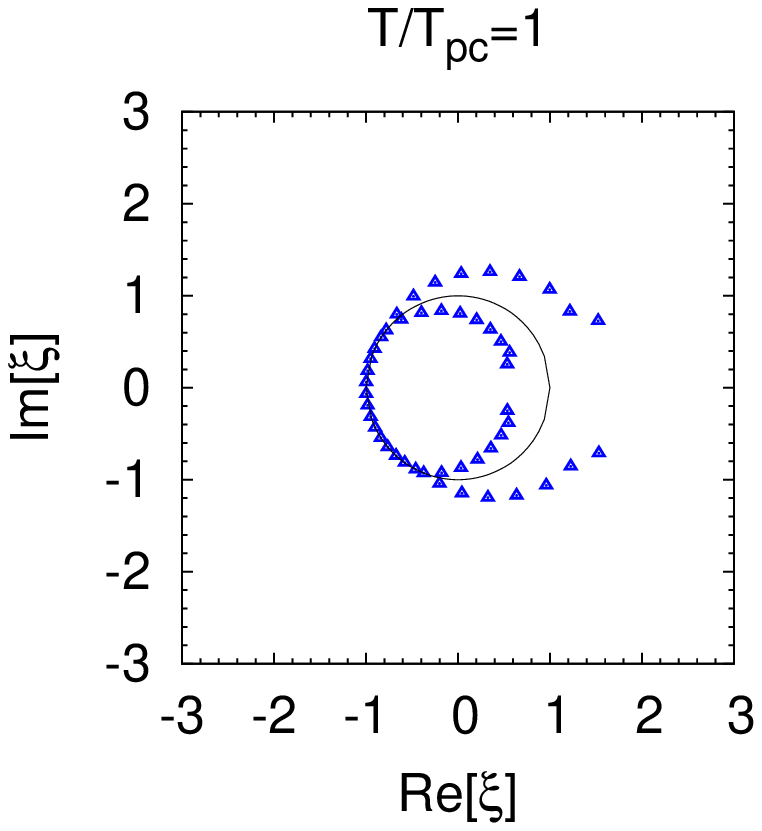}
\includegraphics[width=6.5cm]{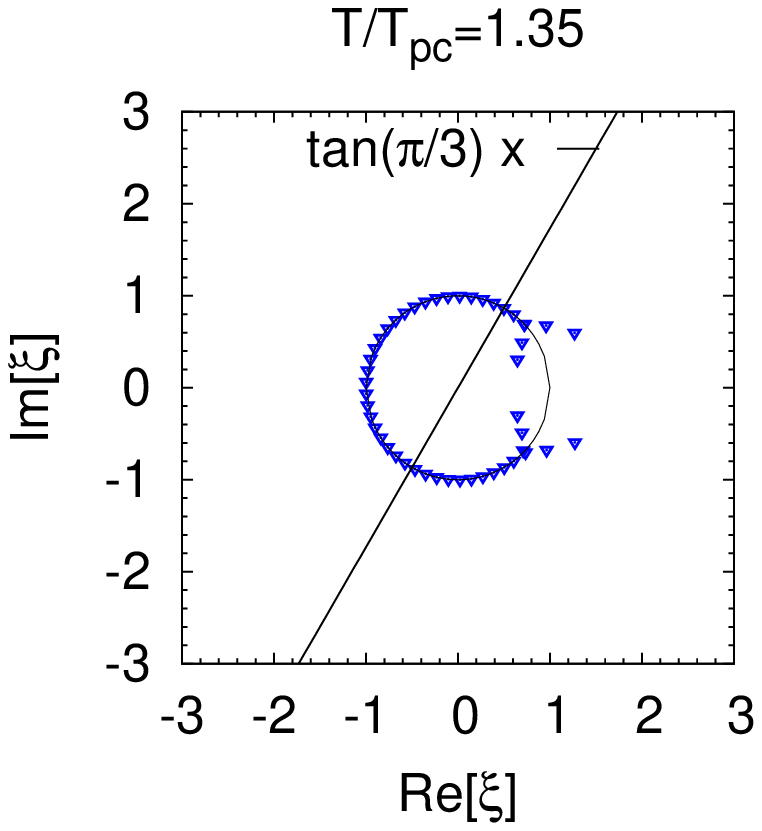}
\caption{The distribution of Lee-Yang zeros, which are roots of 
the fugacity Polynomial shown in Fig.\ref{Fig:2012Feb03fig1}.
The roots are obtained for the leading order terms for $M=24$.
Solid lines are the unit circle.}
\label{Fig:2012Feb03fig2}
\end{center}
\end{figure} 
Accordingly, the Lee-Yang zero trajectory can be reproduced via the truncated polynomial 
on the assumption on $Z_n$. 
Now, we consider the numerical result. 
We employ the data of $Z_n$ up to $M=24$, because the signal-to-noise ratio becomes large for 
$n>M$ at $T/\Tpc=0.93$. 
Although this value of $M$ is small, it corresponds to the baryon density 
$\sim$ 2 [fm$^{-3}$]. 
IMSL Library was used for the calculation of the roots of the fugacity polynomial. 
The result is shown in Fig.~\ref{Fig:2012Feb03fig2}. 
Corresponding to the change of $Z_n$, the Lee-Yang zero trajectory also changes its shape 
from high temperatures to low temperatures. 
The zeros are approximately distributed on two circles at $T/\Tpc=0.93$ and one circle with 
two branches at $T/\Tpc=1$ and $1.35$. 
The trajectory is symmetry with regard to the unit circle because of the charge conjugation 
symmetry between the quark and anti-quark.

The results for the Lee-Yang zero trajectories were obtained by using the data of $Z_n$ shown 
in Fig.~\ref{Fig:2012Feb03fig1}, where both the real and imaginary parts 
were considered. We did not use Eq.~(\ref{Eq:2012Mar24eq2}) to obtain 
Fig.~\ref{Fig:2012Feb03fig2}. 
Hence, the trajectories in Fig.~\ref{Fig:2012Feb03fig2} contain the deviation of 
$Z_n$ from Eq.~(\ref{Eq:2012Mar24eq2}).
The two-circle trajectory at low $T$ is similar to a typical behavior of geometric 
series with positive and negative $n$ components. 
If the polynomial is an ordinary geometric series with positive $n$ components, 
then the roots lie on a circle. Adding the negative $n$ components with 
the charge conjugation symmetry, then the roots locate on two circle. 
Similarly, the two-branch trajectory is a typical behavior of the Gaussian distribution.
Therefore, the trajectories obtained from the data of $Z_n$ qualitatively agree 
with the trajectories obtained from Eq.~(\ref{Eq:2012Mar24eq2}), 
which suggests that Eq.~(\ref{Eq:2012Mar24eq2}) is a good approximation for $Z_n$ 
at least for small $n$.

\begin{figure}[htpb] 
\begin{center}                      
\includegraphics[width=6.5cm]{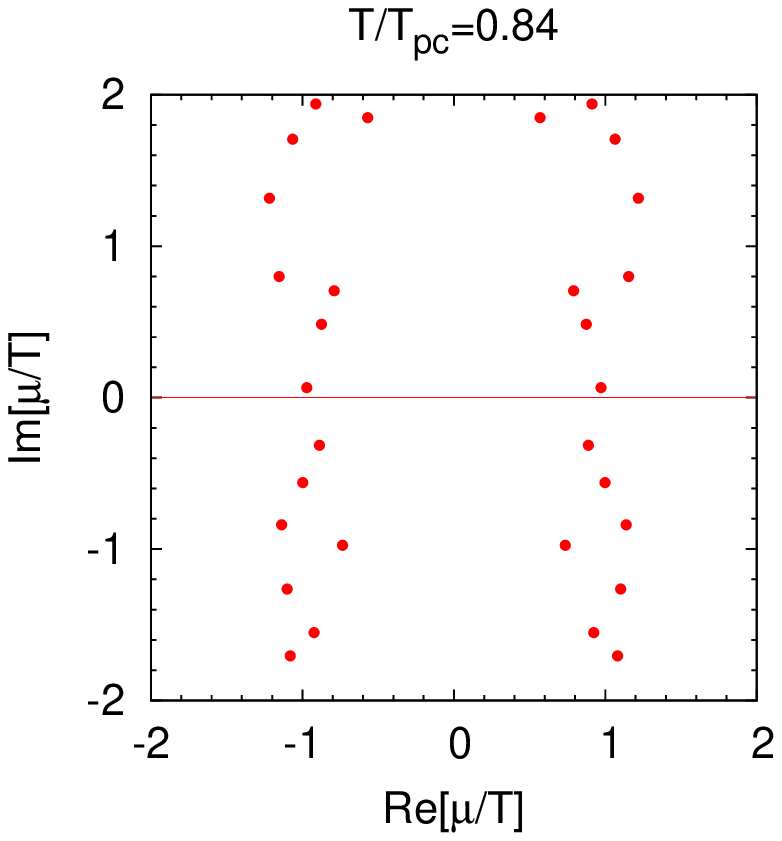}
\includegraphics[width=6.5cm]{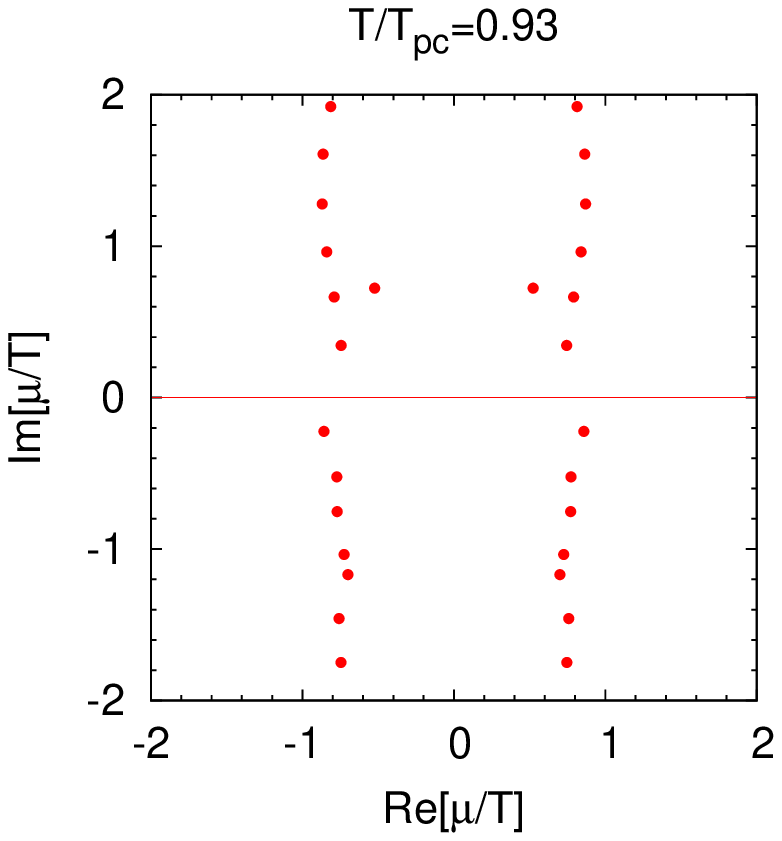}
\caption{The distribution of Lee-Yang zeros on the complex $\mu/T$ plane.}
\label{Fig:2012Mar14fig2}
\end{center}
\end{figure} 
In Fig.~\ref{Fig:2012Mar14fig2}, we have shown the Lee-Yang zero 
distribution for $T/\Tpc=0.84$ and $0.93$ in the complex $\mu/T$ plane. 
The results are invariant under $\mu/T\leftrightarrow -\mu/T$.  

In the Lee-Yang zero theorem, the phase transition point is 
obtained from the finite size scaling analysis of  the zero nearest the positive real axis.
In the present approach, we have made the assumption on $Z_n$. 
If the assumption is valid, then  Lee-Yang zeros are on the same trajectories. 
Then, the zeros would approach to the positive real axis as the volume increases 
in case of a phase transition. 
The phase transition point can be estimated in the canonical formalism, by 
using the Maxwell construction for the S-shape of the $\mu$-$n$ diagram~\cite{Kratochvila:2005mk,Li:2011ee}. 

\begin{figure}[htpb] 
\begin{center}                      
\includegraphics[width=7cm]{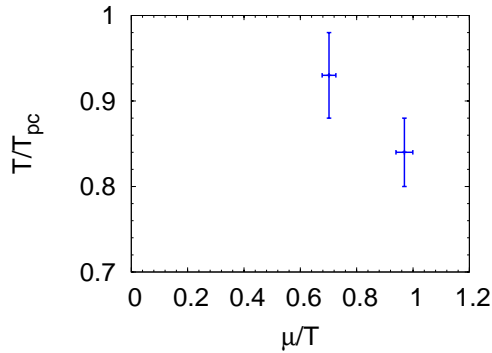}
\caption{A conjectured phase boundary for the first order phase transition. 
The data are obtained from Fig.~(\ref{Fig:2012Mar14fig2}) as the intersection of 
the linear fit and positive real axis with the assumption on $Z_n$.}
\label{Fig:2012Mar24fig1}
\end{center}
\end{figure} 

Following the assumption Eq.~(\ref{Eq:2012Mar24eq2}), we estimate
a phase transition point. 
Here we should note that the overlap problem causes a loss of reliability.  
From the linear fit $\Re[\mu/T]=$ const., we estimate the 
location of the intercept of the trajectory and real positive 
axis, and obtain 
\begin{align}
\mu_c/T=\left\{
\begin{matrix}
0.97(3) & T/\Tpc=0.84 \\
0.70(2) & T/\Tpc=0.93 
\end{matrix}\right.
\end{align}
The result is mapped onto the QCD phase diagram in Fig.~\ref{Fig:2012Mar24fig1}. 
Here the errorbars for $\mu_c/T$ were obtained from $\chi^2$ fit. 
We also estimated the errorbars for $T/\Tpc$, which is taken from 
\cite{Ejiri:2009hq}.
The result is almost consistent with previous studies with 
staggered fermions~\cite{Kratochvila:2005mk} at $\mu/T=0.7$, and 
undershoots those results at $\mu/T=0.97$. 

Here, it is important to consider the applicable limit of the Glasgow method. 
This can be done by using the imaginary chemical potential. 
The pure imaginary chemical potential region is located on the unit circle on the 
complex fugacity plane. 
The Roberge-Weiss (RW) endpoint should be located on $\Im[\mu/T]=\pi/3$. 
It turns out that the singularity on the unit circle appears about 
$\Im[\mu/T]\sim \pi/4$, which is smaller than the value expected from the RW endpoint. 
This would imply that the overlap problem becomes severe for $|\mu /T| >\pi/4$. 
The phase transition points in Fig.~\ref{Fig:2012Mar24fig1} are almost corresponds 
to this limit. In order to determine the phase transition point, the configurations 
should be improved to obtain a better overlap. 

In this section, we have presented an approach for the study of the QCD phase boundary. 
Although our analysis is at fundamental one, 
we found that the canonical distribution and Lee-Yang zero trajectory distinguish 
the crossover behavior at $\Tpc$ and first order behavior at low $T$. 
It is useful to consider the canonical partition function together 
with standard techniques for the phase transition, 
which may provides complementary information to identify the phase 
transition point. 

\begin{figure}[htpb] 
\begin{center}                      
\includegraphics[width=6cm]{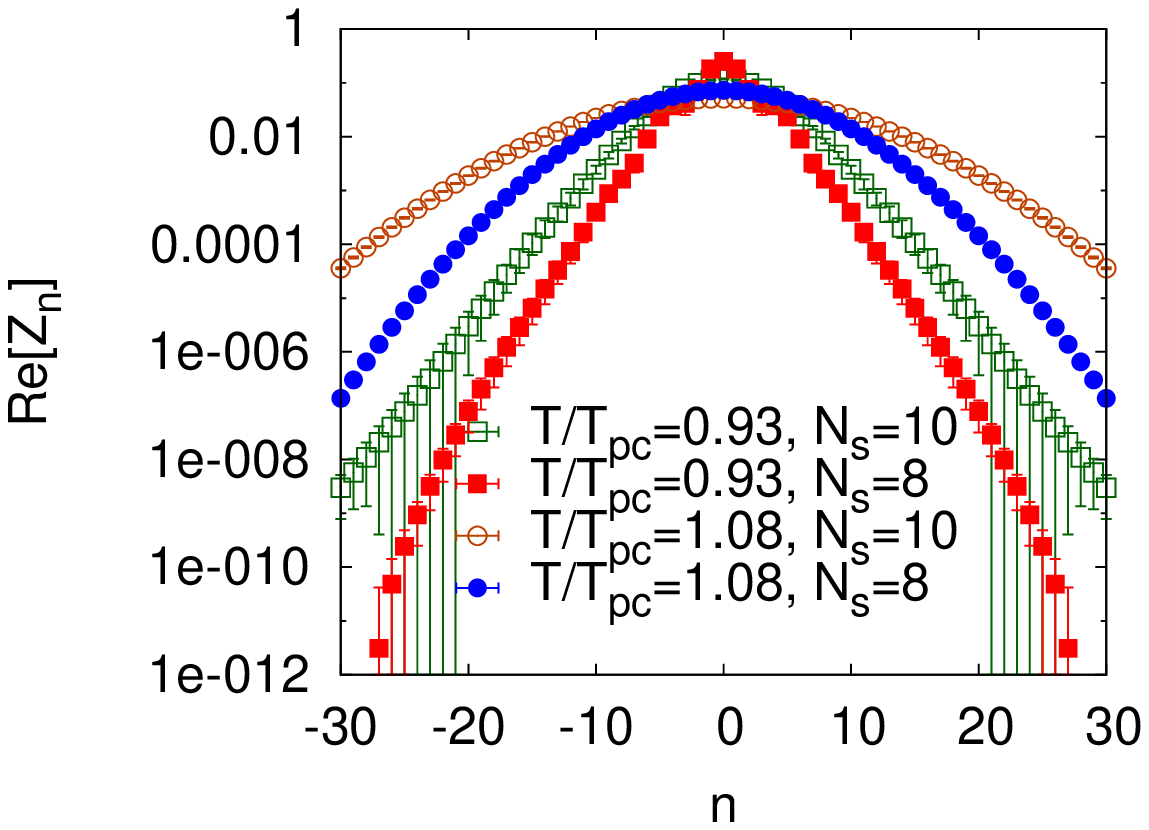}
\includegraphics[width=6cm]{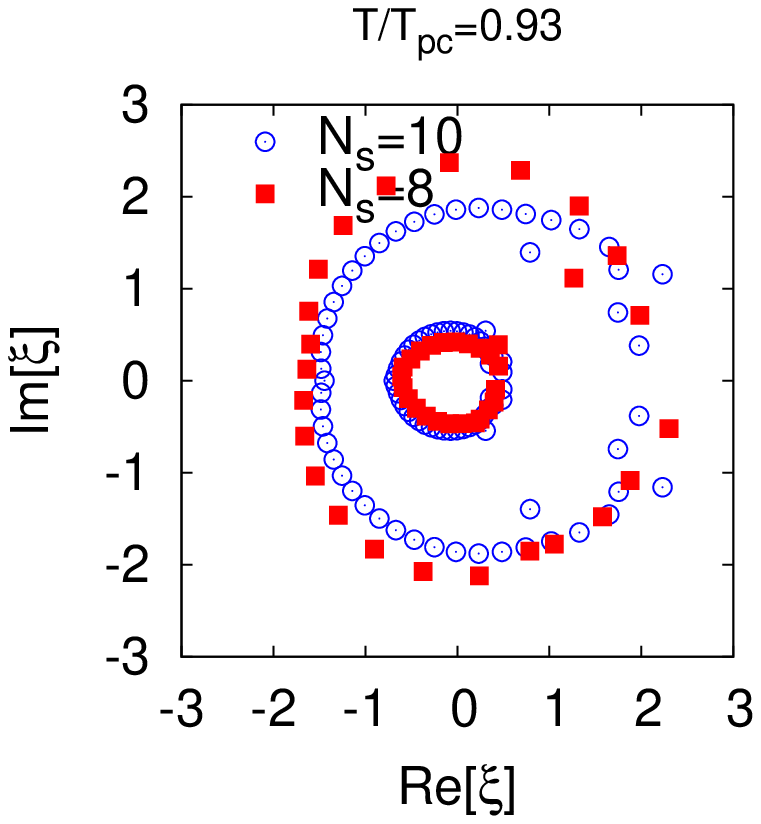}
\caption{Volume dependence of the canonical partition functions (left panel)
and Lee-Yang zero distribution (right panel).}
\label{Fig:2012Jun03fig1}
\end{center}
\end{figure} 
The assumption used in the present analysis should be examined further. 
In particular, the improvement of the overlap and the finite size effect 
on $Z_n$ are important task. 
We showed the results for $N_s=10$ in Fig.~\ref{Fig:2012Jun03fig1}.  
The increasing $N_s$ causes the broadening of the canonical partition 
functions. Although the eigenvalue distribution of the reduced matrix 
is insensitive to $N_s$, canonical partition functions are sensitive 
to $N_s$. 
Since $N_q$ for $N_s=10$ is twice as larger as that for $N_s=8$, 
we employed $M=48$ for $N_s=10$. 
The trajectory of the Lee-Yang zero was not affected largely 
by the increase of $N_s$. 
However, the lattice volume is still small for $N_s=10$, 
and the finite size effect may appear for larger lattices. 
The finite size effect should be investigated for larger lattices. 
Note that in Fig.~\ref{Fig:2012Jun03fig1}, the result for $N_s=8$ 
and $10$ were obtained with the same number of statistics. 
Since the sign problem becomes more severe for lager lattices, 
it is also important to increase the number of the statistics 
to study the finite size effect.

\section{Summary}

We have studied QCD at nonzero chemical potential and temperature
in the lattice QCD simulations. 
We particularly focused on the low temperature regions of the QCD phase diagram, 
and studied several issues with the use of the dimensional reduction formula of 
the fermion determinant. 

In \S~\ref{Sec:secred}, we studied the reduction formula of the Wilson 
fermion determinant and showed several properties of the reduced matrix. 
The reduced matrix is interpreted as the transfer matrices or the generalized 
Polyakov line, and the eigenvalues of the reduced matrix is related to the free 
energy of dynamical quarks. 
The angular distribution of the eigenvalues manifests the $Z_3$ or confinement 
properties of QCD. 
The eigenvalues form the gap, which is related to the pion mass and therefore chiral 
symmetry breaking. 
We found an indication that the eigenvalues of the reduced matrix 
is scaled by the temporal size as $\lambda\sim l^{N_t}$ for $|\lambda|>1$ 
and $\lambda\sim l^{-N_t}$ for $|\lambda|<1$. The $N_t$ scaling law 
controls the temperature dependence of the fermion determinant, and 
therefore is the important finding. 

In section \ref{Sec:2012Mar04sec2}, we studied the property of the fermion 
determinant at low $T$ and finite $\mu$. 
We showed from the lattice simulations that at $T/\Tpc\sim 0.5$ and for $\mu<m_\pi/2$, 
the fermion determinant is insensitive to $\mu$ and 
the average phase factor $\bra \cos \theta \ket$ approaches to one. 
The fluctuation of the reweighting factor, the ratio of the determinant at $\mu=0$ 
and $\mu\neq 0$, exponentially decreases as the temperature decreases. 
Using the $N_t$-scaling law, we also derived the $\mu$-independence of the 
fermion determinant at $T=0$ for $\mu<m_\pi/2$. Hence we concluded that 
the fermion determinant is $\mu$-independence at low $T$ and for $\mu<m_\pi/2$,  
which is the consequence of the $N_t$-scaling law of the eigenvalues of the 
reduced matrix. 

Extending the low temperature studies further, we have considered 
the low-temperature limit of the fermion determinant and of QCD, 
and obtained two expressions of the quark determinant; 
one is for low density and the other is for high density. 
Low density expression corresponds to $\mu$-independence for $\mu<m_\pi/2$. 
The other expression is its high-density counter part. 

We discussed the nature of the high density and low temperature limit of the QCD 
partition function. In the case, QCD approaches to a theory, 
where quarks interacts only in the spatial direction with the ordinary 
Yang-Mills type of the gauge action. 
The corresponding partition function is $Z_3$ invariant even in the 
presence of dynamical quarks. 
Furthermore, the fermion determinant becomes real and the theory is free 
from the sign problem. 

In \S~\ref{Sec:canonical}, we studied the canonical formalism and 
Lee-Yang zero theorem. 
We have shown that the canonical distribution and Lee-Yang zero trajectory 
show characteristic changes when $T$ is varied from high to low temperatures. 
The canonical distribution is similar to a Gaussian function of the quark number 
$n$ at high $T$, and to a geometric series with negative $n$ components. 
The Lee-Yang zero trajectory is a circle with two-branch at high $T$ and 
two circle at low $T$. 
The eigenvalue distribution was almost independent of the lattice size in our 
simulations on $8^3$ and $10^3$, which may suggest that the canonical 
partition function is insensitive to the spatial volume. 
Assuming the obtained $n$-dependence holds for larger $n$, we have shown that 
the $T$-dependence of the canonical distribution and Lee-Yang zero trajectory 
are related to the phase transition. 
The canonical distribution and Lee-Yang zero trajectory distinguish 
the crossover and first order phase transition. 
Hence the investigation of the canonical partition function and 
the Lee-Yang zeros may provides a way to distinguish the phase transition 
and fictitious signals caused from sign problem. 
It would be interesting to combine the present approach 
and the ordinary techniques such as the finite size scaling of the 
Lee-Yang zero, Maxwell construction of the canonical formalism, etc.

For confirmation of the results found in the present work, 
several points should be clarified further. 
Our analysis was performed on the small and coarse lattices with heavy quark 
mass. It is important to remove these lattice artifacts on the 
eigenvalues of the reduced matrix. 
The gap of the eigenvalues is sensitive to the quark mass. 
The tail of the eigenvalue distribution is sensitive to the quark mass 
and lattice spacing. 
These quantities are related to the pion mass and highest energy level, 
which determine the critical values of $\mu$ for the low temperature limits. 
It is also important to examine the $N_t$-scaling law for larger temporal 
lattice size. 

It is also important to investigate the lattice artifacts on the canonical 
partition function. In addition, the finite-size scaling and the improvement 
of the overlap are necessary task to identify the phase transition point. 
In particular, the large quark number sector of the canonical partition function 
plays an important role in the phase transition. 
For the improvement of the overlap, a multi-ensemble reweighting or histogram 
method may be useful~\cite{Saito:2011fs,Nakagawa:2011eu}.

With further confirmations, it will be interesting to study the thermodynamical 
properties of QCD at low temperatures. 

\section*{Acknowledgements}
We thank S. Ejiri, Ph. de Forcrand, S. Hashimoto, M. Hanada and H. Matsufuru for 
valuable comments and stimulating discussions. 
Parts of this work were done during the stay at the workshop ``New Type of Fermions 
on Lattice'' held at YITP. We thank to A. Ohinishi, T. Misumi, D. Adams, C. Hoelbling
for stimulating discussions and valuable information.
KN specially thanks to T. Hatsuda for the hospitality and encouragement during 
the stay at Tokyo University. 
AN acknowledges the hospitality and discussions to M. Yahiro at Kyushu university. 
This work was supported by Grants-in-Aid for Scientific Research 20340055, 20105003, 
23654092 and 20105005. 
The simulation was performed on NEC SX-8R at RCNP, NEC SX-9 at CMC, Osaka University,  
and System A and System B at KEK. 

\appendix
\section{Miscellaneous for reduction formula}

\subsection{Pair nature of the eigenvalues}
\label{Sec:2012Mar03sec1}

We show a proof of a pair nature of the eigenvalues of the reduced matrix.
The pair nature is a consequence of the symmetry of the reduced matrix, 
which was shown in Ref.~\cite{Alexandru:2010yb}.
Here we present a brief proof. 

Substituting  Eq.~(\ref{Nov292011eq1}) into the $\gamma_5$-hermiticity relation 
leads to 
\begin{align}
 (\xi^*)^{-\frac{\Nred}{2}} \prod_{n=1}^{\Nred}(\lambda_n^* + \xi^*)
= (\xi^*)^{\frac{\Nred}{2}}\prod_{n=1}^{\Nred}(\lambda_n + (\xi^*)^{-1}). 
\label{Jan1111eq2}
\end{align}
Since the $\gamma_5$ hermiticity holds for $\forall \mu\in \mathbb{C}$, 
Eq.~(\ref{Jan1111eq2}) also holds for $\forall \xi\in \mathbb{C}$.

Let $\xi_0$ a value of fugacity which set l.h.s of Eq.~(\ref{Jan1111eq2}) 
to be zero, i.e., $\xi_0 = -\lambda_n$.
If l.h.s is zero, then r.h.s also must be zero. 
Hence, an eigenvalue for l.h.s $=0$ must exist for $\xi=\xi_0$. 
This is satisfied by $\lambda + (\xi_0^*)^{-1}=0$. 
Thus, two eigenvalues
\begin{align}
\lambda_n, \frac{1}{\lambda_n^*} \nn
\end{align}
appears at the same time. 
This procedure is applied to all the eigenvalues 
$\lambda_n, (n=1, \cdots \Nred)$. Thus, the pair nature of
the eigenvalues of the transfer matrix is proved. 

\subsection{Z$_3$ Properties of Fugacity coefficients}
Next, we consider the property of the fugacity coefficients $c_n$
under center transformation $\ZNc$, 
\begin{align}
U_4(\vec{x},t_i) \to \omega U_4(\vec{x},t_i),  \exists t_i, \forall \vec{x},
\end{align} 
where $\omega = \exp( 2\pi i k / 3), \;\; (k = \pm 1)$ is an element of $Z_3$. 
In Eq.~(\ref{Eq:2012Jan01eq5}), $C_0$ is invariant under $\ZNc$, since it 
does not contain temporal link variables. 
From Eqs.~(\ref{Eq:2012Jan01eq3}) and (\ref{Eq:2012Feb21eq1})
$Q$ transforms as $Q\to \omega Q$ under $Z_3$. 
The fermion determinant is transformed as 
\begin{align}
\det \Delta(\mu) & = C_0 \xi^{-\Nred/2} \det ( \omega Q + \xi) , \nn \\
                 & = C_0 \xi^{-\Nred/2} \det ( Q + \omega^{-1}\xi), \nn \\
                 & = C_0 \xi^{-\Nred/2} \sum_{n=0}^{\Nred} c_n \omega^{-n} \xi^n, \nn\\
                 & = C_0 \sum_{n=-\Nred/2}^{\Nred/2} c_n \omega^{-n} \xi^n
\label{Eq:2012Mar31eq1}
\end{align}
Then, the fugacity coefficients $c_n$ transforms $c_n \to c_n \omega^{-n}$ 
under the $Z_3$ transformations. 
$c_n$ is invariant for the triality sector $n=3m(m\in\mathbb{Z})$ 
and covariant for the triality nonzero sector $n= 3m + 1, 3m + 2$.

\section{Properties of canonical partition function}
\label{Sec:App_triality}

\subsection{RW invariance and Triality}

Consider the Roberge-Weiss transformation~\cite{Roberge:1986mm}. Temporal links are transformed under $Z_3$
\begin{align}
U_4(\vec{x}, t_i) &\to U_4^\prime(\vec{x}, t_i) = \omega U_4 (\vec{x}, t_i),  \;\; ^\exists t_i, ^\forall\vec{x}, 
\end{align}
and the chemical potential is shifted in the imaginary direction as 
\begin{align}
\frac{\mu}{T} \to \frac{\mu^\prime}{T} = \frac{\mu}{T} - \frac{2\pi i k}{3},  \\
\end{align}
which acts on the fugacity as the rotation $(\xi  \to \xi \omega)$.
It is obvious from Eq.~(\ref{Eq:2012Mar31eq1}) that the transformations 
for the link variables and chemical potential cancel. 
The fermion determinant is invariant under the RW transformation,
\begin{align}
\det \Delta(\mu, \{U\}) &\to \det \Delta(\mu^\prime, \{U\}^\prime) =\det \Delta(\mu, \{U\})
\end{align}
The grand partition function is also invariant  under the RW transformation,
\begin{align}
Z(\mu) \to Z(\mu^\prime) &= \int DU^\prime \det \Delta(\mu^\prime, \{U\}^\prime) e^{-S_G(U^\prime)}, \nn \\
 &= \int DU \det \Delta(\mu, \{U\}) e^{-S_G(U)} \nn, \\
  &= Z(\mu). 
\label{Eq:2012Mar01eq1}
\end{align}

An important property on the canonical partition function is deduced from 
this invariance~\cite{Barbour:1991vs,Kratochvila:2005aaa}. 
For simplicity we set $\mu$ to be pure imaginary. 
Expressing the fermion determinant as the fugacity polynomial, 
the grand partition function is given as
\begin{align}
Z_{GC}(i\mu_I) = \int DU \sum_{n=-M}^M c_n \xi^n e^{S_g},
\label{Jan2911eq1}
\end{align}
and 
\begin{align}
Z_{GC}(i \muit + i \frac{2k\pi}{3})  = \int DU \sum_{n=-M}^M c_n \xi^n \omega^n  e^{S_g}.
\end{align}
Here note that the maximum quark number $M$ is proportional to the spatial 
volume $V$ and diverges at thermodynamical limit, which may spoils this 
discussion near phase transition points. 
Assuming the analiticity, Eq.~(\ref{Eq:2012Mar01eq1}) leads to
\begin{align}
\left( 1 -  \exp( i \frac{2n k\pi}{3} ) \right) \int DU c_n  e^{S_g} =0 .
\end{align}
Thus, we obtain
\begin{align}
Z(n)=\int DU c_n  e^{S_g} =0  \;\; (\mod(n,3)\neq 0).
\label{Eq:2012Mar01eq2}
\end{align}
Hence the canonical partition function must vanish for the triality nonzero sector. 
The above argument is applied to arbitrary number of flavors, which 
can be obtained by changing $M$. 

\subsection{Canonical partition function for triality nonzero sectors}
We have seen the properties of the canonical partition function 
Eq.~(\ref{Eq:2012Mar01eq2}). 
However, we showed that the canonical partition functions do not 
vanish for the triality nonzero sector, see Fig.~\ref{Fig:2012Feb03fig1}.

This small paradox is caused by the importance sampling at $\mu=0$, where 
configurations for one $Z_3$ sector are collected in the presence 
of the quarks. Effects of the non-zero triality sector was investigated in 
Ref.~\cite{Kratochvila:2006jx}.

Here, we consider this point. 
First, we classify the configuration space into three regions according to 
the location of the Polyakov loop $(L)$ on the complex $L$ plane;
\begin{align}
R_1&=\{U | -\pi/3 \le \arg(L) < \pi/3\}, \\
R_2&=\{U | \;\; \pi/3 \le \arg(L) < \pi \},  \\
R_3&=\{U |-\pi \le \arg(L) < -\pi/3\}. 
\end{align}
The grand partition function is written as 
\begin{align}
Z(\mu) = \left(\int_{R_1} + \int_{R_2} + \int_{R_3}\right) DU \det \Delta(\mu) e^{-S_g}
\end{align}
As we have already seen, whole the $Z(\mu)$ is invariant under the 
Roberge-Weiss transformation. 
Let us consider the transformation property of each component. 
Let $I_j$ to be one of $Z_3$ component of the partition function, 
\begin{align}
I_j = \int_{R_j} DU \det \Delta(\mu) e^{-S_g}.
\end{align}
Under the Roberge-Weiss transformation, the Polyakov loop $L$ is rotated in the 
complex $L$ plane, then the configurations $\{U\}$ move from one of $R_i$ 
sector to another $R_{i+1}$. Therefore, $I_j$ transforms as
\begin{align}
I_j(\mu) = I_{j+1}(\mu^\prime), (I_4 = I_1).
\end{align}
Therefore, each $I_j$ is not invariant under the Roberge-Weiss transformation. 

In the importance sampling at $\mu=0$ in the presence of the 
quarks, configurations for one $Z_3$ sector are extensively collected, 
which cover a part of the configuration space, e.g. $R_1$. 
This causes the non vanishing of the triality nonzero sectors in the 
canonical partition function. 

In order to satisfy the triality is to include other $Z_3$ sectors by using 
the RW transformation,
\begin{align}
Z(\mu) &= I_1 (\mu) + I_2(\mu) + I_3(\mu) \nn\\
       &= I_1 (\mu) + I_1(\mu+2\pi i T /3) + I_1(\mu + 4\pi i T /3) \nn \\
   &= \int_{R_1} DU \sum_n c_n \xi^n (1 + \omega^{-n} + \omega^{-2n}) e^{-S_g} \nn \\
 &= \int_{R_1} DU \sum_{n,\mod(n,3)=0} 3 c_n \xi^n  e^{-S_g},
\label{May2311eq1}
\end{align}
which contains only $\mod(n,3)=0$ terms. 
\section{Calculation of coefficients $c_n$}\label{App:WideRangeNum}

The fugacity coefficients $c_n$ of the fermion determinant can be obtained
in the reduction formula,
\be
\det\Delta (\mu) = C_0 \xi^{-N_{red}/2} \prod (\lambda_n + \xi)
= C_0 \sum_{n=-N_{red}/2}^{N_{red}/2} c_n \xi^n .
\label{Eq:AppCn01}
\ee
Their values  vary 
from order one to order $10^{900}$ even on the small $4^4$ lattice. They 
cannot be handled in the double precision. 
 
The simplest way is to use an  arbitrary accuracy libraries.  
This is more than necessary.
To express $c_k$, we need wide range of
floating numbers, but we do not need very high
precision.
In other words, we need wide range of the exponent, but
we do not need very huge significant numbers.  

We express each real and imaginary parts of $c_k$ in a form of
\be
a \times b^L ,
\ee
where
\be
1 \le |a| < b ,
\ee
and $a$ is a double precision real and $L$ is an integer.
We employ the ``module'' in Fortran 90, which allows us to define
a new type of data and mathematical operation among them.
The base $b$ can be any number, and we set it to be 8.

There are several way to get $c_n$ in Eq.(\ref{Eq:AppCn01}).
The simplest way is to use in a recursive way:
\be
\sum_{k=0}^M C_k'\xi^k
= 
(B_0 + B_1 \xi ) \sum_{k=0}^{M-1} C_k\xi^k
\ee
and
\bea
C_0' &=& B_0 C_0
\nn \\
C_k' &=& B_{k-1} C_k + B_k C_{k-1}
\quad  (k = 1, 2, \cdots, M-1)
\nn  \\
C_M' &=& B_1 C_{M-1}
\label{Eq:recursive}
\eea
We calculate several
cases by this method, and by a high accuracy library, FMLIB\cite{Web:FMLIB}.
We got the same results.

A smarter way is a divide-and-conquer method. See (\ref{Algo-DC}).
Here for simplicity we assume $N=2^M$, but this can be loosened
$c \leftarrow  c1 \times c2$ is an operation to determine $c$ from $c_1$
and $c_2$ where
\bea
&&(c(0) + c(1)*x + ... + c(N_3)*x^N_3)
\leftarrow
\nn
\\
&&(c_1(0) + c_1(1)*x + ... + c_1(N_1)*x^{N_1})
\\
&&  \times (c_2(0) + c_2(1)*x + ... + c_2(N_2)*x^{N_2})
\label{Eq:AppCn2}
\eea
and $N_3=N_1+N_2$.

\begin{algorithm}[h]
\caption{Divide-Conquer calculation for the coefficients $c_n$ for
$\prod_{i=1}^{N} (a(i)+x) = \sum_{k=0}^N c(k)*x^k$
}
\label{Algo-DC}
\begin{algorithmic}
\STATE Input $a$, Output $c$)
\STATE Set N
\STATE read a
\STATE CALL DandQ ( a, c, N)
\STATE Recursive SUBROUTINE DandQ ( a, c, N)
\STATE \quad   IF N corresponds the Bottom, RETURN
\STATE \quad   CALL DandQ (a(1), c1, N/2)
\STATE \quad   CALL DandQ (a(N/2+1), c2, N/2)
\STATE \quad   $c \leftarrow  c1 \times c2$
\STATE \quad RETURN
\STATE \quad END SUBROUTINE  
\end{algorithmic}
\end{algorithm}

When $N$ is large (i.e., we are near to the ``top'' of the recursive level) , 
Eq.(\ref{Eq:AppCn2}) is easy to vectorize or
parallelize.
For small $N$ (i.e., we are near to the ``bottom'' of the recursive level) , 
we handle many calculations of the type of Eq.(\ref{Eq:AppCn2}) , then
it is also easy to vectorize or parallelize.

  


\end{document}